\documentclass[a4paper,fleqn,usenatbib]{mnras}

\usepackage{newtxtext,newtxmath}
\usepackage[T1]{fontenc}
\usepackage{ae,aecompl}

\usepackage{graphicx}	% Including figure files
\usepackage{amsmath}	% Advanced maths commands
\usepackage{amssymb}	% Extra maths symbols

\newcommand{\kms}{ km s$^{-1}$}

\newcommand{\II}{~{\sc ii}}
\newcommand{\I}{~{\sc i}}
\newcommand{\III}{~{\sc iii}}
\newcommand{\IV}{~{\sc iv}}

\title{Quasar 2175 \AA$ $ dust absorbers I: metallicity, depletion pattern, and kinematics}

\author[Jingzhe Ma, Jian Ge, Yinan Zhao, J. Xavier Prochaska, Shaohua Zhang, Tuo Ji, Donald P. Schneider]{Jingzhe Ma$^{1}$\thanks{E-mail:jingzhema@ufl.edu (JM)}, Jian Ge$^{1}$, Yinan Zhao$^{1}$, J. Xavier Prochaska$^{2}$, Shaohua Zhang$^{3}$, 
\newauthor Tuo Ji$^{3}$,  Donald P. Schneider$^{4,}$$^{5}$\\
$^{1}$Department of Astronomy, University of Florida, 211 Bryant Space Science Center, Gainesville, 32611, USA\\
$^{2}$Department of Astronomy and Astrophysics, UCO/Lick Observatory, University of California, 1156 High Street, Santa Cruz, 95064, USA\\
$^{3}$Polar Research Institute of China, 451 Jinqiao Road, Pudong, Shanghai 200136, China\\
$^{4}$Department of Astronomy and Astrophysics, The Pennsylvania State University, University Park, PA 16802, USA\\
$^{5}$Institute for Gravitation and the Cosmos, The Pennsylvania State University, University Park, PA 16802, USA\\
}

\begin{document}
\maketitle

\begin{abstract}

We present 13 new 2175 \AA$ $ dust absorbers at $z_{abs}$ = 1.0 - 2.2 towards background quasars from the Sloan Digital Sky Survey.  These absorbers are examined in detail using data from the Echelle Spectrograph and Imager (ESI) on the Keck II telescope.  Many low-ionization lines including Fe\II, Zn\II, Mg\II, Si\II, Al\II, Ni\II, Mn\II, Cr\II, Ti\II, and Ca\II{} are present in the same absorber which gives rise to the 2175 \AA$ $ bump.  The relative metal abundances (with respect to Zn) demonstrate that the depletion patterns of our 2175 \AA$ $ dust absorbers resemble that of the Milky Way clouds although some are disk-like and some are halo-like. The 2175 \AA$ $ dust absorbers have significantly higher depletion levels compared to literature Damped Lyman-$\alpha$ absorbers (DLAs) and subDLAs. The dust depletion level indicator [Fe/Zn] tends to anti-correlate with bump strengths. The velocity profiles from the Keck/ESI spectra also provide kinematical information on the dust absorbers. The dust absorbers are found to have multiple velocity components with velocity widths extending from $\sim$ 100 to $\sim$ 600 \kms, which are larger than those of most DLAs and subDLAs. Assuming the velocity width is a reliable tracer of stellar mass, the host galaxies of 2175 \AA$ $ dust absorbers are expected to be more massive than DLA/subDLA hosts. Not all of the 2175 \AA$ $ dust absorbers are intervening systems towards background quasars. The absorbers towards quasars J1006+1538 and J1047+3423 are proximate systems that could be associated with the quasar itself or the host galaxy.

\end{abstract}


\section{Introduction}

Quasars have been used as a powerful tool to probe the absorbing gas in clouds and galaxies along the lines of sight.  Quasar absorption line spectroscopy enables a study of the physical conditions, chemical abundances, and gas kinematics in the absorption systems.  

Most absorption line systems studied in the past are single-element absorbers due to transitions of a certain element in the gas phase, such as Mg\II{} absorbers, Damped Lyman-$\alpha$ systems (DLAs), C\IV{} absorbers, Ca\II{} absorbers etc. Thanks to the large surveys of quasars in this era (e.g., multiple data releases of the Sloan Digital Sky Survey (SDSS)), a new population of quasar absorbers, which is not due to gas absorption but dust, has been identified and selected. We observe a broad absorption feature in the quasar spectra centered around 2175 \AA$ $ in the absorber's rest-frame. This bump feature is ubiquitously seen in the Milky Way's interstellar extinction curves. The extinction curves along the sight lines of the Large Magellanic Cloud (LMC) also exhibits such a bump feature but the absorption is less significant than in the MW, whereas the canonical Small Magellanic Cloud (SMC) extinction curve is free of the bump feature. The different environmental conditions may affect the formation and destruction of the 2175 \AA$ $ extinction bump. The origin and the carriers of the bump have been a mystery for decades since its first discovery over fifty years ago \citep{Stecher1965}, although some candidates, such as polycyclic aromatic hydrocarbons (PAHs; \citealt{Draine2003}), were proposed. Dust is a critical component in galaxy formation and evolution, and these dust absorbers may provide insights of how galaxies evolve to the Galaxy we live in today, and even to life. 

Previous detections are mostly found in individual intervening absorbers towards background quasars including Mg\II{} absorbers (e.g., \citealt{Wang2004, Srianand2008, Zhou2010, Jiang2011, Zhang2015}) and DLAs (e.g., \citealt{Wucknitz2003, Junkkarinen2004, Wang2012}), or towards gamma-ray burst (GRB) afterglows (e.g., \citealt{Prochaska2009, Eliasdottir2009, Zafar2012}). \cite{Ledoux2015} report in a study of neutral atomic-carbon (C\I) gas that the 2175 \AA$ $ bump feature is detected in 30\% of the C\I{} absorbers. 

Our searches through the SDSS through Data Release 12 \citep{Alam2015} result in hundreds of 2175 \AA$ $ dust absorber candidates (Zhao et al. 2017 in prep). We have conducted follow-up observations on a subset with higher resolution spectrographs including the Echelle Spectrograph and Imager (ESI) on the Keck-II telescope and the Ultraviolet and Visible Echelon Spectrograph (UVES) on the Very Large Telescope (VLT) to characterize their physical and chemical properties. A comprehensive study of one of the strong 2175 \AA$ $ dust absorbers, J1211+0833, shows that neutral carbon, neutral chlorine, and carbon monoxide are simultaneously present in the dust absorber \citep{Ma2015}. This absorber is characterized as  a metal-rich and highly depleted absorption system, which supports the scenario that the presence of the 2175 \AA$ $ bump requires an evolved stellar population. 

In this paper, we present 13 new 2175 \AA$ $ dust absorbers at $z$ = 1.0 - 2.2 observed with the Keck/ESI. This is a pilot sample study on the chemical abundances, depletion patterns, and gas kinematics of quasar 2175 \AA$ $ dust absorbers. We start using the acronym ``2DAs" for 2175 \AA$ $ dust absorbers hereafter. Section \ref{sec:observations} describes target selection, observations, and data reduction. In Section \ref{sec:bump} we introduce the methodology and algorithms that we used to extract bump parameters, e.g. bump strength and bump width.  We present the measurements of gas-phase column densities and relative metal abundances in Section \ref{sec:metal}. Comments on individual absorbers are given in Section \ref{sec:comments}.  We discuss the results and compare with the literature in Section \ref{sec:discussion}. The conclusions are summarized in Section \ref{sec:conclusion}. In a companion paper (Paper II; Ma et al. 2017, in prep), we will specifically discuss the absorbers  where Lyman-$\alpha$ and C\I{} absorption lines are detected, including a few sources from this sample. 

\section{Target selection, observations, and data reduction}
\label{sec:observations}

\begin{table*}
\centering
\caption{Journal of Observations}
\begin{tabular}{ccccccccccc}
\hline\hline
QSO        & R.A.        & Dec.       & $r$ mag    & $z_{em}$  & $z_{abs}$  &  UT Date     & Instr  & Slit     & $t_{exp}$ & S/N           \\
           & (J2000)     & (J2000)    & (AB)     &           &            &       &        & (arcsec) & (s)       & (pixel$^{-1}$)\\
(1)        & (2)         & (3)        & (4)      & (5)       & (6)        & (7)       & (8)    & (9)      & (10)      & (11)          \\
\hline
J0745+4554 &07 45 28.09  &45 54 06.28 & 19.81    &  2.1998   & 1.8612     &2013 Mar 9  &ESI    & 0.75     & 3600      & 12            \\  
J0850+5159 &08 50 42.24  &51 59 11.65 & 18.96    &  1.8939   & 1.3269     &2013 Mar 9  &ESI    & 0.75     & 2400      & 23            \\      
J0901+2044 &09 01 22.67  &20 44 46.53 & 17.76    &  2.0934   & 1.0191     &2013 Mar 8,9&ESI    & 0.75     & 5400      & 55            \\
J0953+3225 &09 53 27.95  &32 25 51.56 & 17.36    &  1.5748   & 1.2375     &2013 Mar 8,9&ESI    & 0.75     & 1200      & 32            \\
J1006+1538 &10 06 30.36  &15 38 43.88 & 19.96    &  2.1950   & 2.2062     &2013 Mar 9  &ESI    & 0.75     & 3600      & 10            \\
J1047+3423 &10 47 00.96  &34 23 37.31 & 19.14    &  1.6800   & 1.6685     &2013 Mar 8,9&ESI    & 0.75     & 3000      & 15            \\
J1127+2424 &11 27 21.09  &24 24 17.14 & 17.79    &  2.0787   & 1.2110     &2013 Mar 8,9&ESI    & 0.75     & 1200      & 23            \\
J1130+1850 &11 30 47.93  &18 50 55.86 & 20.61    &  2.7536   & 2.0119     &2013 Mar 8,9&ESI    & 0.75     & 3600      & 4             \\
J1138+5245 &11 38 36.93  &52 45 11.39 & 18.88    &  1.6337   & 1.1788     &2013 Mar 8,9&ESI    & 0.75     & 2400      & 13            \\
J1211+0833 &12 11 43.42  &08 33 49.71 & 19.47    &   2.4828  & 2.1166	  &2013 Mar 9 	&ESI     &	0.75     & 3300     &    9              \\
                     &                     &                    &             &                 &                &2014 Mar 25, Apr 24 & UVES & 0.90 & 17400 & 6           \\
J1212+2936 &12 12 19.87  &29 36 22.84 & 18.11    &  1.3919   & 1.2202     &2013 Mar 8,9&ESI    & 0.75     & 1500      & 27            \\ 
J1321+2135 &13 21 40.49  &21 35 49.07 & 20.69    &  2.4113   & 2.1253     &2013 Mar 8,9&ESI    & 0.75     & 4200      & 8             \\
J1531+2403 &15 31 31.52  &24 03 33.16 & 19.87    &  2.5256   & 2.0022     &2013 Mar 9  &ESI    & 0.75     & 1500      & 8             \\
J1737+4406 &17 37 04.86  &44 06 03.43 & 19.26    &  1.9564   & 1.6135     &2013 Mar 9  &ESI    & 0.75     & 1200      & 10            \\
\hline
\end{tabular}
\label{tab:obs}
\end{table*}

This sample comprises 14 2DAs (13 new absorbers from this work and one from \citealt{Ma2015}) selected from the SDSS-I/II \citep{York2000, Schneider2010} and SDSS-III/BOSS (\citealt{Eisenstein2011, Ross2012, Dawson2013}) using a similar searching algorithm described in \cite{Jiang2011} and \cite{Ma2015}. This is a subset of a larger Mg\II{} absorber sample (\citealt{Zhu2013}; Zhao et al. 2017 in prep). Briefly, we employ a composite quasar spectrum \citep{Jiang2011} as the intrinsic quasar spectrum, which is then reddened by a parametrized extinction curve to form a model spectrum. The bump fitting was performed on the SDSS spectra which cover a wavelength range of $\sim$ 3800 -- 9200 \AA$ $ for SDSS-I/II or  $\sim$ 3600 -- 10000 \AA$ $  for BOSS with a spectral resolution of R $\sim$ 2000. 

Follow-up observations were conducted with Keck/ESI (\citealt{Sheinis2002}) on the Keck II telescope. The 14 quasar absorbers selected for the Keck observations satisfy the following criteria: (1) evenly spaced in the redshift range of $\sim$ 1 - 2, (2) bump strengths $A_{\rm bump}$ $\geq$ $\sim$ 0.2 $\mu$m$^{-1}$, (3) observable by Keck. Table \ref{tab:obs} presents an observation log describing the 14 quasars in our sample. Column 1 gives the quasar name based on its SDSS coordinate. Columns 2 and 3 list the R.A. and Dec. of the quasar in J2000. Column 4 is the AB $r$ magnitude of the quasar obtained from the SDSS imaging. Columns 5 and 6 give the emission and absorption redshift of the quasar, respectively. The absorption redshift is determined by the Mg\II{} doublet and Fe\II{} absorption lines.  Column 7 lists the UT date of the observation. Columns 8 and 9 provide the instrument and slit used in the observation. Columns 10 and 11 give the exposure time and median signal-to-noise (S/N) ratio per pixel measured around $\sim$ 6500 \AA$ $. 

Throughout the ESI observations, we used the 0.75" slit corresponding to a resolution of FWHM $\sim$ 44 \kms (R $\approx$ 7000). ESI covers the wavelength range $\lambda$ = 3900 -- 11715 \AA$ $. All the spectra were reduced using the XIDL package developed by J.X. Prochaska and J.F. Hennawi, which is publicly available \footnote{http://www.ucolick.org/$\sim$xavier/IDL/}. We performed continuum fitting using the $x\_continuum$ routine within XIDL. 

For the absorber towards J1211+0833, we also obtained high resolution ($R$ $\sim$ 54000) VLT/UVES spectrum. The observational details are provided in \cite{Ma2015}.

\section{Quasar 2175 \AA$ $ dust absorbers}
\label{sec:bump}

\subsection{Composite quasar spectrum method}

This sample was initially selected using the searching algorithm described in  \cite{Jiang2011} and \cite{Ma2015}.  An observed quasar spectrum is modeled by an intrinsic quasar spectrum, reddened by a parameterized extinction curve at the absorber's redshift. We employ the SDSS DR7 composite quasar spectrum \citep{Jiang2011} as the intrinsic spectrum, and we adopt the parameterized extinction curve of \cite{FM1990} and \cite{FM2007} (FM). The FM parameterization is given in the following form, 

\begin{equation}
 A(\lambda)=c_1+c_2x+c_3D(x,x_0,\gamma),   
\end{equation}

\noindent where $x$ = $\lambda^{-1}$ in units of $\micron^{-1}$. Note that a second-order term in the original FM parameterization is not included here. The above parameterized extinction curve comprises two components and five parameters: (1) an underlying UV linear component set by the slope $c_2$ and intercept $c_1$; (2) a Drude profile scaled by the parameter $c_3$, representing the 2175 \AA$ $ bump. The Drude profile can be expressed as               

\begin{equation}
D(x,x_0,\gamma)=\frac{x^2}{(x^2-x_0^2)^2+x^2\gamma^2},
\end{equation}

\noindent  where $x_0$ and $\gamma$ are the bump peak position and full width at half maximum (FWHM) of the Drude profile, respectively. The strength of the bump is defined as the area underneath the curve above the linear component: $A_{\rm bump}=\pi c_3/(2\gamma)$\footnote{$A_{\rm bump}$ defined here is related to $A_{\rm bump}^*$ defined in \cite{FM2007} via this conversion: $A_{\rm bump}$ = $E(B-V)$$A_{\rm bump}^*$. }.  

The observed spectra are first corrected for Galactic extinction using the dust map of \cite{Schlegel1998} and converted to the absorbers' rest-frame. The best-fit bump parameters are derived by a $\chi^2$ minimization approach. Strong emission lines such as Ly$\alpha$, Si\IV, C\IV, C\III, and Mg\II{} and known absorption lines are masked to optimize the fitting. The initial values used in the Drude profile are based on the parameters of the MW extinction curves with a tolerance range of 4.4 $\micron^{-1}$ $\leq$ $x_0$ $\leq$ 4.8 $\micron^{-1}$ for the central/peak position, and 0.5 $\micron^{-1}$ $\leq$ $\gamma$ $\leq$ 2.7 $\micron^{-1}$ for the bump width. The boundaries are set by the distribution of the bump parameters of the MW and LMC2 supershell region (\citealt{Gordon2003,FM2007}). 

The intrinsic variation of the quasar continuum and pseudo-continuum composed of many Fe\II{} emission lines  can sometimes mimic a bump \citep{Pitman2000}. To gauge the significance of the bump, we construct a control sample of quasar spectra (sample size $\sim$ 1000) at a similar redshift to the quasar of interest ($z_{\rm qso}$ -0.05 $<$ $z$ $<$ $z_{\rm qso}$ + 0.05) and i-band S/N $\geq$ 5 \citep{Jiang2011}. We perform the same fitting procedure for each quasar spectrum in the control sample. We define 3$\sigma$ of the bump strength distribution as the bump detection threshold. All the bumps in this sample are detected at $>$ 3$\sigma$. The best-fit extinction curves can be used to estimate the total amount of extinction at V band, A$_V$. We normalize the extinction curves by forcing extinction to be zero at $\lambda$ $\to$ +$\infty$, which is physically reasonable. 

\subsection{Two-step Method}

The above searching algorithm is efficient and sufficiently fast to select bump candidates from over 100,000 quasar spectra. To better quantify the uncertainties in the derived bump parameters, we use the PyMC package \footnote{https://pymc-devs.github.io/pymc/} \citep{Patil2010, Fonnesbeck2015} to implement the Bayesian approach via Markov Chain Monte Carlo (MCMC) sampling for this sample. Priors are given to the 5 parameters according to the distribution of the MW and LMC bumps. Simultaneously fitting all the 5 parameters generally yields good solutions except in the case where there are not enough continuum data covering both sides of the bump. This approach would result in extremely wide and shallow bumps (i.e., almost the entire observed spectrum considered as a bump) or wide and extremely deep bumps (i.e., steep linear component) due to the continuum not being constrained blueward of the bump. To minimize the uncertainties induced by this procedure, we adopt a two-step method: first we fit $c_1$, $c_2$, and $c_3$ by fixing $x_0$ and $\gamma$ to the MW values then the best-fit $c_1$, $c_2$, $c_3$ values determined from the first step are fixed, and we derive the best-fit for $x_0$ and $\gamma$ in the second step. We refer the reader to Zhao et al. 2017 in prep for more discussion on bump fitting algorithms. 

\subsection{Comparisons with the MW and LMC bumps}

\begin{figure}
\centering
{\includegraphics[width=8.7cm, height=6.5cm]{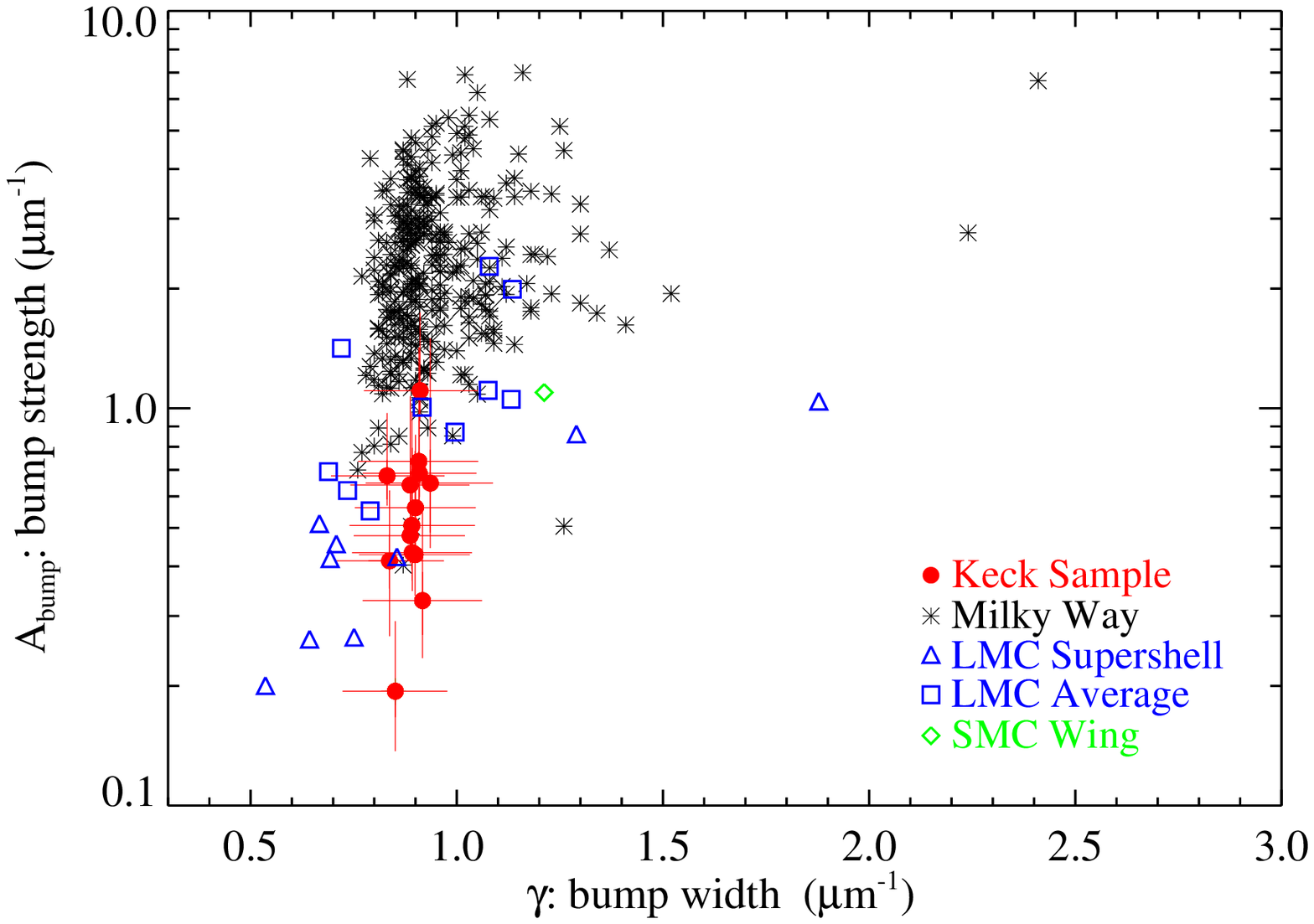}} 
{\includegraphics[width=8.7cm, height=6.5cm]{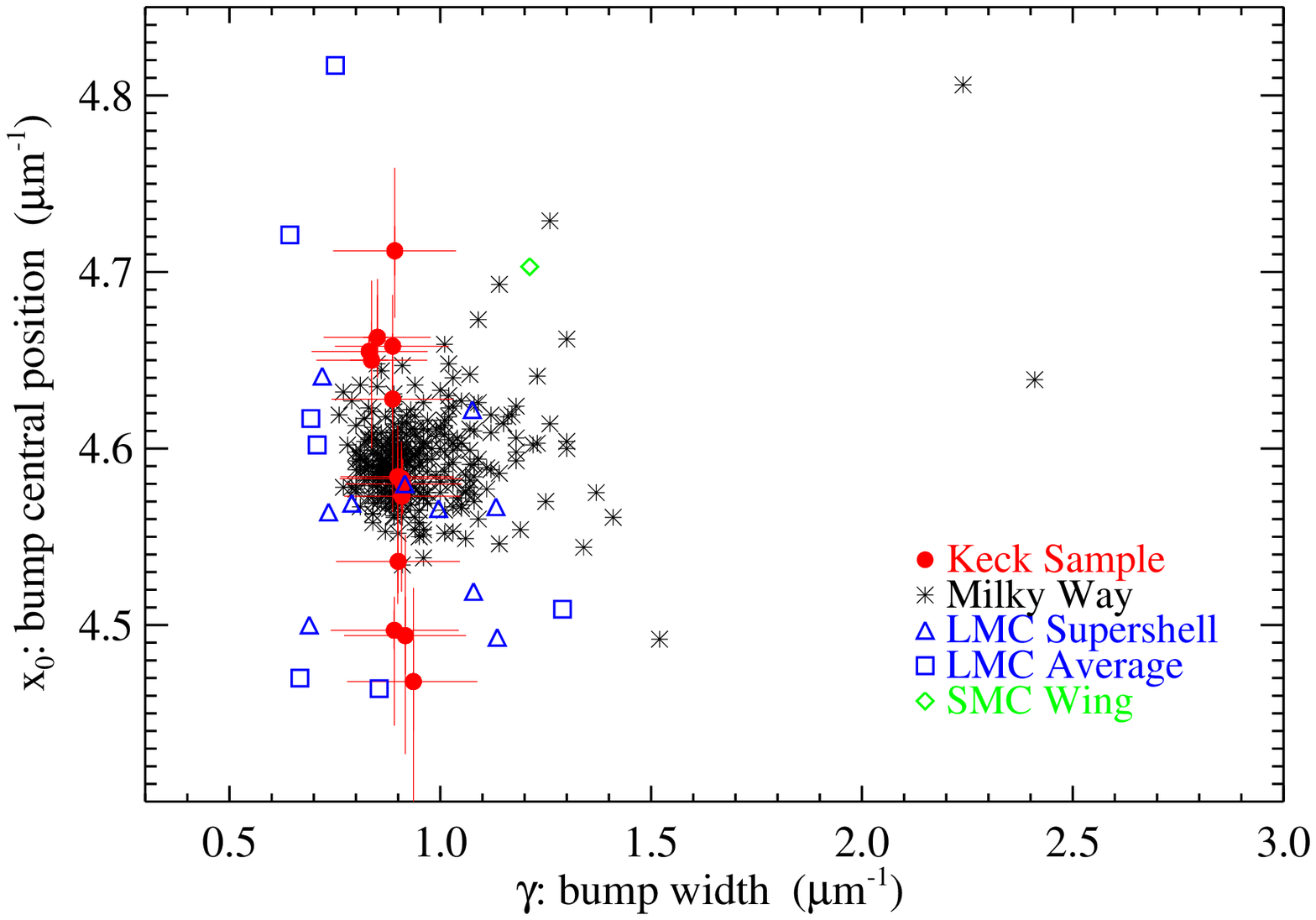}} 
\caption{Top panel: 2175 \AA$ $ bump strength versus bump width. The black crosses are the MW bumps \citep{FM2007} and the blue squares and triangles are the average LMC bumps and the bumps in the LMC supershell \citep{Gordon2003}. The green diamond is the single sightline toward the SMG wing whose extinction curve contains a 2175 \AA$ $ bump. The red circles are the 2175 \AA$ $ bumps in this sample.  Bottom panel: central position versus bump width. The error bars include the statistical uncertainties from the MCMC and systematic uncertainties induced by choice of blue and red templates and priors on the bump width.  }
\label{fig:bump}
\end{figure}

The parameterization of the extinction curve allows for comparisons between the 2DAs and the MW, LMC sight lines where prominent and weaker bump features have been observed, respectively. The comparisons are presented in Figure \ref{fig:bump} with 328 extinction curves towards Galactic stars from \cite{FM2007} and 19 LMC sight lines from \cite{Gordon2003}. The bumps towards the LMC2 supershell region are much weaker than those in the MW and other sight lines towards the LMC. 

The red circles are the bumps in this sample with $\gamma$ = 0.89 $\micron^{-1}$ (peak value of the MW distribution) as the initial value or prior in the first step. The derived bump strength $A_{bump}$  significantly depends on the prior imposed on the bump width $\gamma$. We account for this effect by adding error bars to the bump strengths induced by variations in the bump width priors. In any case, most 2175 \AA$ $ features in our sample are LMC-like bumps. The bottom panel of Figure \ref{fig:bump} displays the central position of the bump versus bump width. The bump positions spread over the range of the MW and LMC bumps.

\subsection{Blue and red template fitting}

The composite quasar spectrum method described above relies on the assumption that the composite is a good substitute for the intrinsic quasar spectrum. However, the intrinsic spectrum is essentially unknown, and we consider systematic uncertainties introduced by using bluer or redder quasar templates. These templates are generated in the same manner as the composite spectrum while only combining the bluest 20\% and reddest 20\% quasar spectra according to their spectral indices. In most cases, the composite quasar spectrum remains the best substitute for the intrinsic quasar spectrum. The derived parameters based on these templates set the upper and lower limits accordingly.  Occasionally a bluer or redder template provides a better fit to the observed spectra and therefore we take the parameters derived from the blue or red template as the best-fit values.

The best-fit parameters of the extinction curves and the uncertainties are listed in Table \ref{tab:bump}. The total error bars on the fitted parameters as well as $A_V$ incorporate the statistical uncertainties using the composite spectrum and the systematic uncertainties due to the red and blue templates. We stress that the total extinction $A_V$ towards individual sight lines can be highly uncertain due to the essentially unknown intrinsic quasar spectrum as the primary source of uncertainty \citep{Krogager2016} and potential flux calibration errors \citep{Harris2016}.

\begin{table*}
\centering
\caption{Best-fit parameters of the extinction curves. The reported error bars incorporate statistical uncertainties from the fitting as well as systematic uncertainties. $\chi^2_{\nu}$ is the reduced $\chi^2$.  }
\begin{tabular}{cccccccc}
\hline\hline
QSO        & $c_1$      & $c_2$    & $c_3$    & $x_0$  & $\gamma$  &  $\chi^2_{\nu}$     & $A_V$          \\
\hline
J0745+4554 & -0.901 $^{+ 0.000}_{- 3.200}$  & 0.537 $^{+ 0.033}_{- 0.316}$ & 0.358 $^{+ 0.111}_{- 0.097}$   &  4.655 $^{+ 0.009}_{- 0.009}$  & 0.831 $^{+ 0.139}_{- 0.136}$    & 1.09    & 0.95 $^{+ 0.09}_{- 0.18}$    \\[0.07cm]   
J0850+5159 & -0.984 $^{+ 0.035}_{- 3.186}$  & 0.474 $^{+ 0.033}_{- 0.360}$ & 0.287 $^{+ 0.133}_{- 0.102}$  &  4.497 $^{+ 0.054}_{- 0.012}$  & 0.891 $^{+ 0.153}_{- 0.151}$     & 1.13   & 0.82 $^{+ 0.10}_{- 0.20}$       \\  [0.07cm]    
J0901+2044 & -1.073 $^{+ 0.079}_{- 3.271}$ & 0.262 $^{+ 0.076}_{- 0.465}$ & 0.386 $^{+ 0.202}_{- 0.145}$   &  4.467 $^{+ 0.095}_{- 0.053}$  & 0.936 $^{+ 0.152}_{- 0.157}$    & 1.54    & 0.22 $^{+ 0.39}_{- 0.20}$    \\[0.07cm]
J0953+3225 & -1.320 $^{+ 0.012}_{- 3.224}$  & 0.110 $^{+ 0.041}_{- 0.354}$& 0.270 $^{+ 0.088}_{- 0.075}$   & 4.658 $^{+ 0.094}_{- 0.007}$  & 0.887 $^{+ 0.133}_{- 0.137}$    & 2.74  & 0.03 $^{+ 0.25}_{- 0.16}$   \\[0.07cm]
J1006+1538 & 1.175 $^{+ 0.035}_{- 3.199}$  & 0.229 $^{+ 0.031}_{- 0.284}$& 0.246 $^{+ 0.103}_{- 0.081}$  &  4.712 $^{+ 0.047}_{- 0.038}$  & 0.892 $^{+ 0.145}_{- 0.146}$     & 1.14  & 0.41 $^{+ 0.07}_{- 0.17}$         \\[0.07cm]
J1047+3423 & -0.397 $^{+ 0.004}_{- 3.239}$ & 0.338 $^{+ 0.041}_{- 0.308}$& 0.105 $^{+ 0.033}_{- 0.028}$   &  4.663 $^{+ 0.033}_{- 0.023}$ & 0.851 $^{+ 0.126}_{- 0.128}$    & 1.14   & 0.61 $^{+ 0.08}_{- 0.18}$       \\[0.07cm]
J1127+2424 & -0.524 $^{+ 0.011}_{- 3.123}$  & 0.143 $^{+ 0.019}_{- 0.347}$& 0.220 $^{+ 0.074}_{- 0.062}$   &  4.650 $^{+ 0.050}_{- 0.045}$   & 0.837 $^{+ 0.132}_{- 0.131}$    & 1.57  & 0.27 $^{+ 0.03}_{- 0.21}$         \\[0.07cm]
J1130+1850 & 1.471 $^{+ 0.003}_{- 3.126}$  & 0.274 $^{+ 0.016}_{- 0.310}$& 0.362 $^{+ 0.136}_{- 0.114}$   &  4.627 $^{+ 0.010}_{- 0.007}$  & 0.887 $^{+ 0.144}_{- 0.145}$     & 1.23   & 0.49 $^{+ 0.04}_{- 0.18}$     \\[0.07cm]
J1138+5245 & -0.224 $^{+ 0.067}_{- 3.215}$  & 0.298 $^{+ 0.041}_{- 0.390}$& 0.425 $^{+ 0.202}_{- 0.154}$    &  4.583 $^{+ 0.064}_{- 0.021}$   & 0.908 $^{+ 0.144}_{- 0.147}$     & 1.16  & 0.22 $^{+ 0.39}_{- 0.13}$         \\[0.07cm]
J1211+0833 & -0.051 $^{+ 0.028}_{- 3.182}$ & 0.442 $^{+ 0.028}_{- 0.301}$& 0.642 $^{+ 0.216}_{- 0.179}$   &   4.580 $^{+ 0.014}_{- 0.017}$ & 0.911 $^{+ 0.140}_{- 0.136}$     & 1.19  & 0.80 $^{+ 0.05}_{- 0.18}$          \\[0.07cm]
J1212+2936 & -1.019 $^{+ 0.020}_{- 3.262}$  & 0.211 $^{+ 0.047}_{- 0.359}$& 0.245 $^{+ 0.101}_{- 0.081}$   &  4.584 $^{+ 0.072}_{- 0.031}$  & 0.899 $^{+ 0.133}_{- 0.136}$     & 1.02 & 0.27 $^{+ 0.20}_{- 0.18}$           \\ [0.07cm]
J1321+2135 & 1.587 $^{+ 0.019}_{- 3.175}$ & 0.202 $^{+ 0.026}_{- 0.288}$ & 0.191 $^{+ 0.072}_{- 0.059}$    & 4.494 $^{+ 0.057}_{- 0.067}$ & 0.917 $^{+ 0.144}_{- 0.145}$     & 1.01   & 0.39 $^{+ 0.03}_{- 0.18}$     \\[0.07cm]
J1531+2403 & 0.796 $^{+ 0.013}_{- 3.157}$  & 0.233 $^{+ 0.022}_{- 0.305}$& 0.322 $^{+ 0.106}_{- 0.090}$    &  4.536 $^{+ 0.013}_{- 0.018}$   & 0.900 $^{+ 0.146}_{- 0.147}$     &1.42  & 0.44 $^{+ 0.03}_{- 0.19}$            \\[0.07cm]
J1737+4406 & 0.141 $^{+ 0.001}_{- 3.200}$  & 0.280 $^{+ 0.033}_{- 0.316}$& 0.397 $^{+ 0.123}_{- 0.103}$   &  4.573 $^{+ 0.014}_{- 0.006}$  & 0.909 $^{+ 0.139}_{- 0.137}$   & 1.19   & 0.50 $^{+ 0.07}_{- 0.19}$     \\[0.07cm]
\hline
\end{tabular}
\label{tab:bump}
\end{table*}

\section{Column densities and relative metal abundances}
\label{sec:metal}

\subsection{Column density measurements}

This subsection describes the column density measurements for the 2DAs using the Apparent Optical Depth Method (AODM; \citealt{Savage1991}). This method is based upon the connection between optical depth and column density, involving converting observed absorption line profiles into apparent optical depths. The total column density can be obtained by integrating the optical depth over the velocity range of the transition. A typical integration interval of [-300, 300] \kms{} is chosen; this width is large enough to include the transition  yet also avoids significant contamination. When multiple transitions of the same species are available, the AODM can reveal unresolved saturated lines by comparing the apparent column density derived from the stronger lines to that of the weaker lines. We adopt the wavelengths and oscillator strengths of \cite{Morton2003} (except C\I{} from \citealt{Jenkins2009}) and the photospheric solar abundances of \cite{Asplund2009} listed in Appendix \ref{appendixA}. Tables in Appendix \ref{appendixB} show for each transition the equivalent width and column density from the AODM. The errors only reflect statistical uncertainty using standard error propagation. One should take a lower limit of 0.05 dex from systematic error (continuum placement). The adopted column densities are the weighted mean of the column densities from unsaturated lines. For non-detections and saturated lines, we report upper and lower limits at the 3$\sigma$ level.  

At the resolution of ESI, Zn\II{} $\lambda$2026 and Zn\II{} $\lambda$2062 are blended with Mg\I{} $\lambda$2026 and Cr\II{} $\lambda$2062. We use Mg\I{} $\lambda$2852 to derive the Mg\I{} column density and estimate its contribution to the blend at 2026 \AA$ $. In all cases, Mg\I{} contributes approximately 10\% -- 20\% to the total equivalent widths, i.e., Zn\II{} dominates the absorption seen at 2026 \AA$ $. The contribution of Cr\II{} $\lambda$2062 to the blend at 2062 \AA$ $ is also removed in the same manner.  We report the contamination-subtracted equivalent widths and column densities for Zn\II{} in Appendix \ref{appendixB}. 

The measurements are independently checked by the Voigt profile fitting method using the VPFIT package \footnote{http://www.ast.cam.ac.uk/~rfc/vpfit.html}. We calculate the total column densities by summing over all the velocity components.  Multiple transitions of a species and different low-ionization lines are simultaneously fit assuming they are kinematically associated. For each velocity component, the redshift and Dopper $b$ parameter are tied together. For the unsaturated lines, the column density measurements derived from VPFIT and AODM are generally in good agreement.

\subsection{Relative metal abundances} 

All the 2DAs demonstrate various metal absorption lines. For our sample, Zn is the best metallicity estimator because the two transitions at 2026 \AA$ $ and 2062 \AA$ $ are covered and not saturated in all the spectra, lying outside the Ly$\alpha$ forest.  Using Zn abundance as a proxy for metal abundance is based on the fact that Zn is a non-refractory element and very mildly depleted.  We use Zn to derive the relative metal abundances and estimate depletions of other elements relative to Zn. The relative abundances $X$/Zn ($X$ = \{Si, Mn, Fe, Cr, Ni, Ti, and Ca\} ) are expressed relative to the solar values, [$X$/Zn] $\equiv$ log($X$/Zn) - log($X$/Zn)$_{\sun}$.

%%%%%%%%%%%%%%%%%%%%%%%%%%%%%%%%%%%
% Comments on individual absorbers
%%%%%%%%%%%%%%%%%%%%%%%%%%%%%%%%%%%
\section{Individual systems}
\label{sec:comments}

\begin{figure}
\centering
{\includegraphics[width=8.6cm, height=6.3cm]{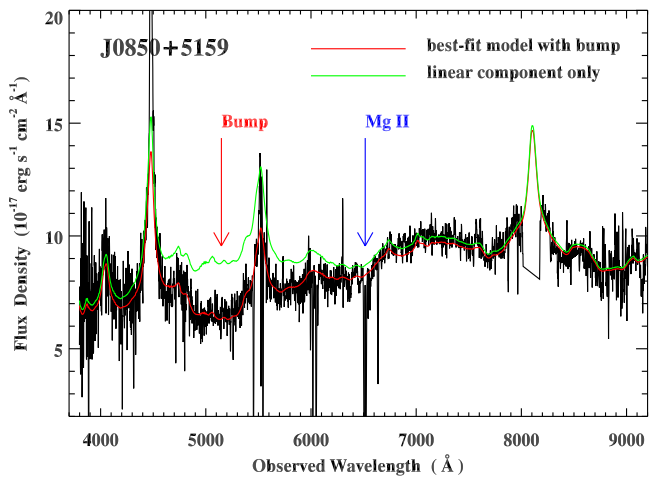}} 
\caption{The bump fitting result for the absorber towards J0850+5159. }
\label{fig:J0850bump}
\end{figure}

\begin{figure}
\centering
{\includegraphics[width=8.6cm, height=6.3cm]{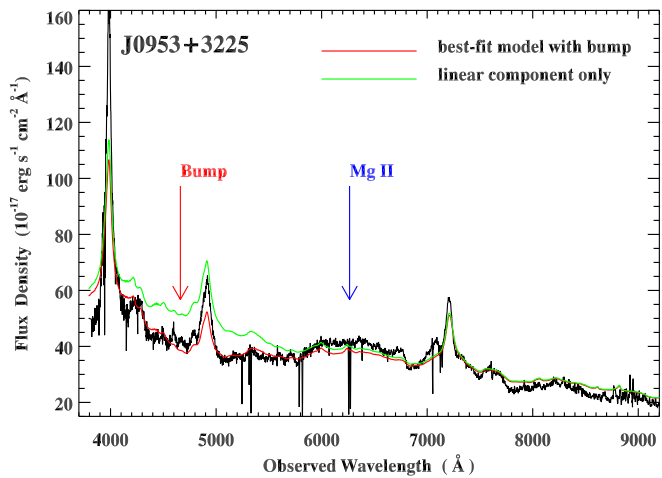}} 
\caption{The bump fitting result for the absorber towards J0953+3225.}
\label{fig:J0953bump}
\end{figure}

\begin{figure}
\centering
{\includegraphics[width=8.6cm, height=6.3cm]{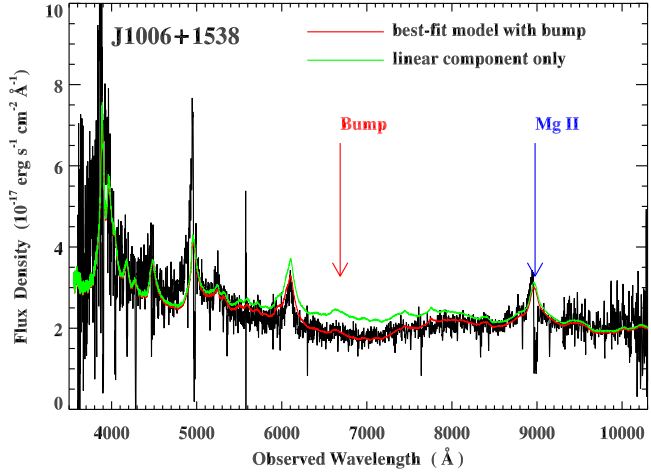}} 
\caption{The bump fitting result for the absorber towards J1006+1538.}
\label{fig:J1006bump}
\end{figure}

\begin{figure}
\centering
{\includegraphics[width=8.6cm, height=6.3cm]{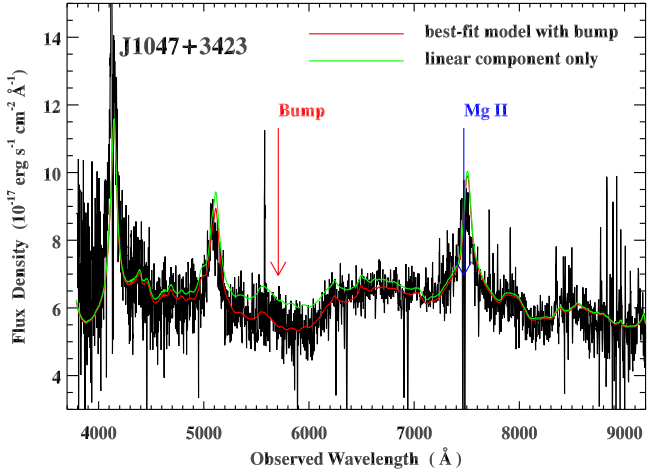}}
\caption{The bump fitting result for the absorber towards J1047+3423.}
\label{fig:J1047bump}
\end{figure}

\begin{figure}
\centering
{\includegraphics[width=8.6cm, height=6.3cm]{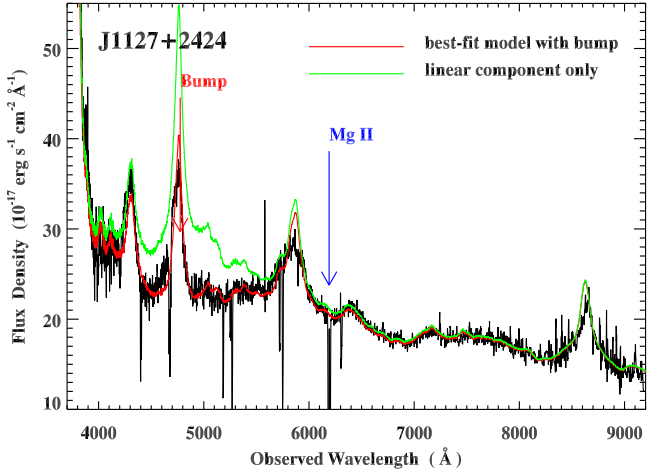}} 
\caption{The bump fitting result for the absorber towards J1127+2424.}
\label{fig:J1127bump}
\end{figure}

\begin{figure}
\centering
{\includegraphics[width=8.6cm, height=6.3cm]{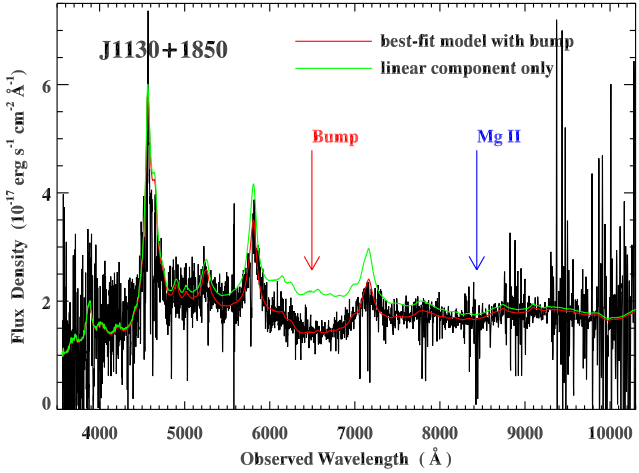}} 
\caption{The bump fitting result for the absorber towards J1130+1850.}
\label{fig:J1130bump}
\end{figure}

\begin{figure}
\centering
{\includegraphics[width=8.6cm, height=6.3cm]{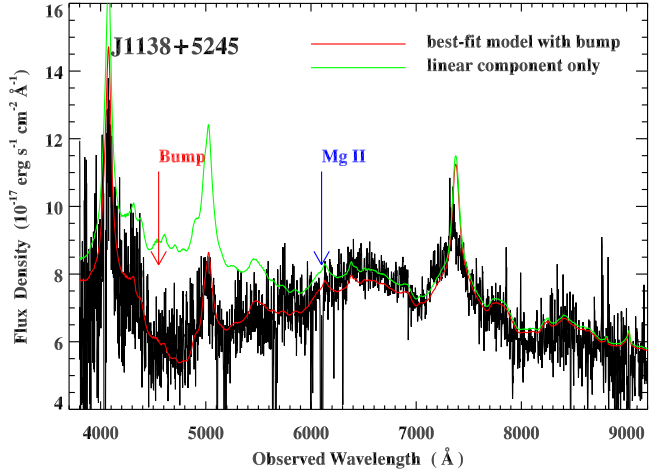}} 
\caption{The bump fitting result for the absorber towards J1138+5245.}
\label{fig:J1138bump}
\end{figure}

\begin{figure}
\centering
{\includegraphics[width=8.6cm, height=6.3cm]{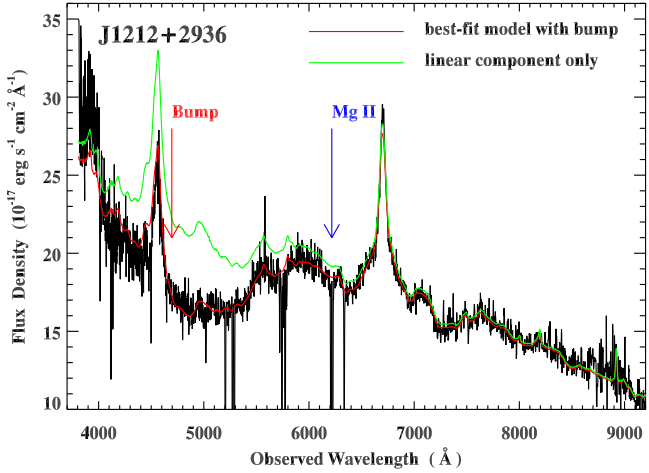}} 
\caption{The bump fitting result for the absorber towards J1212+2936.}
\label{fig:J1212bump}
\end{figure}

\begin{figure}
\centering
{\includegraphics[width=8.6cm, height=6.3cm]{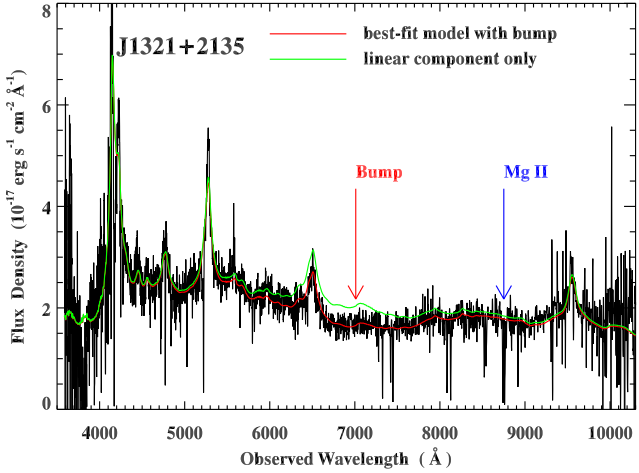}} 
\caption{The bump fitting result for the absorber towards J1321+2135.}
\label{fig:J1321bump}
\end{figure}

\begin{figure}
\centering
{\includegraphics[width=8.6cm, height=6.3cm]{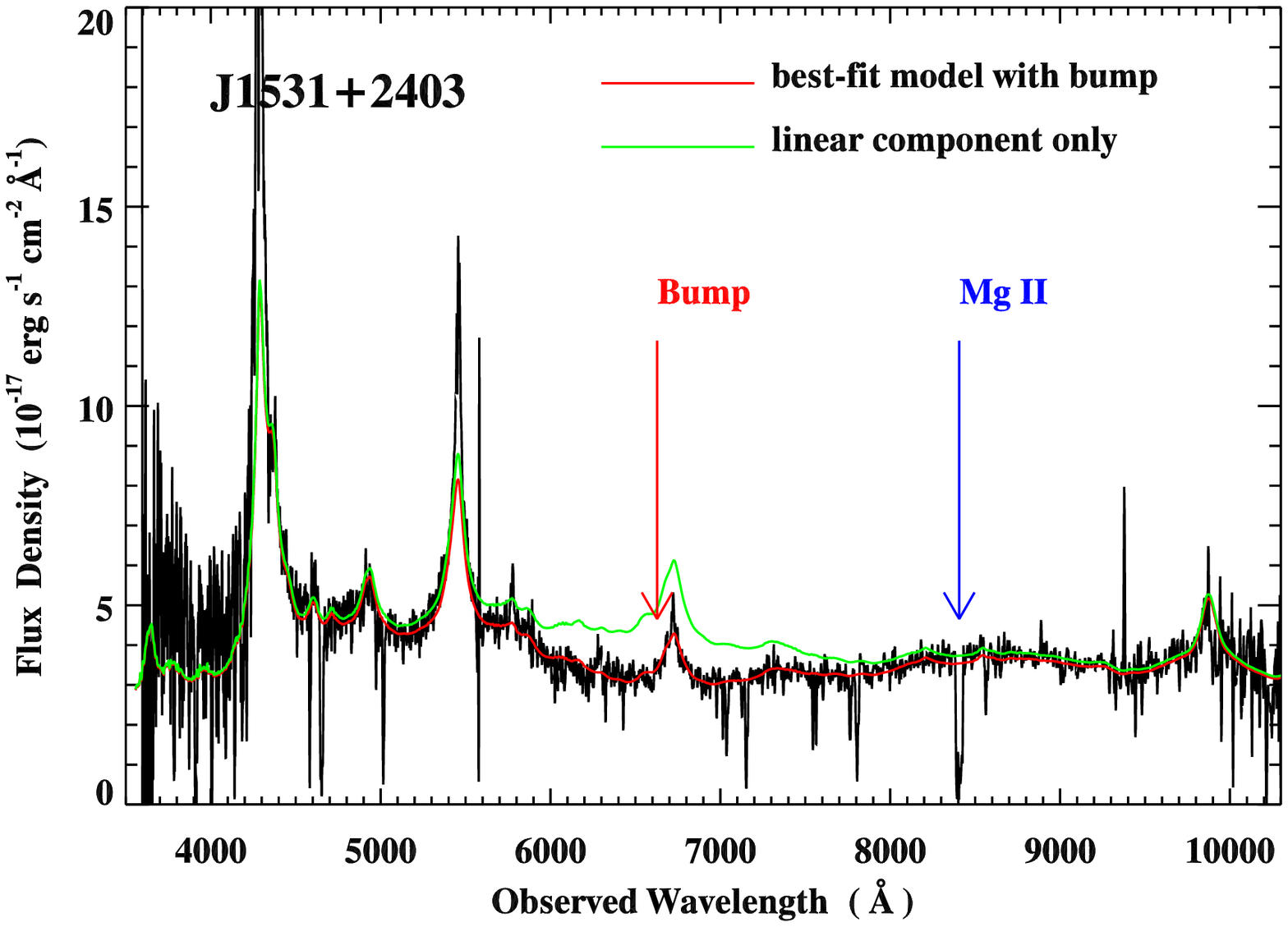}} 
\caption{The bump fitting result for the absorber towards J1531+2403.}
\label{fig:J1531bump}
\end{figure}

\begin{figure}
\centering
{\includegraphics[width=8.6cm, height=6.3cm]{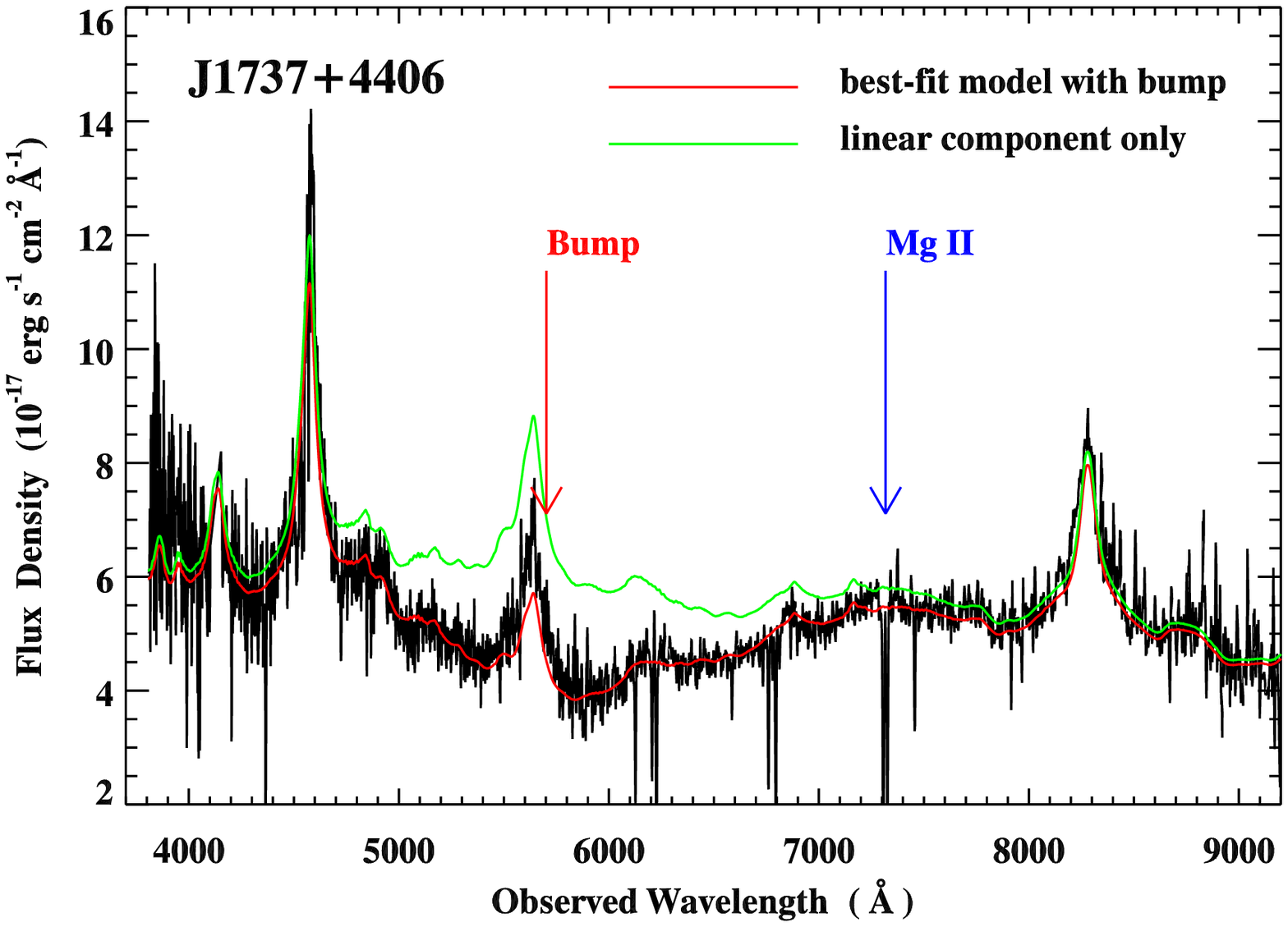}} 
\caption{The bump fitting result for the absorber towards J1737+4406.}
\label{fig:J1737bump}
\end{figure}

This section gives the details of the metal absorption lines covered in the Keck/ESI spectra, column density measurements of the individual systems as well as the properties of the 2175 \AA$ $ bump. Figures \ref{fig:J0745bump}-\ref{fig:1737vpfit} show the quasar spectra with bump fitting results, velocity profiles and Voigt profile fits for individual systems. We present the total column densities derived from both methods in Tables \ref{tab:0745N}-\ref{tab:1737N} and adopt the AODM-derived values unless inconsistence occurs. The final adopted column densities are based on careful examination of the results from both methods. 

\subsection{J0745+4554 $z_{qso}$ = 2.1998, $z_{abs}$ = 1.8612}

\begin{figure}
\centering
{\includegraphics[width=8.7cm, height=6.5cm]{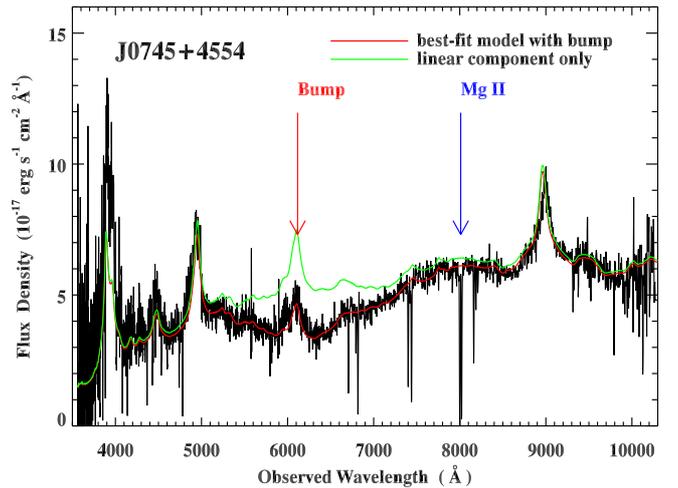}} 
\caption{The bump fitting result for the absorber towards J0745+4554. The black spectrum is the original observed SDSS spectrum smoothed by 3-pixel box car for demonstration purpose. The red curve is the best-fit model with an extinction bump whose central position is denoted by the red arrow. The blue arrow points to the position of the Mg\II{} doublet. The green curve represents the model reddened by the linear component only.  }
\label{fig:J0745bump}
\end{figure}

\begin{figure}
\centering
{\includegraphics[width=8.7cm, height=20.4cm]{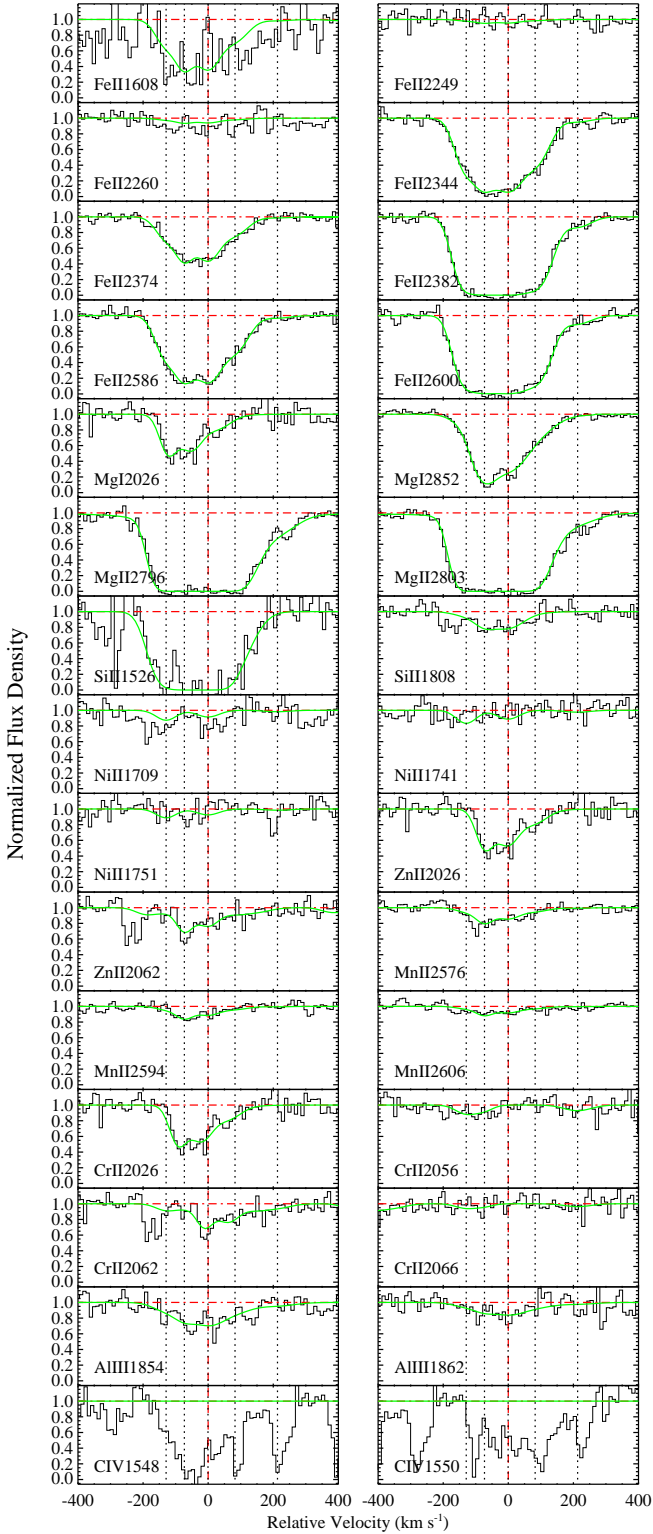}}  
\caption{Velocity profiles of the metal absorption lines in the Keck/ESI spectrum for the absorber at $z$=1.8612 towards J0745+4554. The black curves are the normalized spectra, and the green curves are the VPFIT fits. The vertical dotted lines denote the positions of the velocity components. }
\label{fig:0745vpfit}
\end{figure}

\begin{table}
\centering
\caption{Total column densities for the $z_{abs}$ =1.8612 absorber toward J0745+4554 }
\begin{tabular}{lccc}
\hline\hline
Ion       &  log$N_{\rm AODM}$    &log$N_{\rm FIT}$    &log$N_{\rm adopt}$ \\
 \hline
C\IV         & 14.98 $\pm$ 0.01   &                                &    14.98 $\pm$ 0.01                         \\
Mg\I         &  13.38 $\pm$ 0.03  & 13.39 $\pm$ 0.09    &     13.38 $\pm$ 0.03                         \\
Mg\II        &  $>$ 15.22               &   $>$ 15.97                          &     $>$ 15.97                        \\
Al\II          &  $>$ 14.45               &  $>$ 14.44                          &    $>$ 14.45                           \\
Al\III         &  13.50 $\pm$ 0.04  & 13.46 $\pm$ 0.10    &    13.50 $\pm$ 0.04                         \\
Si\II          &  15.77 $\pm$ 0.04  & 15.71 $\pm$ 0.13    &     15.77 $\pm$ 0.04                          \\
Cr\II         &  $<$ 13.90       &  13.54 $\pm$ 0.23    &     13.54 $\pm$ 0.23                        \\
Mn\II        &   13.11 $\pm$ 0.05  &  13.23 $\pm$ 0.08    &    13.11 $\pm$ 0.05                         \\
Fe\II         &  15.09 $\pm$ 0.09  &  14.95 $\pm$ 0.07    &      15.09 $\pm$ 0.09                       \\
Ni\II          &  $<$ 14.95   &  14.03 $\pm$ 0.07    &    14.03 $\pm$ 0.07                         \\
Zn\II         &  13.56 $\pm$ 0.03  &   13.60 $\pm$ 0.09   &    13.56 $\pm$ 0.03                           \\
\hline
\end{tabular}
\label{tab:0745N}
\end{table}

The observed quasar spectrum of J0745+4554 clearly requires a significant absorption bump (Figure \ref{fig:J0745bump}). The red curve is the best-fit model with an absorption bump while the green curve represents the model reddened by the linear component only.  The best-fit parameters of the extinction curve derived using the two-step method are $c_1$ = -0.901, $c_2$ = 0.537, $c_3$ = 0.358, $x_0$ = 4.655, and $\gamma$ = 0.831. The corresponding bump strength is 0.676 $\micron^{-1}$.

The Keck/ESI spectrum for this absorber covers a set of Fe\II{} lines $\lambda$1608, 1611, 2249, 2260, 2344, 2374, 2382, 2586, 2600, and multiple other low-ionization transitions including Si\II{} $\lambda$ 1526, 1808, Zn\II{} $\lambda$2026, 2062, Al\II{} $\lambda$1670, Ni\II{} $\lambda$1709, 1741, 1751, Cr\II{} $\lambda$2026, 2056, 2062 and Mn\II{} $\lambda$2576, 2594, 2606, and the saturated Mg\II{} doublet $\lambda$2796, 2803. The intermediate and high-ionization transitions Al\III{} $\lambda$1854, 1862 and C\IV{} $\lambda$1548, 1550 are also present in the spectrum. The strong Mg\I{} $\lambda$2852 line guides the subtraction of Mg\I's contribution to the Zn\II{} + Mg\I{} $\lambda$ 2026 blend.  To independently check the AODM-derived column densities, we simultaneously fit all the low-ionization lines using VPFIT as the velocity structures are expected to be similar.  As the individual components are not fully resolved, we start with two velocity components and gradually add more components judging from the goodness-of-fit. As a result, five components at $v$ $\sim$ -129, -73, -1, +82, +213 \kms  are required to yield a reasonably good fit.  All the column densities are consistent (i.e. within the error bars) with each other except for Ni\II{} where the VPFIT fits are inconsistent with the observed spectrum. The high-ionization lines C\IV{} $\lambda$1548, 1550 are generally consistent with having five velocity components, however, the relative strengths of the two components are $v$ = +94, +241 \kms are significantly higher than the low-ionization lines. 

The resulting total column densities of Fe and Zn are logN(Fe\II{}) = 15.09 $\pm$ 0.09 and logN(Zn\II{}) = 13.56 $\pm$ 0.03, which yield a dust depletion indicator of [Fe/Zn] = -1.41 $\pm$ 0.09. The velocity profiles are shown in Figure \ref{fig:0745vpfit} and the column densities derived from two methods are summarized in Table \ref{tab:0745N}.

\subsection{J0850+5159 $z_{qso}$ = 1.8939, $z_{abs}$ = 1.3269}

\begin{figure}
\centering
{\includegraphics[width=8.7cm, height=20.4cm]{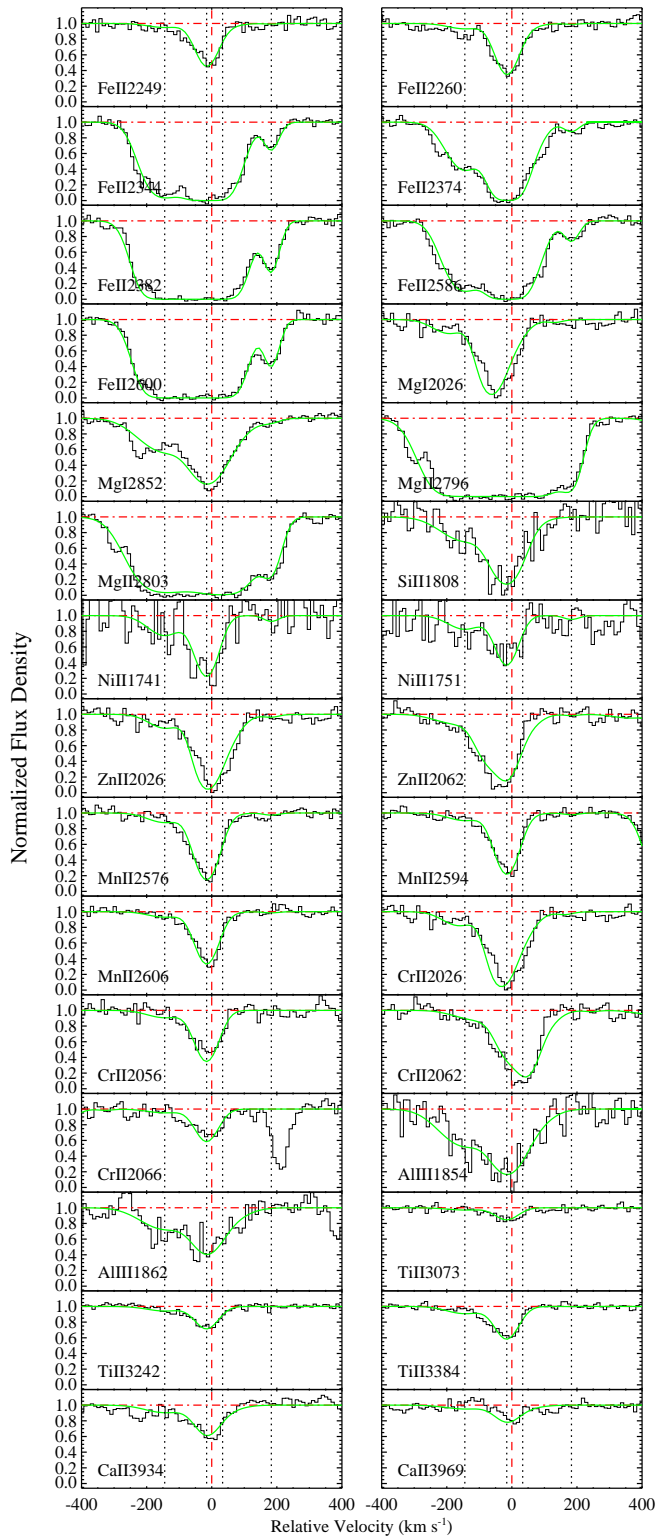}} 
\caption{Velocity profiles of the metal absorption lines in Keck/ESI for the absorber  at $z$=1.3269 towards J0850+5159.  }
\label{fig:0850vpfit}
\end{figure}

\begin{table}
\centering
\caption{Total column densities for the $z_{abs}$ =1.3269 absorber toward J0850+5159 }
\begin{tabular}{lccc}
\hline\hline
Ion       &  log$N_{\rm AODM}$    &log$N_{\rm FIT}$    &log$N_{\rm adopt}$ \\
\hline
Mg\I         &  13.38 $\pm$ 0.01  & 13.38 $\pm$ 0.02    &    13.38 $\pm$ 0.01                         \\
Mg\II        &   $>$ 15.24             &   $>$ 14.81             &    $>$ 15.24                         \\
Al\III         &  14.24 $\pm$ 0.01  & 14.07 $\pm$ 0.05    &    14.24 $\pm$ 0.01                         \\
Si\II          &  16.51 $\pm$ 0.02  & 16.44 $\pm$ 0.10    &     16.51 $\pm$ 0.02                        \\
Ca\II        &  13.00 $\pm$ 0.02  &  13.00 $\pm$ 0.04    &    13.00 $\pm$ 0.02                         \\
Ti\II          &  13.24 $\pm$ 0.02  &  13.27 $\pm$ 0.02    &   13.24 $\pm$ 0.02                           \\
Cr\II         &  14.15 $\pm$ 0.02  &  14.34 $\pm$ 0.02    &    14.15 $\pm$ 0.02                         \\
Mn\II        &  13.84 $\pm$ 0.01  &  13.88 $\pm$ 0.02    &    13.84 $\pm$ 0.01                         \\
Fe\II         &  15.88 $\pm$ 0.01  &  15.85 $\pm$ 0.02    &      15.88 $\pm$ 0.01                       \\
Ni\II          &  14.24 $\pm$ 0.06  &  14.93 $\pm$ 0.08    &     14.24 $\pm$ 0.06                        \\
Zn\II         &  14.04 $\pm$ 0.01  &   14.10 $\pm$ 0.04   &       14.04 $\pm$ 0.01                        \\
\hline
\end{tabular}
\label{tab:0850N}
\end{table}

The best-fit parameters for this dust absorber are $c_1$ = -0.984, $c_2$ = 0.474, $c_3$ = 0.287, $x_0$ = 4.497, $\gamma$ = 0.891 with a bump strength of 0.507.

This spectrum covers a set of Fe\II{} transitions at $\lambda$2249, 2260, 2344, 2374, 2382, 2586, 2600. In addition to the commonly observed low-ionizations lines (i.e. Fe\II, Zn\II, Mg\II, Si\II, Ni\II, Cr\II, Mn\II{}), Ti\II{} $\lambda$3073, 3242, 3384 and Ca\II{} $\lambda$3934, 3969 are also covered and well detected in the spectrum.  All the low-ionization lines as well as the intermediate-ionization Al\III{} lines at $\lambda$1854, 1862 are simultaneously fit. Most strong lines are well fit with four velocity components at $v$ =  -145, -16, +32, +185 \kms while for the weaker lines only the central component contributes (Figure \ref{fig:0850vpfit}). Adding more components would probably yield a better fit. However due to the limited resolution, the individual components are heavily blended and hard to be identified.  

The total column densities of Fe\II{} and Zn\II{} are logN(Fe\II{}) = 15.88 $\pm$ 0.01 and logN(Zn\II{}) = 14.04 $\pm$ 0.01, both the highest among the sample. We also generated a curve of growth to check if the Zn\II{} lines are saturated. It turns out that the Zn\II{} lines are still on the linear part of the curve although almost reaching the logarithmic part. The relative abundance [Fe/Zn] is -1.10 $\pm$ 0.01. The Ti\II{} and Ca\II{} column densities are logN(Ti\II{}) = 13.24 $\pm$ 0.02 and logN(Ca\II{}) = 13.00 $\pm$ 0.02. Table \ref{tab:0850N} lists all the column density measurements.

\subsection{J0901+2044 $z_{qso}$ = 2.0934, $z_{abs}$ = 1.0191}

\begin{figure}
\centering
{\includegraphics[width=8.6cm, height=6.3cm]{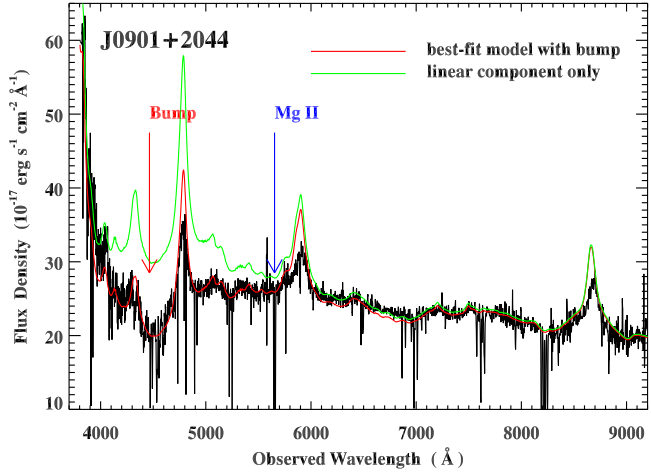}} 
\caption{The bump fitting result for the absorber towards J0901+2044.}
\label{fig:J0901bump}
\end{figure}

\begin{figure}
\centering
{\includegraphics[width=8.7cm, height=17.7cm]{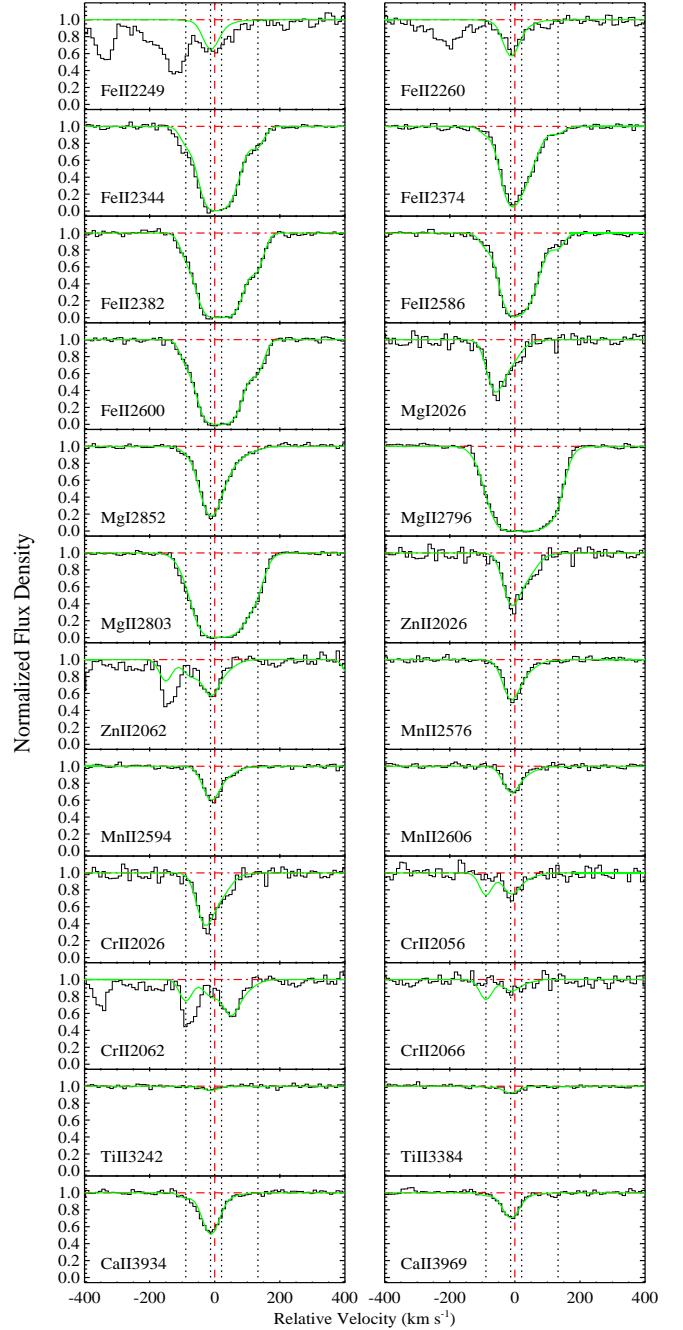}}  
\caption{Velocity profiles of the metal absorption lines in Keck/ESI for the absorber at $z$=1.0191 towards J0901+2044.  }
\label{fig:0901vpfit}
\end{figure}

\begin{table}
\centering
\caption{Total column densities for the $z_{abs}$ =1.0191 absorber toward J0901+2044 }
\begin{tabular}{lccc}
\hline\hline
Ion       &  log$N_{\rm AODM}$    &log$N_{\rm FIT}$    &log$N_{\rm adopt}$ \\
\hline
Mg\I         &  13.00 $\pm$ 0.01  & 13.08 $\pm$ 0.03    &   13.00 $\pm$ 0.01                          \\
Mg\II        &   $>$ 14.74            &   $>$ 14.62                          &     $>$ 14.74                         \\
Ca\II        &  12.85 $\pm$ 0.01  &  12.90 $\pm$ 0.04    &      12.85 $\pm$ 0.01                        \\
Ti\II          &  12.33 $\pm$ 0.09  &  12.27 $\pm$ 0.17    &      12.33 $\pm$ 0.09                       \\
Cr\II         &  13.65 $\pm$ 0.05  &  13.59 $\pm$ 0.22   &     13.65 $\pm$ 0.05                        \\
Mn\II        &  13.27 $\pm$ 0.01  &  13.32 $\pm$ 0.04    &     13.27 $\pm$ 0.01                        \\
Fe\II         &  15.54 $\pm$ 0.01  &  15.45 $\pm$ 0.06    &     15.45 $\pm$ 0.06                         \\
Zn\II         &  13.16 $\pm$ 0.02  &   13.47 $\pm$ 0.12   &       13.47 $\pm$ 0.12                        \\
\hline
\end{tabular}
\label{tab:0901N}
\end{table}

This absorber has the lowest redshift in this sample. The continuum blueward of the 2175 \AA$ $ bump is not covered by the observed spectrum, therefore different quasar templates lead to large variations in the best-fit bump strength without being ruled out by the data. In any case, a prominent extinction bump is required to represent the observed spectrum. We show the result using the composite quasar spectrum in Figure \ref{fig:J0901bump}. The corresponding best-fit parameters are $c_1$ = -1.073, $c_2$ = 0.262, $c_3$ = 0.386, $x_0$ = 4.468, $\gamma$ = 0.936 with a bump strength of 0.648. 

This Keck/ESI spectrum is of the highest S/N  among all the sources in this sample. This low-redshift absorption system contains many low-ionization transitions including Zn\II{} $\lambda$2026, 2062, Mn\II{} $\lambda$2576, 2594, 2606, Cr\II{} $\lambda$ 2056, Ti\II{} 3384, Ca\II{} $\lambda$3934, 3969, and  seven Fe\II{} lines $\lambda$2249, 2260, 2344, 2374, 2382, 2586, 2600. The velocity profiles are best fit with four components at $v$ = -89, -13, +21, +133 \kms (Figure \ref{fig:0901vpfit}).  
The Fe\II{} $\lambda$2249, 2260 lines are contaminated according to the simultaneous fitting, although the interval is restricted to [-80 \kms, +80 \kms] when applying the AODM. VPFIT overpredicts a velocity component at $v$ = -89 \kms that is not seen in the stronger Cr\II{} $\lambda$ 2056 line due to the contamination/unknown absorption seen in the Zn\II{} + Cr\II{} $\lambda$2062 blend. We exclude this component in the reported $N_{\rm FIT}$. No high-ionization transitions are covered in the spectrum for this low-redshift absorber.

We adopt the column density of Fe\II{} derived from VPFIT logN(Fe\II{}) = 15.45 $\pm$ 0.06.  The Fe-to-Zn relative abundance is [Fe/Zn] = -0.96 $\pm$ 0.13 with a Zn\II{} column density of logN(Zn\II{}) = 13.47 $\pm$ 0.12. The Ti\II{} and Ca\II{} column densities are 12.33 $\pm$ 0.09 and 12.85 $\pm$ 0.01. All the column density measurements are listed in Table \ref{tab:0901N}.

\subsection{J0953+3225 $z_{qso}$ = 1.5748, $z_{abs}$ = 1.2375}

\begin{figure}
\centering
{\includegraphics[width=8.7cm, height=19.3cm]{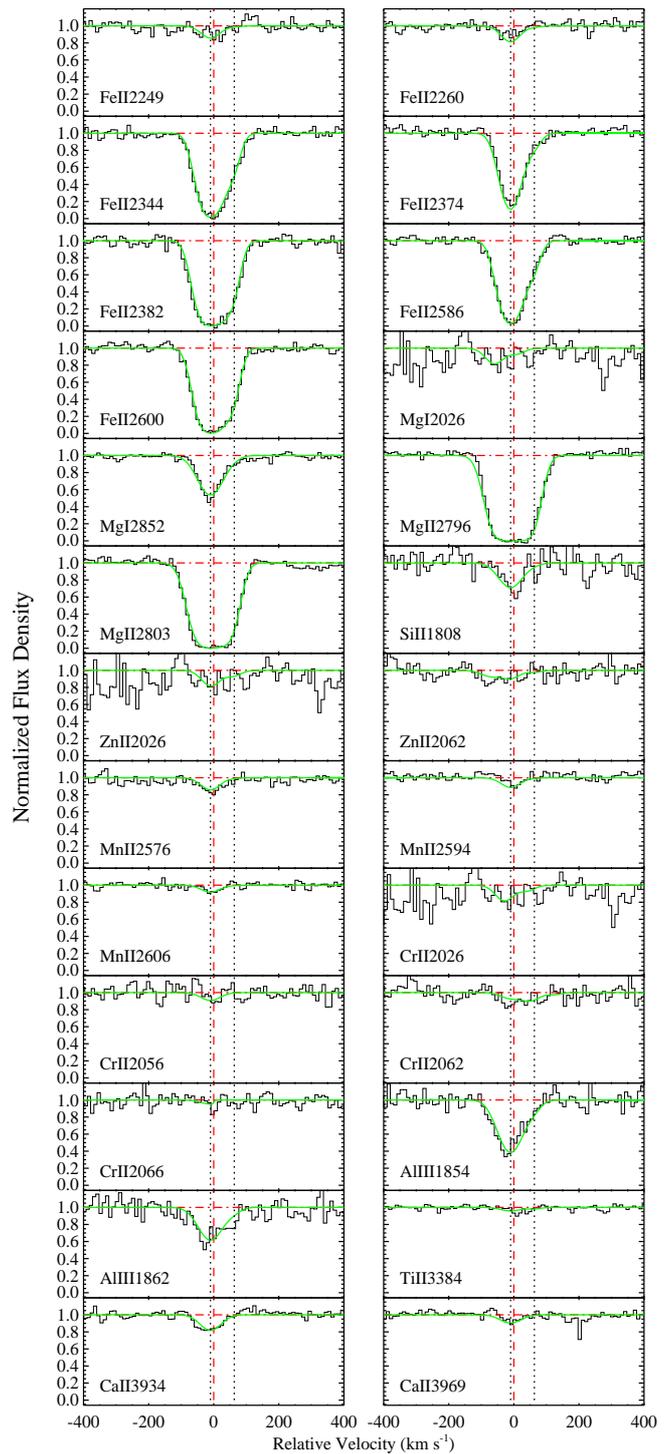}} 
\caption{Velocity profiles of the metal absorption lines in Keck/ESI for the absorber at $z$=1.2375 towards J0953+3225.  }
\label{fig:0953vpfit}
\end{figure}

\begin{table}
\centering
\caption{Total column densities for the $z_{abs}$ = 1.2375 absorber toward J0953+3225  }
\begin{tabular}{lccc}
\hline\hline
Ion       &  log$N_{\rm AODM}$    &log$N_{\rm FIT}$    &log$N_{\rm adopt}$ \\
\hline
Mg\I         &  12.64 $\pm$ 0.02  & 12.63 $\pm$ 0.02    &    12.64 $\pm$ 0.02                          \\
Mg\II        &   $>$ 14.57            &   $>$ 14.63                         &     $>$ 14.63                         \\
Al\III         &  13.55 $\pm$ 0.02  & 13.53 $\pm$ 0.04   &        13.55 $\pm$ 0.02                      \\
Si\II          &  15.46 $\pm$ 0.12  & 15.49 $\pm$ 0.13    &        15.46 $\pm$ 0.12                      \\
Ca\II        &  12.28 $\pm$ 0.08  &  12.39 $\pm$ 0.06    &    12.28 $\pm$ 0.08                         \\
Ti\II          &  12.32 $\pm$ 0.15  &  12.10 $\pm$ 0.25    &       12.32 $\pm$ 0.15                       \\
Cr\II         &  $<$ 13.74   &  13.17 $\pm$ 0.18   &        13.17 $\pm$ 0.18                     \\
Mn\II        &  12.62 $\pm$ 0.06  &  12.66 $\pm$ 0.07    &        12.62 $\pm$ 0.06                      \\
Fe\II         &  14.81 $\pm$ 0.08  &  15.03 $\pm$ 0.03    &    14.81 $\pm$ 0.08                          \\
Zn\II         &  12.98 $\pm$ 0.08  &   12.44 $\pm$ 0.28 &       12.98 $\pm$ 0.08                       \\
\hline
\end{tabular}
\label{tab:0953N}
\end{table}

This quasar spectrum does not have enough data for continuum fitting blueward of the 2175 \AA$ $ bump. The intrinsic quasar spectrum may have more variations than the representative composite quasar spectrum as it is not very well fit to the observed continuum. The composite spectrum remains the best substitute compared to the blue and red templates.  The corresponding best-fit parameters are $c_1$ = -1.320, $c_2$ = 0.110, $c_3$ = 0.270, $x_0$ = 4.658, $\gamma$ = 0.887 with a bump strength of 0.478. 

A great number of low-ionization elements are detected in this absorber including Si\II, Zn\II, Mg\II, Mn\II, Cr\II, Fe\II, Ti\II, and Ca\II. The spectrum also covers the intermediate-ionization lines Al\III{} $\lambda$1854, 1862. Two velocity components at $v$ =  -10, +63 \kms are fit to describe the profiles with most transitions appearing only in the primary component at $v$ = -10 \kms (Figure \ref{fig:0953vpfit}). The Zn\II{} $\lambda$2026, 2062 lines are very weak compared with those in other absorbers. The [Fe/Zn] value is -1.11 $\pm$ 0.11.
All the column density measurements are summarized in Table \ref{tab:0953N}.

 \subsection{J1006+1538 $z_{qso}$ = 2.1950, $z_{abs}$ = 2.2062}

 \begin{figure}
\centering
{\includegraphics[width=8.7cm, height=21.7cm]{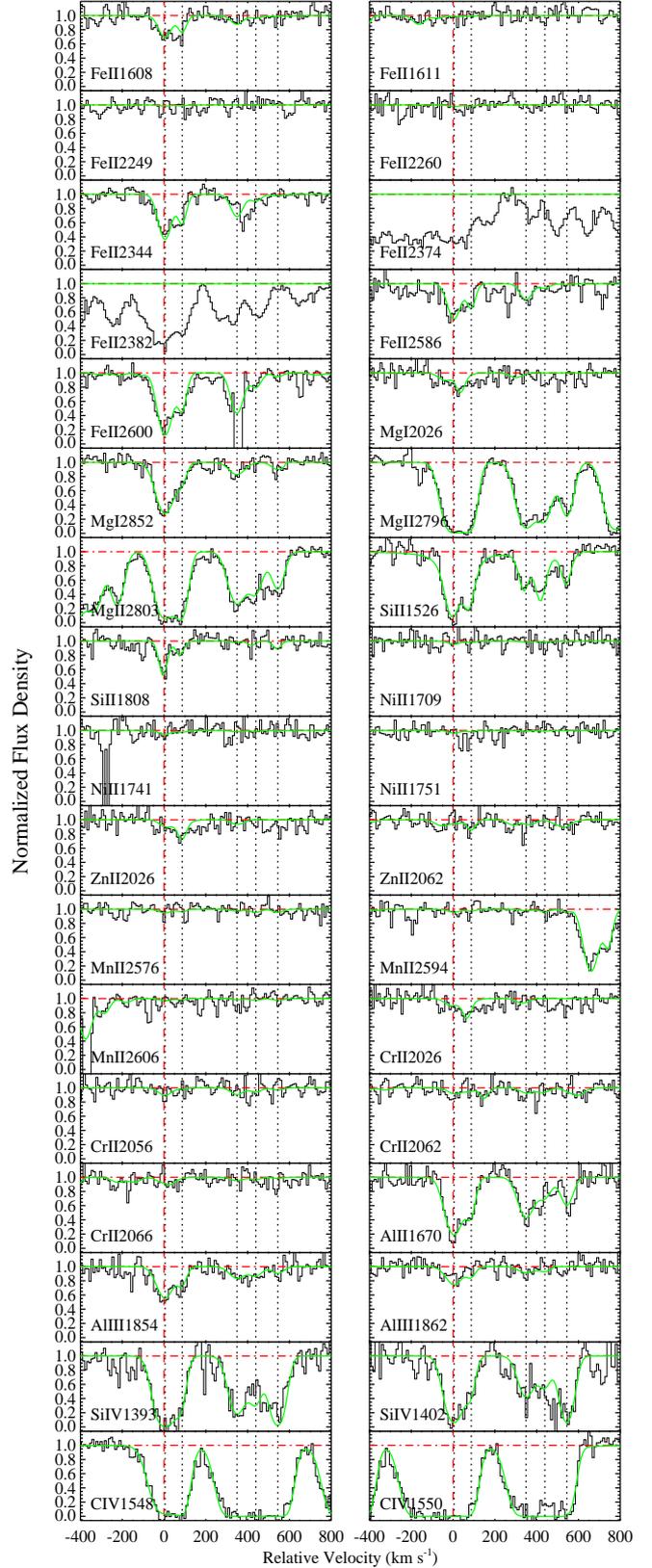}}  
\caption{Velocity profiles of the metal absorption lines in Keck/ESI for the absorber at $z$ $\sim$ 2.2062 towards J1006+1538.  }
\label{fig:1006vpfit}
\end{figure}

\begin{table}
\centering
\caption{Total column densities for the $z_{abs}$ = 2.2062 absorber toward J1006+1538. }
\begin{tabular}{lccc}
\hline\hline
Ion       &  log$N_{\rm AODM}$    & log$N_{\rm FIT}$   & log$N_{\rm adopt}$  \\
\hline
C\IV         &                                &  $>$ 16.23 &$>$   16.23                               \\
Mg\I         &  13.08 $\pm$ 0.02  &13.10 $\pm$ 0.06& 13.10 $\pm$ 0.06                                \\
Mg\II        &   $>$ 14.74            &  $>$ 16.39 &  $>$ 16.39                                                 \\
Al\II        &  13.63 $\pm$ 0.01  &13.72 $\pm$ 0.11  &13.72 $\pm$ 0.11                               \\
Al\III         &  13.60 $\pm$ 0.03  & 13.60 $\pm$ 0.09 &13.60 $\pm$ 0.09                            \\
Si\II           & 15.70 $\pm$ 0.07  & 15.30 $\pm$ 0.19 & 15.30$\pm$ 0.19                               \\
Si\IV         & $>$ 15.00              &  $>$ 15.10   & $>$ 15.10                                \\
Cr\II         &  $<$ 13.65  & $<$13.63& $<$ 13.63                             \\
Mn\II        & $<$ 12.98    & $<$ 13.72 &$<$ 12.98                                \\
Fe\II         & 14.34 $\pm$ 0.02   &  14.36 $\pm$ 0.12 & 14.36 $\pm$ 0.12                              \\
Ni\II          &  $<$ 13.62    &  $<$ 13.70 &$<$ 13.62                                                   \\
Zn\II         & 13.42 $\pm$ 0.05   &  13.21 $\pm$ 0.21 & 13.21 $\pm$ 0.21                               \\
\hline
\end{tabular}
\label{tab:1006N}
\end{table}

The red template provides a better fit than the composite spectrum for this absorber. The best-fit parameters derived from the two-step method are $c_1$ = 1.175, $c_2$ = 0.229, $c_3$ = 0.246, $x_0$ = 4.712, $\gamma$ = 0.892 with a bump strength of 0.433. We should note that simultaneously fitting the 5 parameters would result in a much wider bump, i.e. the entire observed spectrum appears to be a bump while the continuum on both sides of the bump is not constrained due to lack of data. Usually we discard a solution like this, however, in this very system we may not be able to rule out this solution. 
 
The absorber towards J1006+1538 is a unique system in the sense that its absorption redshift is the highest in the sample, even higher than the quasar's emission redshift, which means that the absorber is likely associated with the quasar rather than an intervening system.  Many low-ionization transitions as well as C\IV{} and Si\IV{} lines are present in the system. The two Fe\II{} lines at 2344, 2374 \AA$ $ buried in the sky lines are not included in the simultaneous fitting. In this particular case, we use Si\II{} lines instead of Fe\II{} lines as the reference lines in the simultaneous fitting by tying the redshift and Doppler parameter in each velocity component. The best-fit VPFIT curves comprise of five velocity components at $v$ = -6, +81, +354, +451, +547 \kms (Figure \ref{fig:1006vpfit}). The column density measurements are listed in Table \ref{tab:1006N}. Given the very broad velocity structure, the above-mentioned extremely broad bump, which is derived from simultaneously fitting all the parameters, may not be totally unphysical. The Doppler velocities relative to the quasar at $z$ = 2.1950 are 1050 - 1600 \kms. Such tremendous speeds toward the quasar suggest that the absorbing gas clouds may be falling into the central engine of the quasar  \citep{Lu2007}.  The Fe-to-Zn relative abundance indicates high depletion level with [Fe/Zn] = -1.19 $\pm$ 0.24. We will further discuss the implication of the 2175 \AA$ $ bump appearing in such an associated absorber with the quasar in Section \ref{sec:discussion}.

 \subsection{J1047+3423 $z_{qso}$ = 1.6800, $z_{abs}$ = 1.6685}
 
 \begin{figure}
\centering
{\includegraphics[width=8.7cm, height=20.7cm]{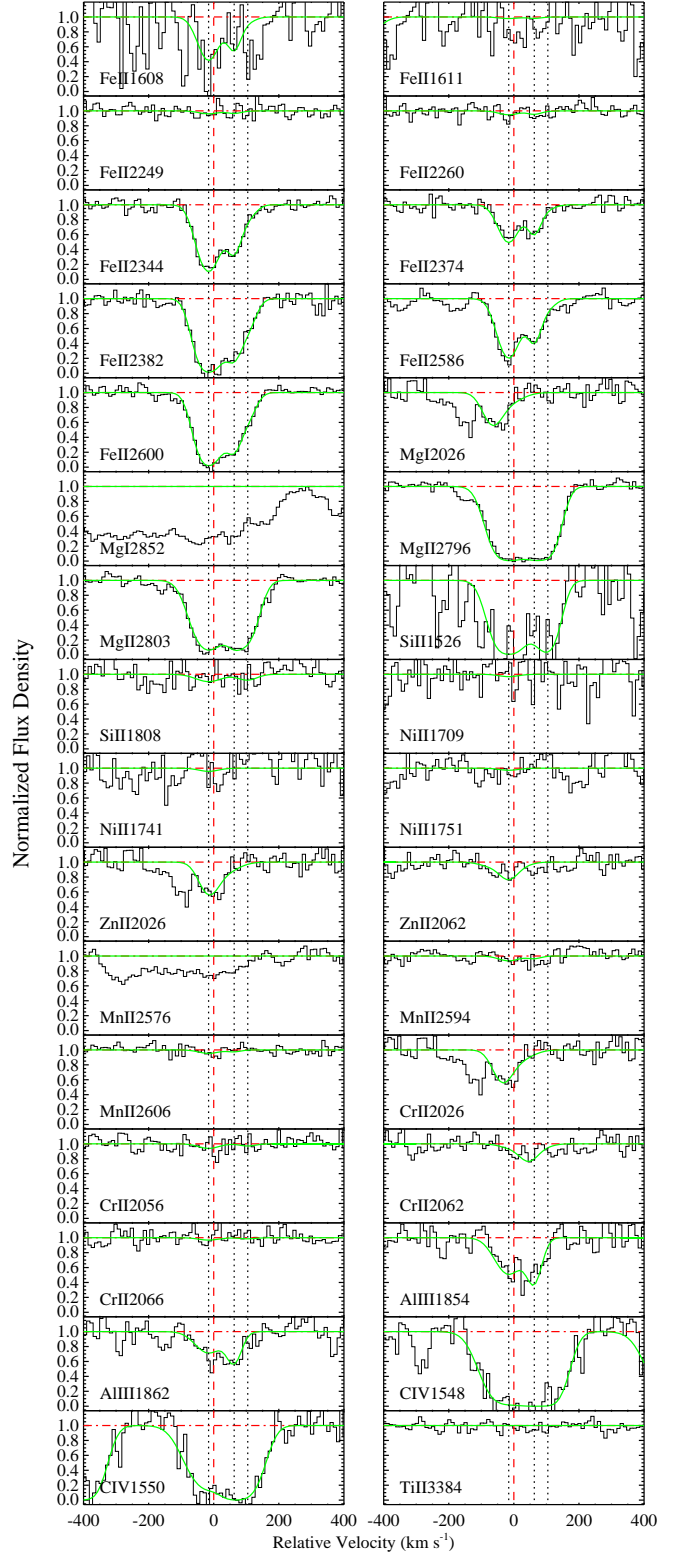}} 
\caption{Velocity profiles of the metal absorption lines in Keck/ESI for the absorber at $z$=1.6685 towards J1047+3423.  }
\label{fig:1047vpfit}
\end{figure}

\begin{table}
\centering
\caption{Total column densities for the $z_{abs}$ = 1.6685 absorber toward J1047+3423  }
\begin{tabular}{lccc}
\hline\hline
Ion       &  log$N_{\rm AODM}$    &log$N_{\rm FIT}$    &log$N_{\rm adopt}$ \\
\hline
C\IV         &  $>$ 15.67              &      $>$ 15.69     &  $>$ 15.69      \\   
Mg\I         &    & 13.36 $\pm$ 0.25      &       13.36 $\pm$ 0.25                   \\
Mg\II        &  $>$ 14.37          &   $>$ 14.38                        &       $>$ 14.38                                   \\
Al\II          & $>$ 13.85           &  13.70 $\pm$ 0.12   &        13.70 $\pm$ 0.12           \\
Al\III         &  13.64 $\pm$ 0.03  & 13.69 $\pm$ 0.07   &      13.64 $\pm$ 0.03                        \\
Si\II          &  15.82 $\pm$ 0.13  & 15.20 $\pm$ 0.22    &       15.82 $\pm$ 0.13                       \\
Ca\II        &  12.28 $\pm$ 0.08  &  12.39 $\pm$ 0.05    &      12.28 $\pm$ 0.08                       \\
Ti\II          &         &  $<$ 12.35    &      $<$ 12.35                       \\
Cr\II         &  $<$ 13.54    &  $<$ 13.08  &         $<$ 13.08                    \\
Mn\II        &  12.76 $\pm$ 0.12  &  12.56 $\pm$ 0.22    &   12.76 $\pm$ 0.12                           \\
Fe\II         &  14.69 $\pm$ 0.02  &  14.66 $\pm$ 0.04    &       14.69 $\pm$ 0.02                         \\
Ni\II          &                                 &   $<$ 13.21              &    $<$ 13.21                               \\
Zn\II         &   $<$ 13.51   &   13.15 $\pm$ 0.09   &        13.15 $\pm$ 0.09                        \\
\hline
\end{tabular}
\label{tab:1047N}
\end{table}

The 2175 \AA$ $  bump in this absorber is the weakest in this sample with $c_1$ = -0.397, $c_2$ = 0.338, $c_3$ = 0.105, $x_0$ = 4.663, $\gamma$ = 0.851, and $A_{\rm bump}$ = 0.194. 

The Keck/ESI spectrum covers C\IV, Mg\I, Mg\II, Al\II, Al\III, Si\II, Ca\II, Ti\II, Cr\II, Mn\II, Fe\II, Ni\II{} and Zn\II{} lines. The absorption redshift is only slightly smaller than the emission redshift of J1047+3423. This is another proximate system in this sample. We simultaneously fit these lines except Mg\I{} $\lambda$2852 and Mn\II{} $\lambda$2576 where they are severely contaminated by sky lines or unknown absorption. The profiles are fit with three velocity components at $v$ = -16, +63, +104 \kms as shown in Figure \ref{fig:1047vpfit}. The corresponding velocities relative to the quasar are -1302, -1223, -1182 \kms, indicating that the absorbing materials may be outflowing from the quasar toward us. 

The column densities are summarized in Table \ref{tab:1047N}. The depletion level  of this absorber is among the highest with [Fe/Zn] = -1.40 $\pm$ 0.09. Detecting a 2175 \AA$ $ bump in quasar outflows requires certain favorable environment. We will further discuss this absorber together with the absorber towards J1006+1538 in Section \ref{sec:discussion}.

\subsection{J1127+2424 $z_{qso}$ = 2.0787, $z_{abs}$ = 1.2110}

\begin{figure}
\centering
{\includegraphics[width=8.7cm, height=18.7cm]{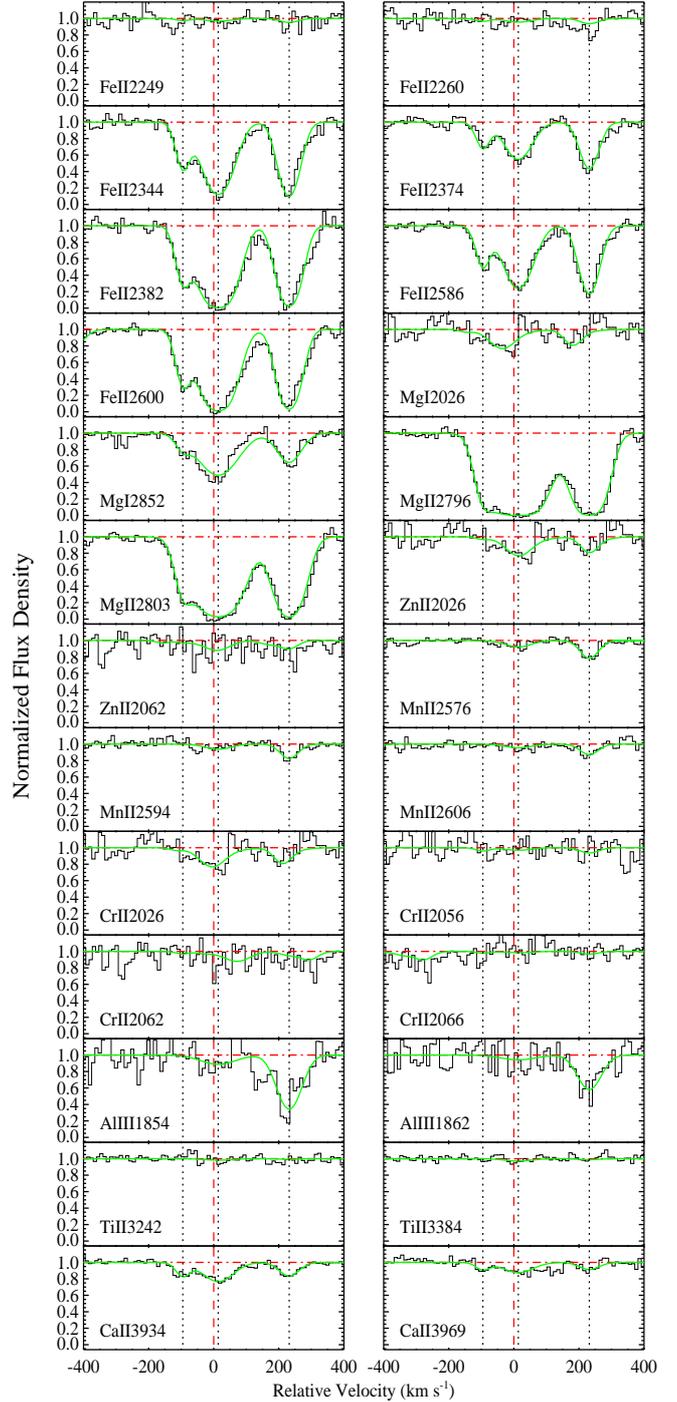}}  
\caption{Velocity profiles of the metal absorption lines in Keck/ESI for the absorber at $z$=1.2110 towards J1127+2424.  }
\label{fig:1127vpfit}
\end{figure}

\begin{table}
\centering
\caption{Total column densities for the $z_{abs}$ = 1.2110 absorber toward J1127+2424  }
\begin{tabular}{lccc}
\hline\hline
Ion       &  log$N_{\rm AODM}$    &log$N_{\rm FIT}$    &log$N_{\rm adopt}$ \\
\hline
Mg\I         &  13.01 $\pm$ 0.01  & 13.03 $\pm$ 0.01   &   13.01 $\pm$ 0.01                          \\
Mg\II        &   $>$ 14.87            &  $>$ 14.63                        &     $>$ 14.87                               \\
Al\III         &  13.64 $\pm$ 0.04  & 13.60 $\pm$ 0.05   &       13.64 $\pm$ 0.04                       \\
Si\II          &  $<$ 17.05              & & $<$ 17.05    \\
Ca\II        &  12.92 $\pm$ 0.02  &  12.91 $\pm$ 0.02    &      12.92 $\pm$ 0.02                         \\
Ti\II          &  $<$ 12.70    &  12.07 $\pm$ 0.23    &        $<$ 12.70                     \\
Cr\II         &  $<$ 13.87  & 13.24 $\pm$ 0.23 &      13.24 $\pm$ 0.23                       \\
Mn\II        &  13.01 $\pm$ 0.04   &  13.01 $\pm$ 0.03   &        13.01 $\pm$ 0.04                       \\
Fe\II         & 14.84 $\pm$ 0.01   &  14.87 $\pm$ 0.01   &     14.84 $\pm$ 0.01                         \\
Zn\II         & 12.87 $\pm$ 0.07   &  13.15 $\pm$ 0.07  &         13.15 $\pm$ 0.07                    \\
\hline
\end{tabular}
\label{tab:1127N}
\end{table}

The best-fit extinction curve gives $c_1$ = -0.524, $c_2$ = 0.143, $c_3$ = 0.220, $x_0$ = 4.650, $\gamma$ = 0.837, and $A_{\rm bump}$ = 0.413. 

The Keck/ESI spectrum covers the commonly observed low-ionization lines of the following species: Zn\II, Fe\II, Mg\II, Mn\II, Cr\II, Ti\II, Ca\II, and the intermediate-ionization element Al\III. The velocity profiles are resolved into three major components at $v$ = -95, +14, +232 \kms as shown in Figure \ref{fig:1127vpfit}. The velocity component at $z$ = 1.2127 is not blended with the other two and is about 200 \kms away. The two components are $v$ = +14 and $v$ = +232 \kms contain comparable amounts of column densities for most species except Al\III{} which contributes mostly in the $v$ = +232 \kms component, almost an order of magnitude higher than in the $z$ = 1.2111 component. This absorber has a [Fe/Zn] value of -1.25 $\pm$ 0.07.  The column densities measurements are listed in Table \ref{tab:1127N}.

\subsection{J1130+1850 $z_{qso}$ = 2.7536, $z_{abs}$ = 2.0119}

\begin{figure}
\centering
{\includegraphics[width=8.7cm, height=21.4cm]{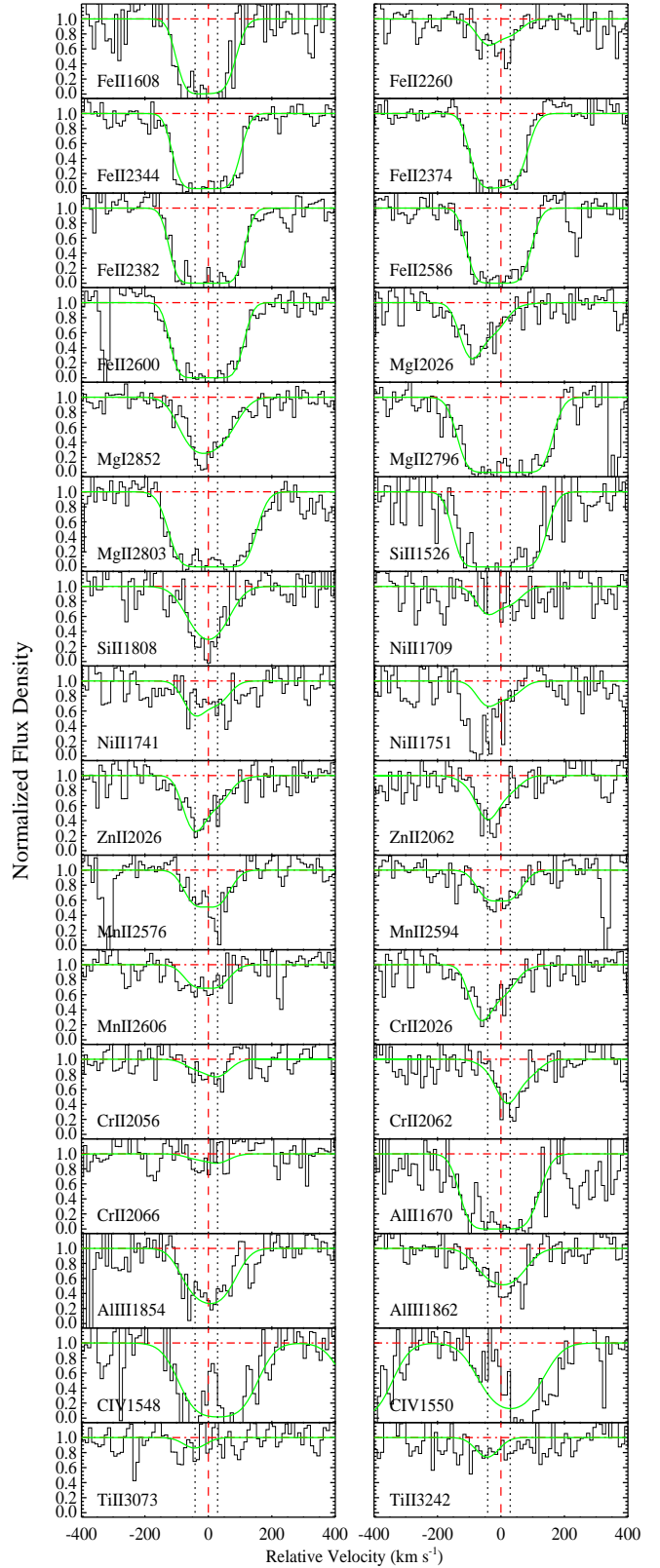}}  
\caption{Velocity profiles of the metal absorption lines in Keck/ESI for the absorber at $z$=2.0119 towards J1130+1850.  }
\label{fig:1130vpfit}
\end{figure}

\begin{table}
\centering
\caption{Total column densities for the $z_{abs}$ = 2.0119 absorber toward J1130+1850  }
\begin{tabular}{lccc}
\hline\hline
Ion       &  log$N_{\rm AODM}$    &log$N_{\rm FIT}$    &log$N_{\rm adopt}$ \\
\hline
C\IV         &     $>$ 15.60                            &  $>$ 14.96  &$>$ 15.60    \\
Mg\I         &  13.26 $\pm$ 0.02  & 13.19 $\pm$ 0.13   &   13.26 $\pm$ 0.02                           \\
Mg\II        &    $>$ 14.91             &  $>$ 15.18                        &         $>$ 15.18                              \\
Al\II         &  $>$ 14.47    & $>$ 14.34   &       $>$ 14.47                      \\
Al\III         & 13.90 $\pm$ 0.02  & 13.87 $\pm$ 0.15   &     13.90 $\pm$ 0.02                        \\
Si\II          &  16.60 $\pm$ 0.02   &         16.14 $\pm$ 0.22  &  16.60 $\pm$ 0.02  \\
Ti\II          &  $<$ 12.70    &  12.07 $\pm$ 0.23    &    12.07 $\pm$ 0.23                          \\
Cr\II         &  13.83 $\pm$ 0.16  & 13.77 $\pm$ 0.29 &     13.83 $\pm$ 0.16                        \\
Mn\II        &  13.71 $\pm$ 0.02   &  13.61 $\pm$ 0.11   & 13.71 $\pm$ 0.02                             \\
Fe\II         & 15.91 $\pm$ 0.01   &  15.56 $\pm$ 0.15   &     15.91 $\pm$ 0.01                       \\
Ni\II         &  14.71 $\pm$ 0.07  & 14.60 $\pm$ 0.17  &    14.60 $\pm$ 0.17                          \\
Zn\II         & 13.70 $\pm$ 0.03   &  13.72 $\pm$ 0.15  &     13.70 $\pm$ 0.03                         \\
\hline
\end{tabular}
\label{tab:1130N}
\end{table}

The continuum on both sides of the 2175 \AA$ $ bump is well constrained. The best-fit parameters are $c_1$ = 1.471, $c_2$ = 0.274, $c_3$ = 0.362, $x_0$ = 4.628, $\gamma$ = 0.887 with a bump strength of 0.641. 

Many low-ionization lines as well as high-ionization lines C\IV{} $\lambda$1548, 1550 and Si\IV $\lambda$1393, 1402 are covered in the Keck/ESI spectrum. The Si\IV{} $\lambda$1393, 1402 lines fall in the Lyman-$\alpha$ forest therefore we do not include them in the simultaneous fitting. The profiles are fit with two velocity components at $v$ = -42, +29 \kms (Figure \ref{fig:1130vpfit}). The [Fe/Zn] for this absorber is -0.73 $\pm$ 0.03. We list the column density measurements in Table \ref{tab:1130N}.

\subsection{J1138+5245 $z_{qso}$ = 1.6337, $z_{abs}$ = 1.1788}

\begin{figure}
\centering
{\includegraphics[width=8.7cm, height=17.5cm]{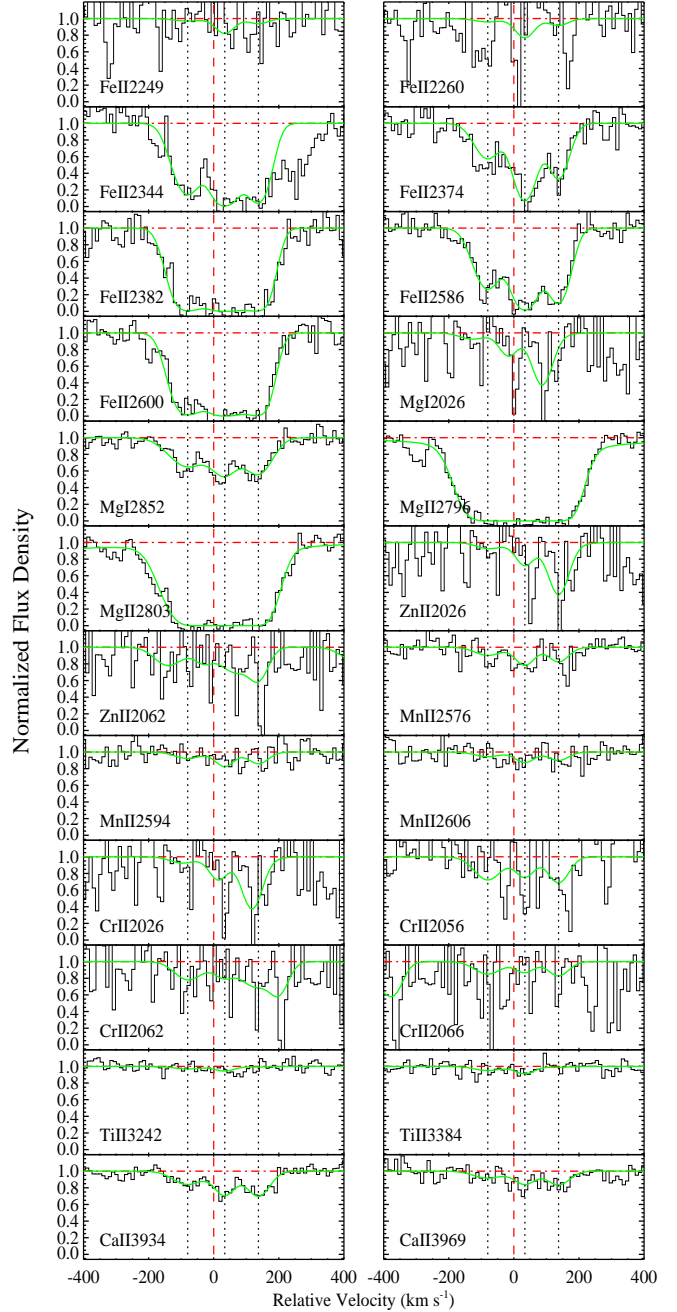}}  
\caption{Velocity profiles of the metal absorption lines in Keck/ESI for the absorber at $z$=1.1788 towards J1138+5245.  }
\label{fig:1138vpfit}
\end{figure}

\begin{table}
\centering
\caption{Total column densities for the $z_{abs}$ = 1.1788 absorber toward J1138+5245  }
\begin{tabular}{lccc}
\hline\hline
Ion       &  log$N_{\rm AODM}$    &log$N_{\rm FIT}$    &log$N_{\rm adopt}$ \\
\hline
Mg\I         &  13.08 $\pm$ 0.01  & 13.09 $\pm$ 0.03   &      13.08 $\pm$ 0.01                        \\
Mg\II        &  $>$ 15.29            &  $>$ 17.85                        &  $>$ 17.85                                 \\
Al\III         &  13.64 $\pm$ 0.04  & $>$ 14.70   &     $>$ 14.70                        \\
Ca\II        &  13.09 $\pm$ 0.03  &  13.07 $\pm$ 0.03    &      13.09 $\pm$ 0.03                       \\
Ti\II          &  $<$ 13.12         &  12.61 $\pm$ 0.12    &    12.61 $\pm$ 0.12                          \\
Cr\II         &  $<$ 14.13    & 14.17 $\pm$ 0.10 &          14.17 $\pm$ 0.10                    \\
Mn\II        &  13.36 $\pm$ 0.06   &  13.24 $\pm$ 0.06   &    13.36 $\pm$ 0.06                         \\
Fe\II         & 15.88 $\pm$ 0.15   &  15.31 $\pm$ 0.07   &       15.31 $\pm$ 0.07                      \\
Zn\II         & $<$ 14.19   &  13.58 $\pm$ 0.12  &     13.58 $\pm$ 0.12                       \\
\hline
\end{tabular}
\label{tab:1138N}
\end{table}

The 2175 \AA$ $ bump in this absorber is relatively both very wide and deep with $c_1$ = -0.224, $c_2$ = 0.298, $c_3$ = 0.425, $x_0$ = 4.583, $\gamma$ = 0.908 and a bump strength of 0.735. 

The Keck/ESI spectrum covers the following elements: Mg\I, Mg\II, Al\III, Ca\II, Ti\II, Cr\II, Mn\II, Fe\II, and Zn\II. However, many of them are affected by the low S/N. We simultaneously fit all the lines using VPFIT with three velocity components at $v$ = -80, +34, +137 \kms as shown in Figure \ref{fig:1138vpfit}. The [Fe/Zn] value for this absorber is -1.21 $\pm$ 0.14. The AODM-derived and VPFIT-derived column densities are summarized in Table \ref{tab:1138N}.

\subsection{J1212+2936 $z_{qso}$ = 1.3919, $z_{abs}$ = 1.2202}

\begin{figure}
\centering
{\includegraphics[width=8.7cm, height=19cm]{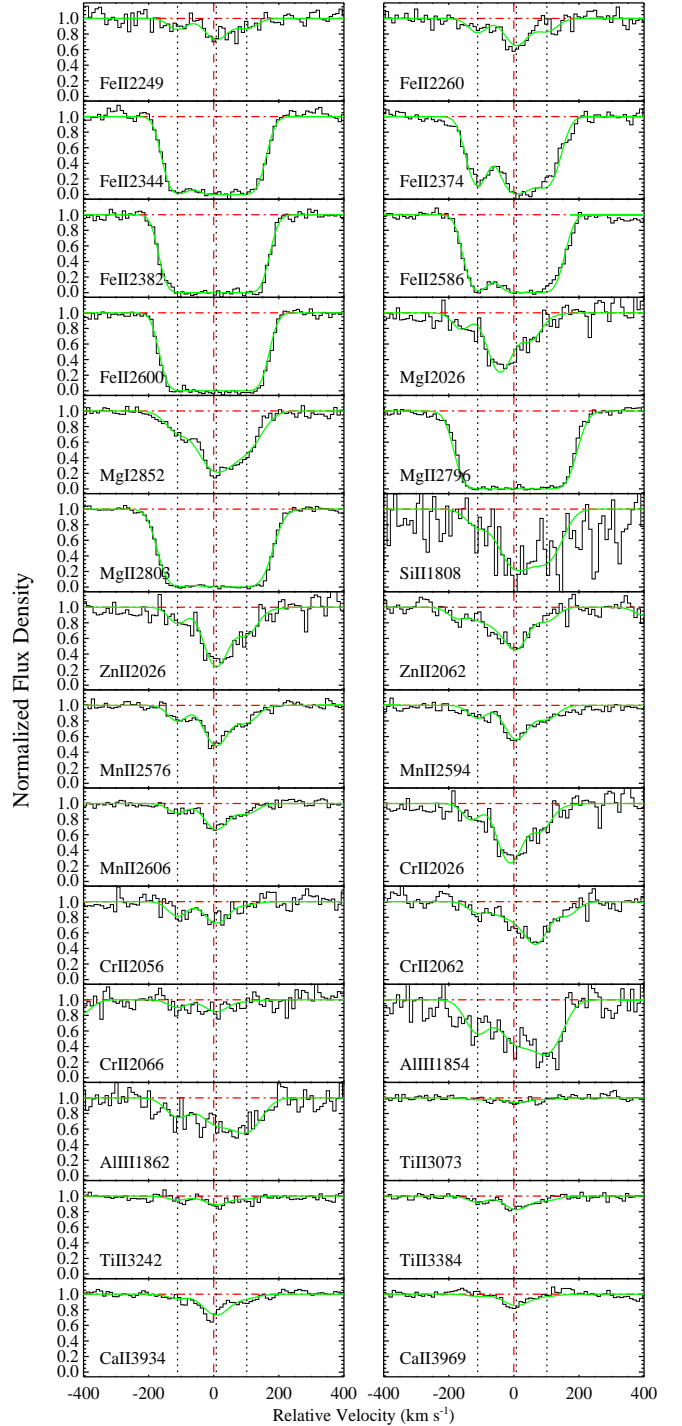}} 
\caption{Velocity profiles of the metal absorption lines in Keck/ESI for the absorber at $z$=1.2202 towards J1212+2936.  }
\label{fig:1212vpfit}
\end{figure}

\begin{table}
\centering
\caption{Total column densities for the $z_{abs}$ = 1.2202 absorber toward J1212+2936  }
\begin{tabular}{lccc}
\hline\hline
Ion       &  log$N_{\rm AODM}$    &log$N_{\rm FIT}$    &log$N_{\rm adopt}$ \\
\hline
Mg\I         &  13.29 $\pm$ 0.01  & 13.29 $\pm$ 0.01   &    13.29 $\pm$ 0.01                         \\
Mg\II        &   $>$ 15.17            &     $>$ 14.91                  &     $>$ 15.17                        \\
Si\II         & 16.93 $\pm$ 0.01   &  16.47 $\pm$ 0.10   &     16.47 $\pm$ 0.10                       \\
Al\III         &  14.01 $\pm$ 0.02  & 13.97 $\pm$ 0.03   &    14.01 $\pm$ 0.02                          \\
Ca\II        &  12.73 $\pm$ 0.03  &  12.77 $\pm$ 0.03    &   12.73 $\pm$ 0.03                           \\
Ti\II          &  13.01 $\pm$ 0.03    &  12.94 $\pm$ 0.03    &     13.01 $\pm$ 0.03                         \\
Cr\II         &  13.88 $\pm$ 0.05 & 13.92 $\pm$ 0.04 &      13.88 $\pm$ 0.05                       \\
Mn\II        &  13.62 $\pm$ 0.04   &  13.61 $\pm$ 0.01   &    13.62 $\pm$ 0.04                         \\
Fe\II         & 15.63 $\pm$ 0.03   &  15.65 $\pm$ 0.02   &        15.63 $\pm$ 0.03                     \\
Zn\II         & 13.78 $\pm$ 0.01   &  13.79 $\pm$ 0.02  &     13.78 $\pm$ 0.01                        \\
\hline
\end{tabular}
\label{tab:1212N}
\end{table}

The quasar spectrum of J1212+2936 resembles that of J0901+2044 with a steep rising slope towards shorter wavelengths. The best-fit extinction curve yields $c_1$ = -1.019, $c_2$ = 0.211, $c_3$ = 0.245, $x_0$ = 4.584, $\gamma$ = 0.899 and a bump strength of $A_{\rm bump}$ = 0.428. 

Multiple transitions are covered in the Keck/ESI spectrum for this absorber: seven Fe\II{} lines, Si\II{} $\lambda$ 1808, Zn\II{} $\lambda$2026, 2062, Mg\II $\lambda$2796, 2803, Mn\II{} $\lambda$2576, 2594, 2606, Al\III{} $\lambda$1854, 1862, Cr\II{} $\lambda$2026. 2056, 2062, Ti\II{} $\lambda$3073, 3242, 3384, Ca\II{} $\lambda$ 3934, 3969.  Three velocity components at $v$ = -111, +8, +101 \kms are fit to all the low-ionization lines and Al\III{} lines (Figure \ref{fig:1212vpfit}). The central component is the strongest one for all species. 

This absorber has both relatively high Fe\II{} and Zn\II{} column densities with logN(Fe\II{}) = 15.63 $\pm$ 0.03 and logN(Zn\II{}) = 13.78 $\pm$ 0.01. The resulting relative abundance [Fe/Zn] is -1.09 $\pm$ 0.03. The column densities of Ti\II{} and Ca\II{} are logN(Ti\II{}) = 13.01 $\pm$ 0.03 and logN(Ca\II{}) = 12.73 $\pm$ 0.03. All the column densities are listed in Table \ref{tab:1212N}.

\subsection{J1321+2135 $z_{qso}$ = 2.4113, $z_{abs}$ = 2.1253}

\begin{figure}
\centering
{\includegraphics[width=8.7cm, height=22.3cm]{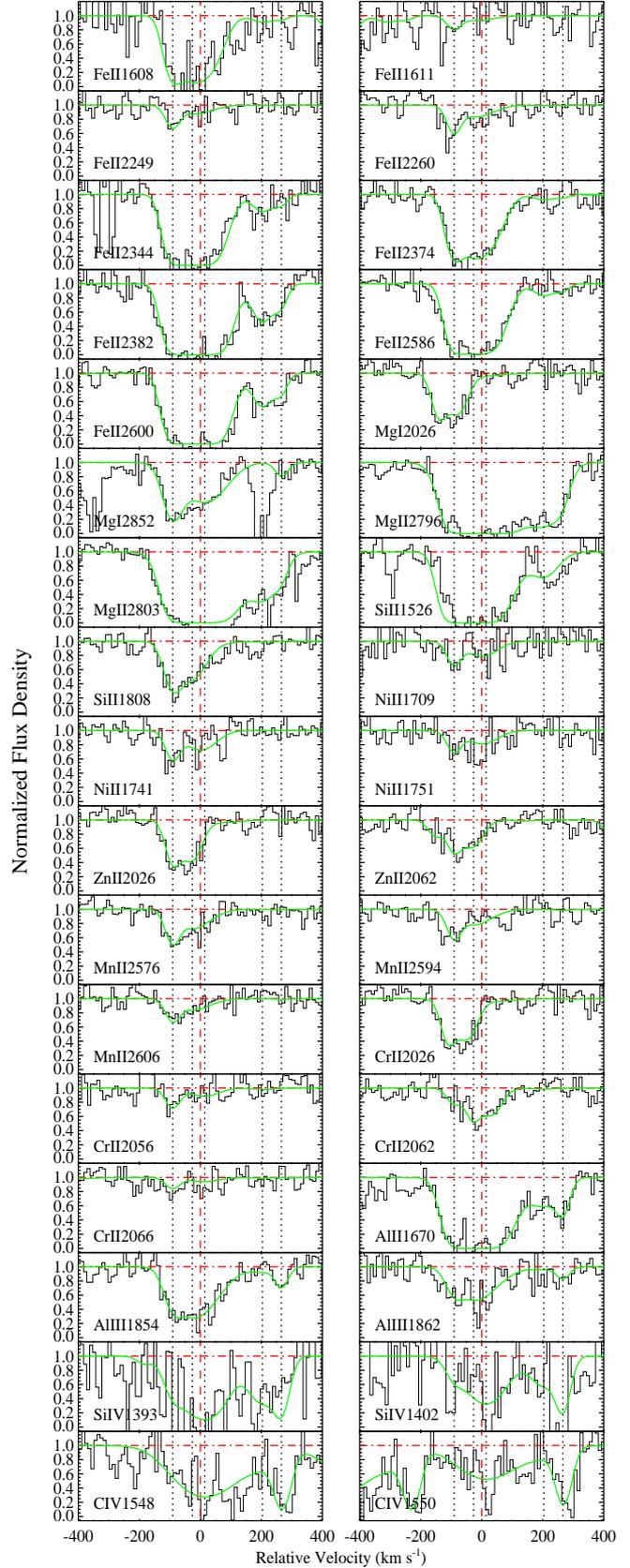}}  
\caption{Velocity profiles of the metal absorption lines in Keck/ESI for the absorber at $z$=2.1253 towards J1321+2135.  }
\label{fig:1321vpfit}
\end{figure}

\begin{table}
\centering
\caption{Total column densities for the $z_{abs}$ = 2.1253 absorber toward J1321+2135 }
\begin{tabular}{lccc}
\hline\hline
Ion       &  log$N_{\rm AODM}$    &log$N_{\rm FIT}$    &log$N_{\rm adopt}$ \\
\hline
C\IV         & 15.11 $\pm$ 0.01   &     14.86 $\pm$ 0.15                           &    15.11 $\pm$ 0.01                            \\
Mg\I         &  13.81 $\pm$ 0.01  & 13.28 $\pm$ 0.04    &         13.28 $\pm$ 0.04                     \\
Mg\II        &  $>$ 15.31            &   $>$ 14.91                         &    $>$ 15.31                            \\
Al\II          &  $>$ 14.45                  &  $>$ 14.72                          &     $>$ 14.72                              \\
Al\III         &  13.99 $\pm$ 0.02  & 13.97 $\pm$ 0.07    &    13.99 $\pm$ 0.02                          \\
Si\II          &   16.26  $\pm$ 0.02 & 16.20 $\pm$ 0.06    &      16.26  $\pm$ 0.02                       \\
Si\IV        &    $>$ 15.27                           & $>$ 14.80 &  $>$ 15.27         \\
Cr\II         &  $<$ 14.35      &  13.81 $\pm$ 0.09    &   13.81 $\pm$ 0.09                         \\
Mn\II        &   13.56 $\pm$ 0.03  &  13.51 $\pm$ 0.07    &       13.56 $\pm$ 0.03                       \\
Fe\II         &  15.84 $\pm$ 0.02 &  15.56 $\pm$ 0.05   &  15.84 $\pm$ 0.02                           \\
Ni\II          &  14.62 $\pm$ 0.06 &  14.55 $\pm$ 0.08    &    14.62 $\pm$ 0.06                          \\
Zn\II         &  13.61 $\pm$ 0.03  &   13.68$\pm$ 0.06   &      13.61 $\pm$ 0.03                         \\
\hline
\end{tabular}
\label{tab:1321N}
\end{table}

The best-fit extinction curve yields $c_1$ = 1.587, $c_2$ = 0.202, $c_3$ = 0.191, $x_0$ = 4.494, $\gamma$ = 0.917 with a bump strength of 0.328. 

A great number of low-ionization lines including Mg\I, Mg\II, Al\II, Al\III, Si\II, Cr\II, Mn\II, Fe\II, Ni\II, and Zn\II{} and high-ionization lines C\IV, Si\IV{} are present in the Keck/ESI spectrum. At least five components are required to fit the velocity profiles with the three components at $v$ = -91, -27, +14 \kms being the primary components while the other two are located at +203 and +266 \kms as shown in Figure \ref{fig:1321vpfit}. The [Fe/Zn] value for this absorber is -0.71 $\pm$ 0.04. We summarize the AODM-derived and VPFIT-derived column densities in Table \ref{tab:1321N}.

\subsection{J1531+2403 $z_{qso}$ = 2.5256, $z_{abs}$ = 2.0022}

\begin{figure}
\centering
{\includegraphics[width=8.7cm, height=22.3cm]{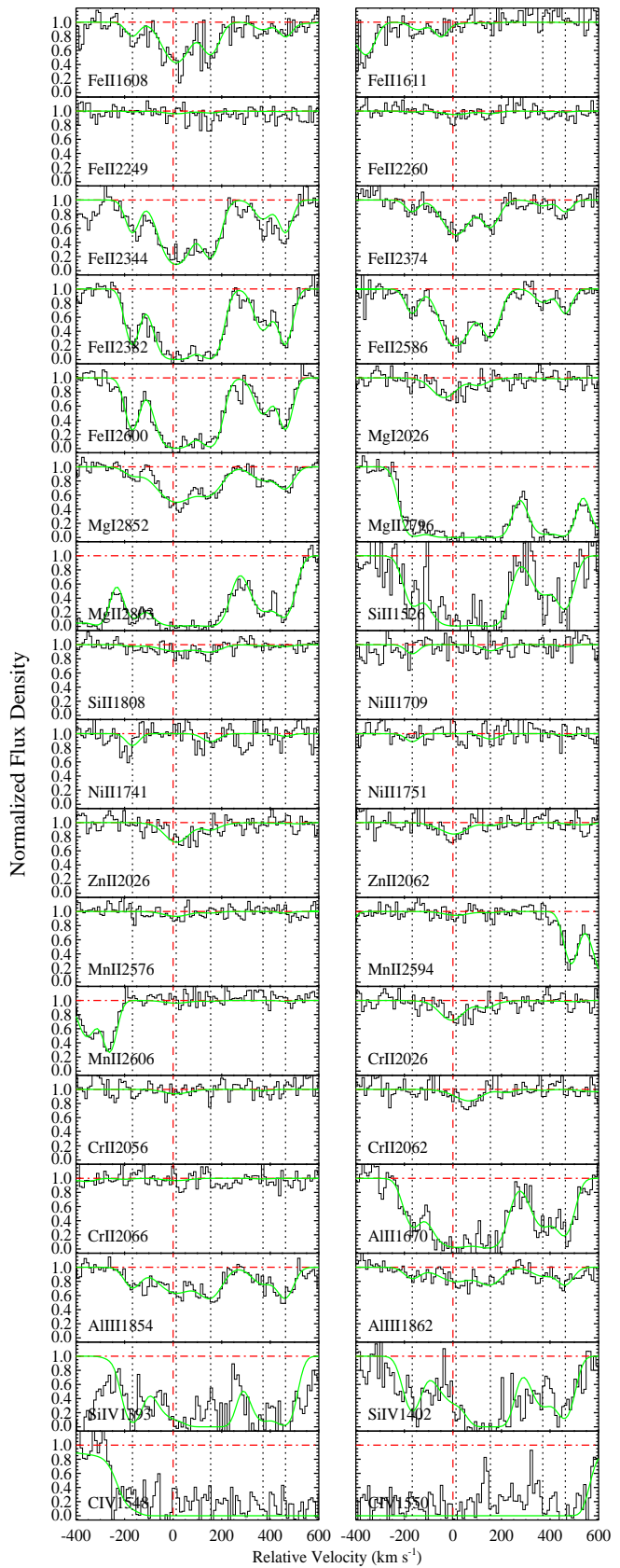}}  
\caption{Velocity profiles of the metal absorption lines in Keck/ESI for the absorber at $z$=2.0022 towards J1531+2403.  }
\label{fig:1531vpfit}
\end{figure}

\begin{table}
\centering
\caption{Total column densities for the $z_{abs}$ = 2.0022 absorber toward J1531+2403 }
\begin{tabular}{lccc}
\hline\hline
Ion       &  log$N_{\rm AODM}$    &log$N_{\rm FIT}$    &log$N_{\rm adopt}$ \\
\hline
C\IV         & $>$ 15.76  &     19.59 $\pm$ 0.19                    &      $>$ 15.76                        \\
Mg\I         &  13.15 $\pm$ 0.01 & 13.23 $\pm$ 0.02    &       13.15 $\pm$ 0.01                      \\
Mg\II        &  $>$ 15.31             &   $>$ 14.94                        &  $>$ 15.31                             \\
Al\II          &  $>$ 14.59                  & 14.27 $\pm$ 0.06                         &     $>$ 14.59                             \\
Al\III         &  13.94 $\pm$ 0.01  & 13.94 $\pm$ 0.03    &    13.94 $\pm$ 0.01                         \\
Si\II          &   15.83 $\pm$ 0.01  & 15.51 $\pm$ 0.07    &      15.83 $\pm$ 0.01                       \\
Si\IV        &      $>$ 15.58                           & $>$ 15.50 &         $>$ 15.58     \\
Cr\II         &  $<$ 13.60     &  13.24 $\pm$ 0.25    &       13.24 $\pm$ 0.25                       \\
Mn\II        &   13.56 $\pm$ 0.03  &  12.69 $\pm$ 0.17    &  13.56 $\pm$ 0.03                             \\
Fe\II         &  14.90 $\pm$ 0.01 &  14.90 $\pm$ 0.01   &   14.90 $\pm$ 0.01                           \\
Ni\II          &  14.15 $\pm$ 0.14 &  14.10 $\pm$ 0.12    &   14.15 $\pm$ 0.14                          \\
Zn\II         &  13.21 $\pm$ 0.06  &   13.20$\pm$ 0.08   &  13.21 $\pm$ 0.06                            \\
\hline
\end{tabular}
\label{tab:1531N}
\end{table}

The continuum on both sides of the 2175 \AA$ $ bump is well fit and the derived extinction curve gives $c_1$ = 0.796, $c_2$ = 0.233, $c_3$ = 0.322, $x_0$ = 4.536, $\gamma$ = 0.900 with a bump strength of 0.562. 

The velocity profiles of all the absorption lines in this absorber (Mg\I, Mg\II, Al\II, Al\III, Si\II, Cr\II, Mn\II, Fe\II, Ni\II, Zn\II, C\IV, Si\IV) are best fit with five components at $v$ =  -168, +13, +155, +370, +463 \kms, spreading over 600 \kms (Figure \ref{fig:1531vpfit}). The Fe-to-Zn relative abundance of [Fe/Zn] = -1.25 $\pm$ 0.06 indicates a high depletion level.  All the column density measurements are shown in Table \ref{tab:1531N}.

\subsection{J1737+4406 $z_{qso}$ = 1.9564, $z_{abs}$ = 1.6135}

\begin{figure}
\centering
{\includegraphics[width=8.7cm, height=22.3cm]{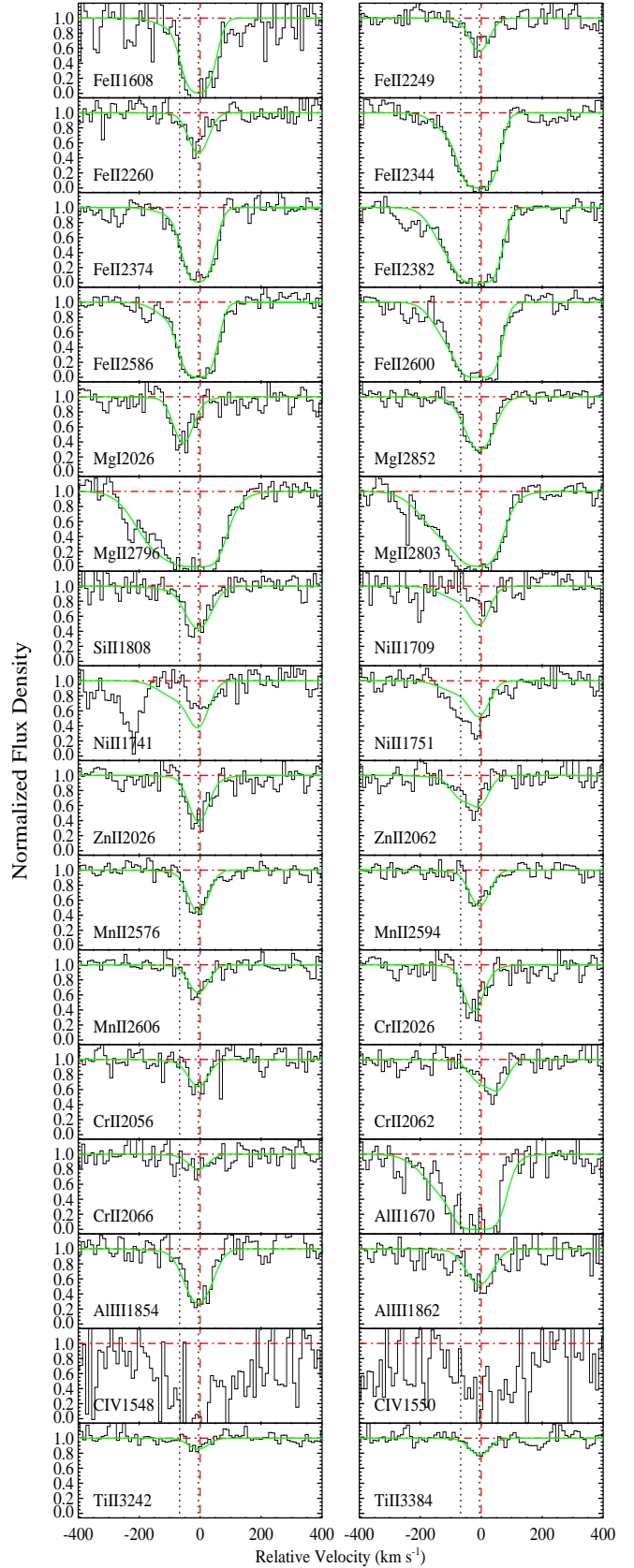}}  
\caption{Velocity profiles of the metal absorption lines in Keck/ESI for the absorber at $z$=1.6135 towards J1737+4406.  }
\label{fig:1737vpfit}
\end{figure}

\begin{table}
\centering
\caption{Total column densities for the $z_{abs}$ = 1.6135 absorber toward J1737+4406  }
\begin{tabular}{lccc}
\hline\hline
Ion       &  log$N_{\rm AODM}$    &log$N_{\rm FIT}$    &log$N_{\rm adopt}$ \\
\hline
C\IV         &  15.44 $\pm$ 0.01 &                           &     15.44 $\pm$ 0.01                      \\
Mg\I         &  12.92 $\pm$ 0.02  & 12.98 $\pm$ 0.04   &     12.92 $\pm$ 0.02                        \\
Mg\II        &   $>$ 15.06           &    $>$ 14.55                         &   $>$ 15.06                            \\
Si\II         & 15.89 $\pm$ 0.06   &  15.94 $\pm$ 0.11   &   15.89 $\pm$ 0.06                         \\
Al\II         &  $>$ 14.43      & $>$ 14.40  &       $>$ 14.43                       \\
Al\III         &  13.64 $\pm$ 0.03  & 13.68 $\pm$ 0.06   &    13.64 $\pm$ 0.03                         \\
Ti\II          &  12.72 $\pm$ 0.07    &  12.81 $\pm$ 0.10    &     12.72 $\pm$ 0.07                        \\
Cr\II         &  13.88 $\pm$ 0.10 & 13.83 $\pm$ 0.11 &   13.88 $\pm$ 0.10                           \\
Mn\II        &  13.42 $\pm$ 0.03   &  13.44 $\pm$ 0.04   &    13.42 $\pm$ 0.03                          \\
Fe\II         & 15.48 $\pm$ 0.07   &  15.63 $\pm$ 0.05   &  15.48 $\pm$ 0.07                          \\
Ni\II          & 14.31 $\pm$ 0.06   &  14.72 $\pm$ 0.05    &      14.31 $\pm$ 0.06                       \\
Zn\II         & 13.54 $\pm$ 0.04   &  13.43 $\pm$ 0.08  &    13.54 $\pm$ 0.04                        \\
\hline
\end{tabular}
\label{tab:1737N}
\end{table}

The quasar spectrum clearly shows the requirement of a significant 2175 \AA$ $ bump. The best-fit parameters are $c_1$ = 0.141, $c_2$ = 0.280, $c_3$ = 0.397, $x_0$ = 4.573, $\gamma$ = 0.909 with a bump strength of 0.686. 

This Keck/ESI spectrum covers the transitions of the following species: C\IV, Fe\II, Zn\II, Mg\II, Si\II, Ni\II, Cr\II, Mn\II, Al\II, Al\III, and Ti\II.  Simultaneous fitting using VPFIT results in two velocity components at $v$ = -67, -5 \kms (Figure \ref{fig:1737vpfit}). There is a significant discrepancy between the observed Ni\II{} profiles and the fits. The discrepancy may be due to noise/contamination or may arise from a different velocity structure than the one from the simultaneous fitting. The VPFIT-derived column densities agree well with the AODM-derived ones expect Ni\II. 

The dust depletion indicator [Fe/Zn] is -1.00 $\pm$ 0.08 with logN(Fe\II) = 15.48 $\pm$ 0.07 and logN(Zn\II) = 13.54 $\pm$ 0.04. All the column density measurements are shown in Table \ref{tab:1737N}.

\section{Discussion}
\label{sec:discussion}

\subsection{Relative abundances}

In summary, Fe\II{} and Zn\II{} are detected in all the 2DAs in this sample and Mg\II{} is saturated in all cases. The low-ionization lines, Si\II, Al\II, Ca\II, Ti\II, Mn\II, Cr\II, and Ni\II, are detected in 11, 8, 7, 7, 13, 12 and 7 absorbers, respectively. The observed relative abundances are the result of both the nucleosynthetic processes and dust depletion. The ratios of depleted elements (X = {Fe, Cr, Ca, Ti, Mn, Ni} etc.) to undepleted (Zn) elements provide a measure of the dust content in the quasar absorbers. Compared to the solar ratio [X/Zn] = 0, the more negative the relative abundance, the larger is element X depleted from its gas phase to the dust phase. These elements are classified into two categories, ``$\alpha$ elements" and ``Fe-peak elements", that have different origins in the history of star formation. $\alpha$-elements (O, N, Mg, Si, S, Ti, Ca, ...) originate mainly from core-collapse Type II supernovae (SNe), whereas Fe-peak elements (V, Cr, Mn, Fe, Co, Ni, ...) are primarily synthesized by Type Ia SNe.

\subsection{Depletion pattern}

\begin{figure*}
\centering
{\includegraphics[width=14cm, height=10cm]{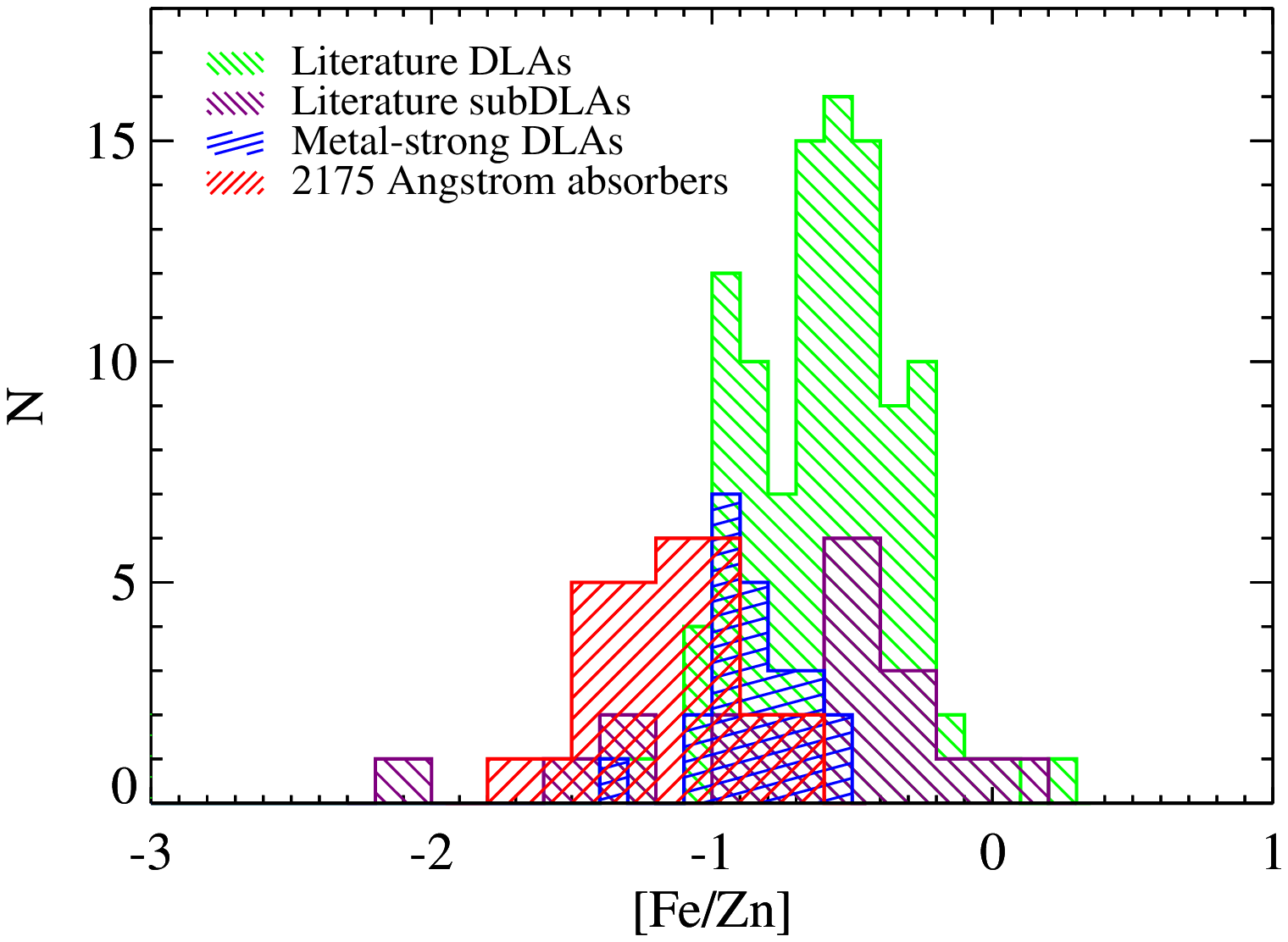}}  
\caption{Comparison of [Fe/Zn] between the 2175 \AA$ $ absorbers and literature DLAs/subDLAs compiled by \citep{Quiret2016}. Metal-strong DLAs are defined as having logN(Zn\II) $\geq$ 13.15 or logN(Si\II) $\geq$ 15.95 \citep{Herbert-Fort2006}. }
\label{fig:hist}
\end{figure*}

[Fe/Zn] is commonly used as an indicator of the depletion level of heavy elements onto dust grains. We compare the histogram of [Fe/Zn] with that of literature DLAs and subDLAs compiled by \cite{Quiret2016} in Figure \ref{fig:hist}. The distributions of DLAs and subDLAs peak around the same [Fe/Zn] value (0.4 - 0.5) although the distribution of subDLAs has a tail toward lower [Fe/Zn]. The 2DAs, however, peak at much lower [Fe/Zn] values $\sim$ -1.2 than DLAs and subDLAs. On average, the 2DAs show higher depletion level, which supports the existence of dust grains and therefore the 2175 \AA$ $ bump. We also show a histogram of metal-strong DLAs, which are defined as having logN(Zn\II) $\geq$ 13.15 or logN(Si\II) $\geq$ 15.95 \citep{Herbert-Fort2006}.  Metal-strong DLAs peak closest to the 2DAs. To quantitatively determine whether two samples are drawn from the same parent population, we perform the two-sample Kolmogorov-Smirnov (KS) test on the 2DAs and DLAs/subDLAs. The KS tests on the distributions of the 2DAs versus DLAs and versus subDLAs result in $p$-values of 2.8 $\times$ 10$^{-5}$ and 4.4 $\times$ 10$^{-4}$ respectively, allowing us to reject the null hypothesis that the two samples are drawn from the same parent population. The $p$-value obtained for the test on metal-strong DLAs with the 2DAs is 8.3 $\times$ 10$^{-3}$.  Metal-strong DLAs have significant overlap with the 2DAs, i.e., they are more likely to contain a 2175 \AA$ $ bump than normal DLAs. The high depletion level tail of the distribution of subDLAs also overlaps with the 2DAs. The 2DAs may be a mix of metal-strong DLAs and subDLAs. We further examine whether there exists a correlation between the bump parameters and [Fe/Zn] in Section \ref{sec:FeZnbump}.

\begin{figure*}
\centering
{\includegraphics[width=16cm, height=12cm]{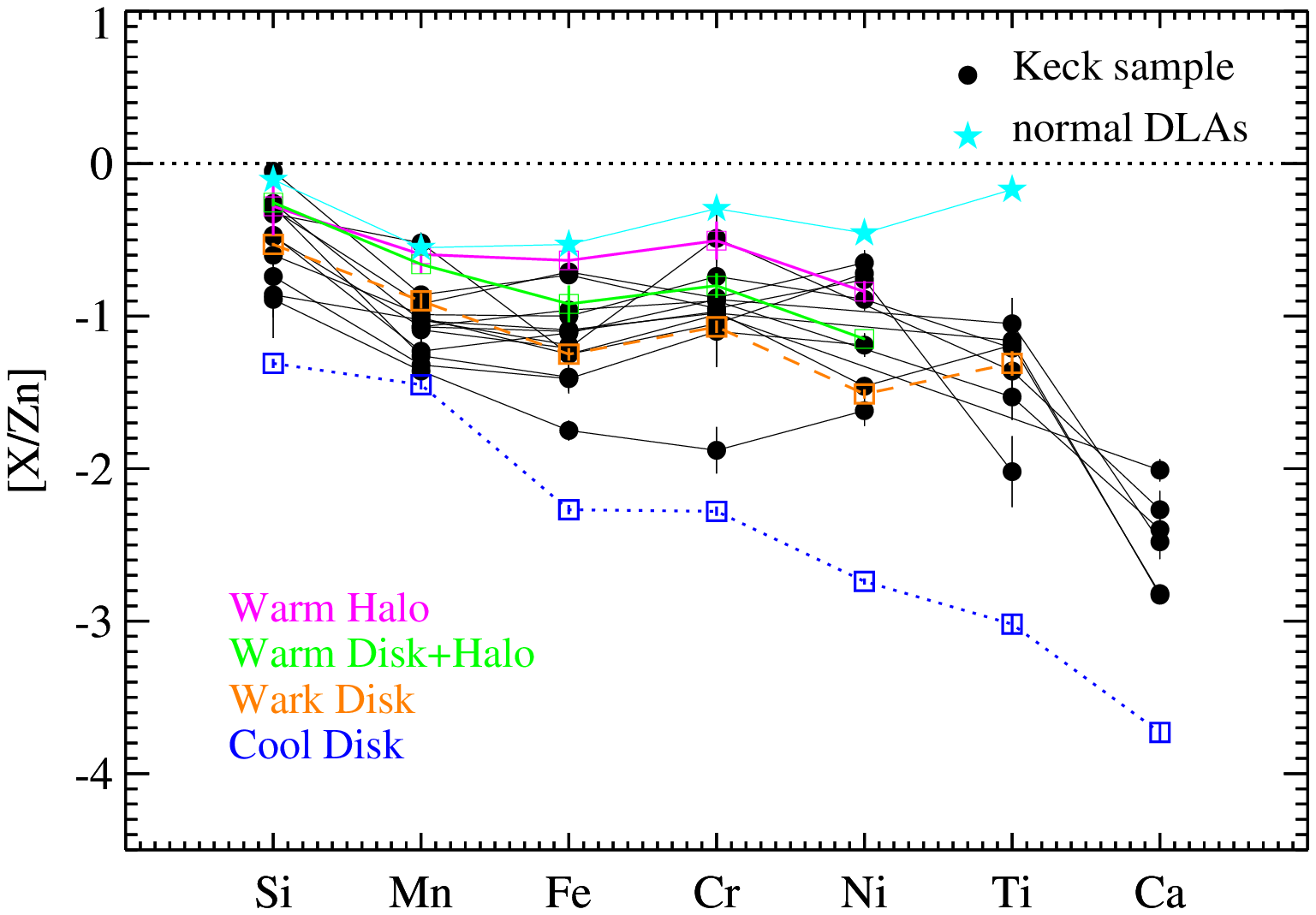}} 
\caption{The depletion patterns, in terms of relative abundances to Zn, of the 2175 \AA$ $ dust absorbers in comparison with the patterns in the MW disk and halo clouds towards $\zeta$ Ophiuchi \citep{Savage1996}. The cyan stars represent the fiducial depletion pattern of normal DLAs. }
\label{fig:depletion}
\end{figure*}

Different elements are locked in dust grains by different amounts, i.e. the depletion pattern, therefore the column densities reflect the chemical composition of the dust. Figure \ref{fig:depletion} shows the depletion patterns of the 2DAs compared with that observed in the Galactic disk and halo clouds \citep{Savage1996}. The four patterns represent the cool disk, warm disk, warm halo blended with disk gas absorption, and warm halo. There is a clear progression toward increasing abundances from the disk to the halo. Refractory elements are more heavily depleted in cool disk clouds than in warm clouds.  Dust destruction in the halo environments is expected to be more severe than in the disk environments, maybe as a result of more frequent and severe shocking of the halo clouds compared with the disk clouds. Therefore, one would expect that dust grains and the 2175 \AA$ $ bump carriers are more readily observable in the disk clouds. The relative abundances of the absorbers, however, exhibit diverse depletion patterns: some resemble the disk pattern and some are more halo-like. It could be due to that some quasar sightlines probe the halo environment while others probe the disk but the velocity components are not well resolved at ESI's resolution, therefore we observe the mixed depletion patterns. Nevertheless, they are all more heavily depleted or contain more dust content compared to the fiducial (average) depletion pattern of normal DLAs in the literature.

\subsection{[Fe/Zn] versus bump strength}
\label{sec:FeZnbump}

\begin{figure}
\centering
{\includegraphics[width=8.6cm, height=6.3cm]{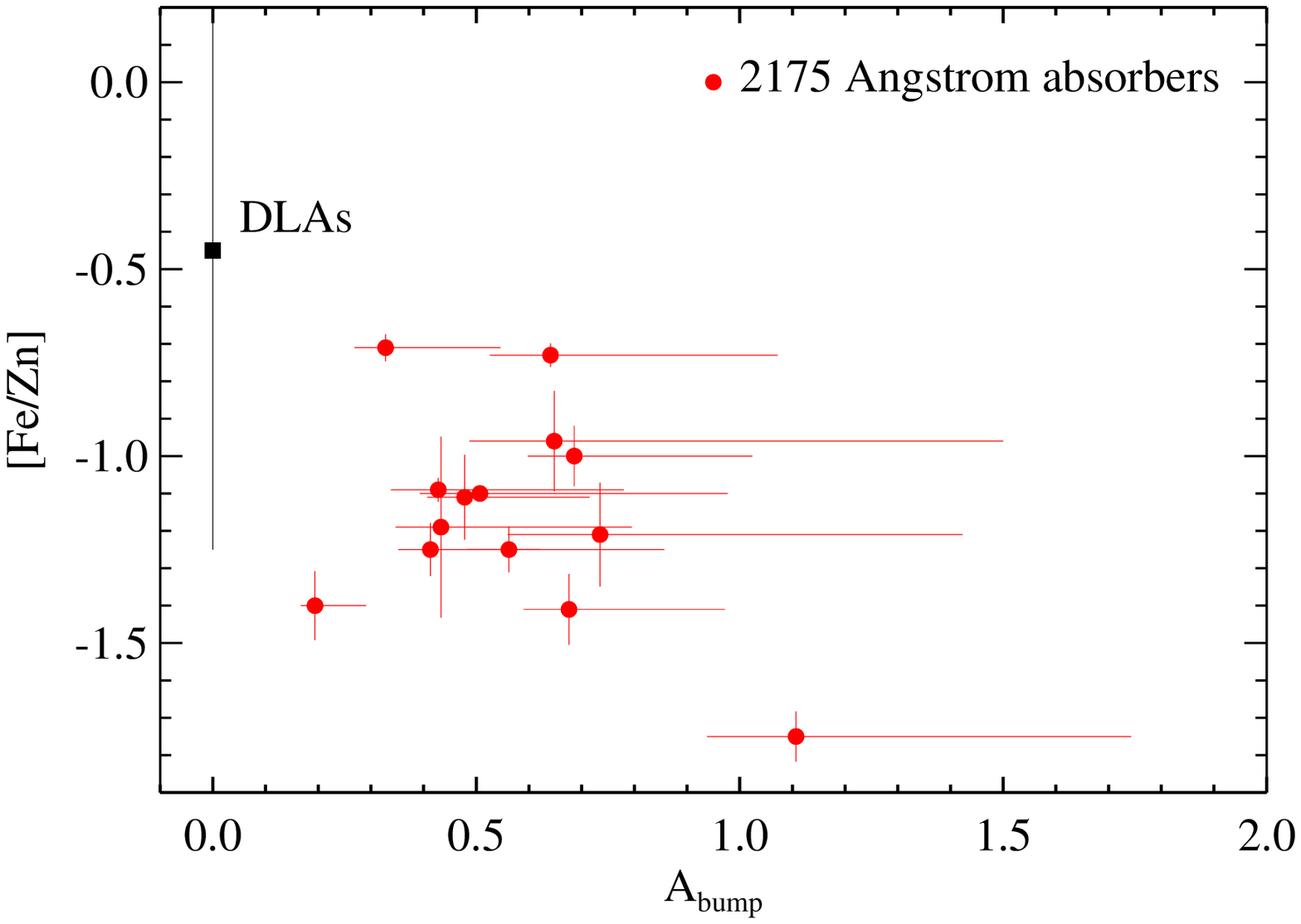}}  
\caption{[Fe/Zn] versus bump strength. The black squares denote the average DLAs with little dust content, i.e. $A_{bump}$ = 0. The data tend to suggest an anti-correlation between [Fe/Zn] and bump strength although with a large scatter.}
\label{fig:FeZn_bump}
\end{figure}

Figure \ref{fig:FeZn_bump} displays [Fe/Zn] versus bump strength (data listed in Table \ref{tab_v90}). Most DLAs in the literature are known to have little dust content, i.e., the bump strength is essentially zero. The dust absorbers tend to have stronger bumps with lower [Fe/Zn] values or higher depletion, which is expected because dust grains form out of metal depletion, although with a large scatter. The anti-correlation extends to reference DLAs with a mean [Fe/Zn] $\sim$ -0.45 and $A_{\rm bump}$ = 0.  A larger sample of 2DAs is required to confirm the observed trend. 

\subsection{$A_{\rm V}$ versus bump strength}

\begin{figure}
\centering
{\includegraphics[width=8.6cm, height=6.3cm]{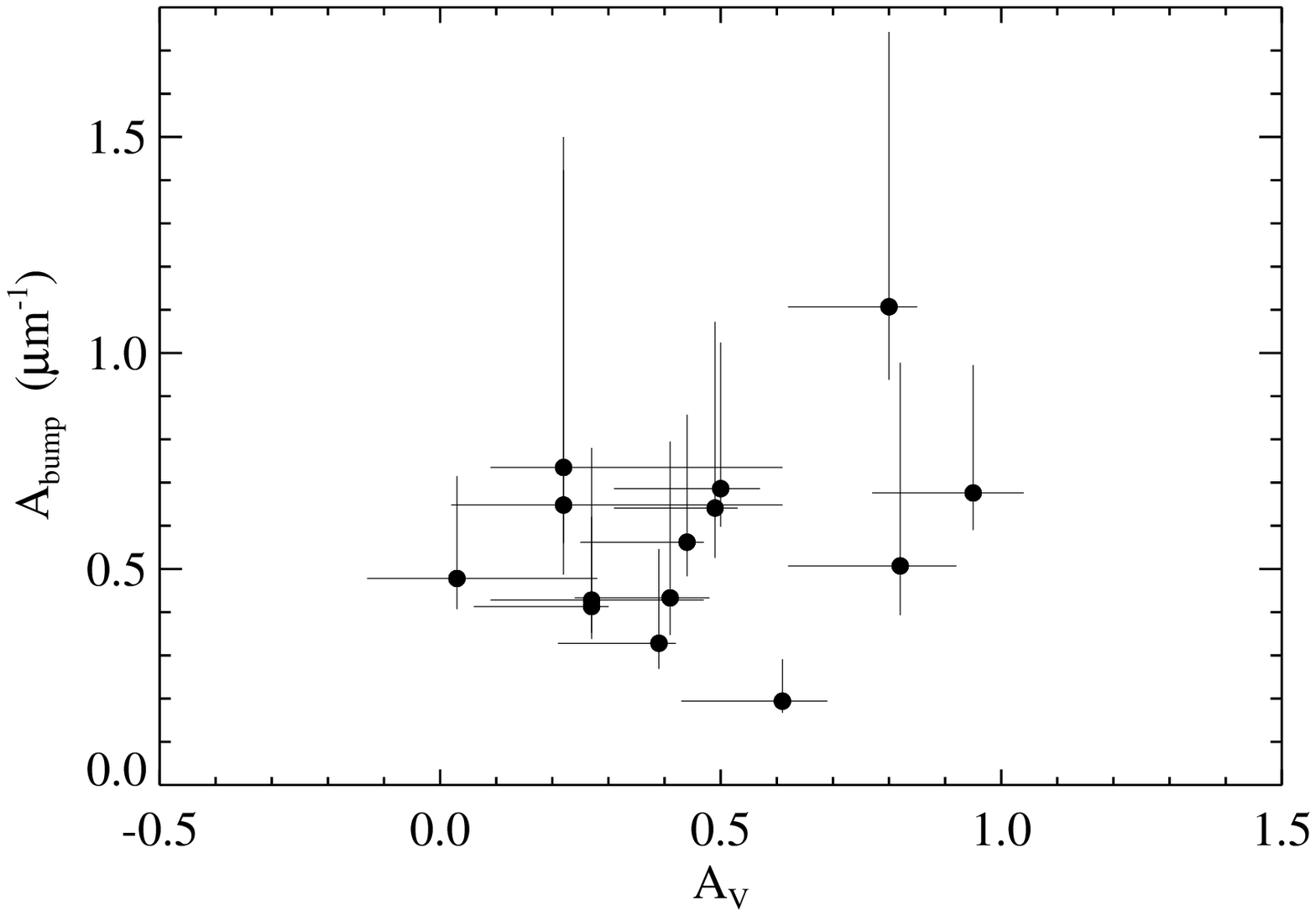}}  
\caption{$A_{\rm V}$ versus bump strength. The data suggest a very weak trend of higher $A_{\rm V}$ and larger $A_{\rm bump}$.}
\label{fig:Av_bump}
\end{figure}

The V-band extinction is a general indicator of how dusty these systems are. We therefore expect a correlation between the bump strength and $A_{\rm V}$ although the bump carriers are made of specific dust grains. Figure \ref{fig:Av_bump} shows the data points in this work but only a very weak trend is observed. Again we caution that due to the intrinsic variations of the individual quasar spectra, the inferred $A_{\rm V}$ can only be taken as an estimate under the assumptions of certain quasar composite templates and normalization at $\lambda$ $\rightarrow$ $+\infty$. A statistical analysis of a large sample of these absorbers' extinction curves would alleviate the uncertainties along individual sight lines.

\subsection{Kinematics}

In addition to the elemental abundances derived from the Keck/ESI spectra, the velocity profiles also provide information on the kinematics of the absorbers. We adopt the definition of the velocity interval $\Delta V_{90}$ by \cite{Prochaska1997} to describe the width of the absorption profile. $\Delta V_{90}$ is defined as the velocity interval that is from 5\% to 95\% of the total integrated optical depth $\tau_{\rm tot}$ = $\int\tau(v)dv$, where $\tau(v)$ is the apparent optical depth. We use Fe\II{} lines to measure $\Delta V_{90}$ as Fe\II{} lines are detected in all the absorbers in our sample and have multiple transitions such that one can choose the unsaturated, unblended ones. We list all the measurements in Table \ref{tab_v90}. We note that ESI spectra would systematically overestimate the velocity width due to the relatively low resolution. The uncertainty in the $\Delta V_{90}$ values for the Keck/ESI data is estimated to be $\sim$ 20 \kms \citep{Prochaska2008}.

\begin{figure*}
\centering
{\includegraphics[width=14cm, height=10cm]{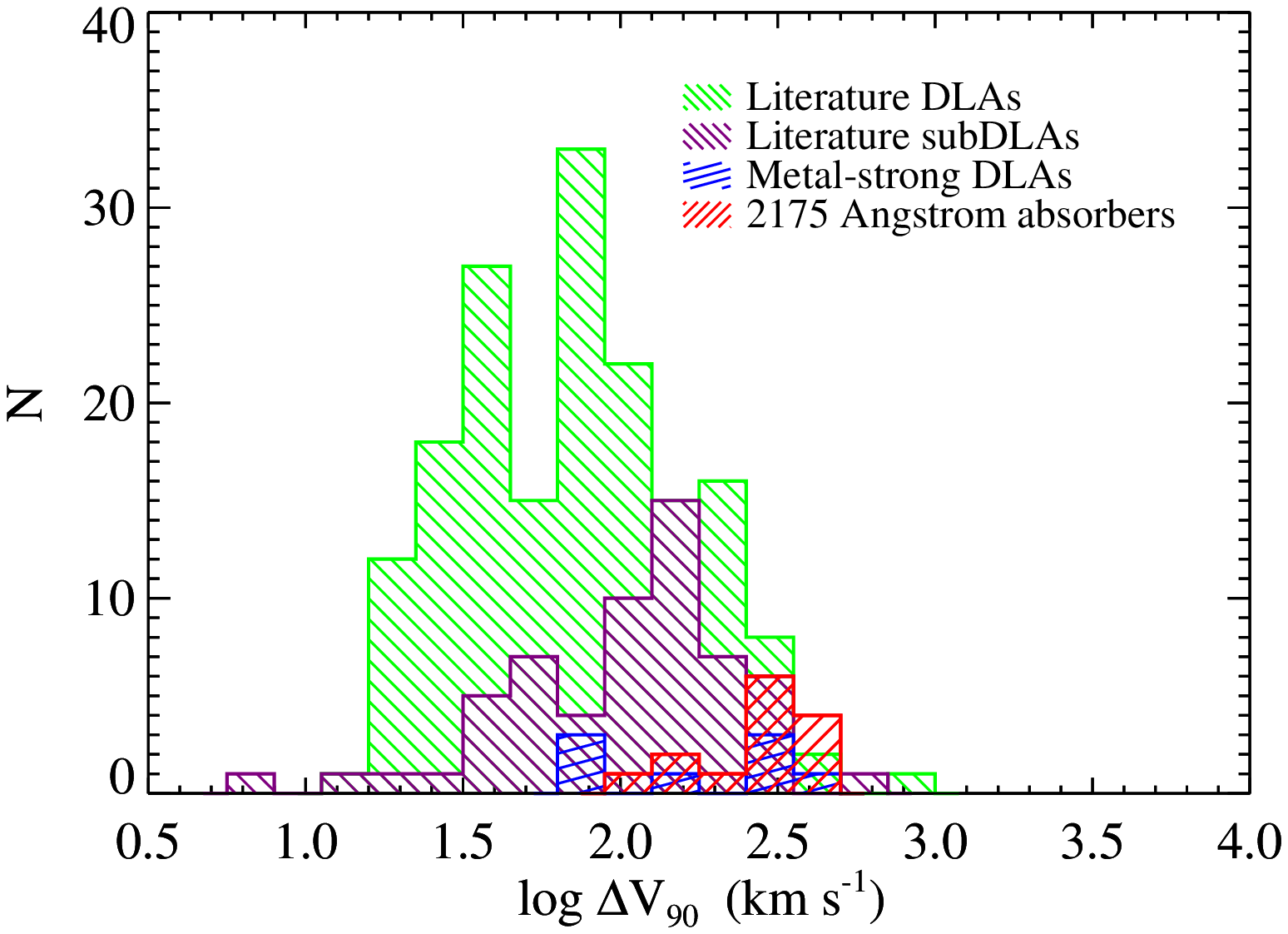}}   
\caption{Comparison of velocity width log$\Delta V_{90}$ between the 2175 \AA$ $ dust absorbers and literature DLAs/subDLAs compiled by \citep{Quiret2016}. Metal-strong DLAs are defined as having logN(Zn\II) $\geq$ 13.15 or logN(Si\II) $\geq$ 15.95 \citep{Herbert-Fort2006}. }
\label{fig:v90hist}
\end{figure*}

Gas kinematics based on the observed $\Delta V_{90}$ have been extensively studied in DLAs and subDLAs (e.g., \citealt{Ledoux2006, Meiring2009, Neeleman2013, Moller2013, Som2015, Quiret2016}). Figure \ref{fig:v90hist} displays a histogram of $\Delta V_{90}$ in logarithmic scale, comparing our measurements to those of literature DLAs and subDLAs. On average, DLAs have the lowest $\Delta V_{90}$ value,  and the 2DAs are the highest in $\Delta V_{90}$ although a small sample.  Most DLAs have $\Delta V_{90}$ values below 200 \kms while there is a tail in the distribution (more emphasized in a linear histogram) at velocity widths significantly above this value. The $\Delta V_{90}$ distribution of the 2DAs lies in the tail of DLAs. We perform the two-sample KS tests on the distributions of 2DAs versus DLAs, subDLAs, and metal-strong DLAs, yielding $p$-values of 3.2 $\times$ 10$^{-7}$, 3.1 $\times$ 10$^{-4}$, and 0.23, respectively. We can again reject the null hypothesis that the 2DAs are drawn from the same parent population as DLAs/subDLAs. The 2DAs are likely drawn from the same population as the metal-strong DLAs, which is consistent with the analysis from the [Fe/Zn] distributions. 

Early simulations suggest that the velocity width $\Delta V_{90}$ reflects the gravitational potential well of the host galaxy responsible for the absorption (e.g., \citealt{Prochaska1997, Haehnelt1998, Pontzen2008}). Similarly, the line profile velocity width has been interpreted as a proxy for mass as the result of the observed velocity-metallicity correlation and an underlying mass-metallicity relation for DLAs' host galaxies (e.g., \citealt{Ledoux2006, Moller2013}). If $\Delta V_{90}$ is a reliable tracer of mass, the 2DAs are expected to be more massive than most DLA and subDLA galaxies. However, this correlation is complicated by the complex gas processes at play and many other factors, therefore a definite conclusion is yet to be drawn \citep{Quiret2016}. Further numerical simulations may shed light upon the distributions of $\Delta V_{90}$ for different populations.

\begin{table}
\centering
\caption{Properties of the 2175 \AA$ $ dust absorbers. The bump strength, $A_{\rm bump}$, is measured using the two-step PyMC method. For the absorber towards J1211+0833, $A_{\rm bump}$ reported here is lower than but consistent (within the error bars) with the measurement from a simultaneous fitting in \citealt{Ma2015}.  The velocity intervals, $\Delta V_{90}$, are based on Fe\II{} lines. The uncertainty on $\Delta V_{90}$ is estimated to be $\sim$ 20 \kms. }
\begin{tabular}{lccc}
\hline\hline
Sourcename & bump strength &  [Fe/Zn] & $\Delta V_{90}$ \\
                      & ($\micron$$^{-1}$)   &       &  (\kms) \\
\hline
J0745+4554 & 0.676 $^{+ 0.296}_{- 0.086}$ &  -1.41 $\pm$ 0.09 & 259.8   \\[0.07cm]
J0850+5159 & 0.507 $^{+ 0.470}_{- 0.114}$ & -1.10 $\pm$ 0.01 & 329.7   \\    [0.07cm]  
J0901+2044 & 0.648 $^{+ 0.852}_{- 0.161}$ &  -0.96 $\pm$ 0.13 &149.9   \\[0.07cm]
J0953+3225 & 0.478 $^{+ 0.237}_{- 0.071}$ & -1.11 $\pm$ 0.11 & 110.0    \\[0.07cm]
J1006+1538 & 0.433 $^{+ 0.362}_{- 0.086}$ & -1.19 $\pm$ 0.24 &490.0   \\[0.07cm]
J1047+3423 & 0.194 $^{+ 0.097}_{- 0.027}$ & -1.40 $\pm$ 0.09 &159.9      \\[0.07cm]
J1127+2424 & 0.413 $^{+ 0.208}_{- 0.061}$ & -1.25 $\pm$ 0.07 & 369.8   \\[0.07cm]
J1130+1850 & 0.641 $^{+ 0.431}_{- 0.115}$ & -0.73 $\pm$ 0.03 & 339.7     \\[0.07cm]
J1138+5245 & 0.735 $^{+ 0.688}_{- 0.175}$ &  -1.21 $\pm$ 0.14 & 339.8  \\[0.07cm]
J1211+0833 & 1.107 $^{+ 0.636}_{- 0.169}$ & -1.75 $\pm$ 0.07 & 190.0 \\[0.07cm]
J1212+2936 & 0.428 $^{+ 0.352}_{- 0.090}$ & -1.09 $\pm$ 0.03 & 269.8      \\ [0.07cm]
J1321+2135 & 0.328 $^{+ 0.218}_{- 0.059}$ & -0.71 $\pm$ 0.04 & 439.8  \\[0.07cm]
J1531+2403 & 0.562 $^{+ 0.295}_{- 0.079}$ & -1.25 $\pm$ 0.06 & 379.6       \\[0.07cm]
J1737+4406 & 0.686 $^{+ 0.338}_{- 0.088}$ & -1.00 $\pm$ 0.08 &  309.7      \\[0.07cm]
\hline
\end{tabular}
\label{tab_v90}
\end{table}

\subsection{2175 \AA$ $ bump in proximate absorbers}

Absorption systems with $z_{qso}$ $\sim$ $z_{abs}$ within $\sim$ 5000 \kms of a background quasar are proximate systems, which are often interpreted to be associated with the quasar host galaxy or its immediate environment (e.g. \citealt{Ellison2010}). The origin of the absorption can be 1) absorption by halo clouds of the quasar host galaxy; 2) absorption by material from a starburst wind in the quasar host galaxy; 3) intrinsic to the black hole region; 4) absorption by external galaxies clustering around the quasar \citep{Lu2007, Ellison2010}. There are two proximate systems in this sample: the absorber toward J1047+3423 ($z_{qso}$ $\gtrsim$ $z_{abs}$) and the absorber toward J1006+1538 ($z_{abs}$ $\gtrsim$ $z_{qso}$).  The error bar on the emission redshift of J1006+1538 does not account for a possible redshift velocity offset. A similar case has been reported in \cite{Pan2017}, which also contains a 2175 \AA$ $ bump. The nature of the system is proposed to be either strong star formation activity in the quasar host or intrinsic to the quasar itself. A controversial question is also raised: whether the bump carriers can survive in harsh quasar environments. Observational evidence exists that small carbonaceous grains such as PAH molecules survive in the close vicinity ($<$ 1 kpc) of an AGN \citep{Tombesi2015} or the innermost region of the torus \citep{Mor2009}. Observational supports for the intrinsic origin for many associated absorbers include their high metallicity close to or above solar (e.g. \citealt{Hamann1997,Ellison2010}), partial coverage, variability, and distance to the quasar.

\section{Conclusions}
\label{sec:conclusion}

We have presented 13 new 2DAs at $z$ $\sim$ 1.0 - 2.2. We have discussed in detail the bump parameters, column density measurements, relative metal abundance, depletion pattern, and kinematics for each absorber, and compared the sample properties with literature DLAs and subDLAs.  The results are summarized in the following. 

\begin{itemize}

\item We perform bump fitting on the SDSS quasar spectra using a composite quasar spectrum reddened by a parametrized extinction curve. The measured 2175 \AA$ $ bumps in this sample resemble the LMC-type bumps and are weaker than most of the MW bumps. 

\item Many low-ionization lines are detected in these absorbers from the Keck/ESI spectra including Fe\II, Zn\II, Si\II, Al\II, Ni\II, Mn\II, Ti\II, Ca\II, and Cr\II. We measure the column densities of elements using both the AODM and VPFIT.  The relative metal abundances (with respect to Zn) reveal the level of different metals being depleted onto dust grains.  Our 2DAs show a diverse depletion pattern: some resemble that of the MW disk clouds, others are more like the MW halo pattern. We compare the dust depletion level indicator, [Fe/Zn], with that of literature DLAs and subDLAs. The 2DAs have significantly lower [Fe/Zn] values or higher depletion than most DLAs and subDLAs, although subDLAs have a low [Fe/Zn] tail in the distribution. The high depletion levels confirm the existence of dust grains and the 2175 \AA$ $ bump. We also find an anti-correlation between [Fe/Zn] and bump strengths,  although with a large scatter. 

\item The velocity profiles derived from the Keck/ESI spectra provide kinematical information of the absorbers. Most absorbers present multiple velocity components and the velocity intervals extend from $\sim$100 to $\sim$ 600 \kms. Compared to literature DLAs and subDLAs, the 2DAs on average have higher $\Delta V_{90}$ values. Invoking the well-established mass-metallicity relation and velocity width as a tracer of stellar mass, the 2DAs are expected to be more massive than the galaxies hosting DLAs and subDLAs. 

\item The two 2DAs towards J1006+1538 and J1047+3423 have the absorption redshifts close to the quasars' emission-line redshifts. These absorbers are likely associated with the quasar host galaxy or its immediate environment. The absorber towards J1006+1538 is even more intriguing as its absorption redshift is larger than its emission redshift, which could be a candidate for an absorbing gas falling into the central engine of the quasar. We still need more observational evidence to reveal the origin of these absorbers.

\end{itemize}

This paper is a pilot sample study of 2DAs with chemical abundance, depletion pattern, and kinematics analyses.  In Paper II, we will present the neutral gas content (H\I{} and C\I) and a correlation analysis of 2DAs including a few sources from the Keck sample.

\section*{Acknowledgments}

We thank Tayyaba Zafar for the constructive and detailed comments which have improved the paper. We thank Peng Jiang for his contribution to the early version of the bump searching algorithm. This work was partially supported by the University of Florida and the UF-UCF SRI program. Jingzhe Ma was funded by the UF alumni fellowship. This work has made use of data obtained by the SDSS-I/II, SDSS-III/BOSS, and Keck.  

Funding for the SDSS and SDSS-II has been provided by the Alfred P. Sloan Foundation, the Participating Institutions, the National Science Foundation, the U.S. Department of Energy, the National Aeronautics and Space Administration, the Japanese Monbukagakusho, the Max Planck Society, and the Higher Education Funding Council for England. The SDSS Web Site is http://www.sdss.org/. The SDSS is managed by the Astrophysical Research Consortium for the Participating Institutions. The Participating Institutions are the American Museum of Natural History, Astrophysical Institute Potsdam, University of Basel, University of Cambridge, Case Western Reserve University, University of Chicago, Drexel University, Fermilab, the Institute for Advanced Study, the Japan Participation Group, Johns Hopkins University, the Joint Institute for Nuclear Astrophysics, the Kavli Institute for Particle Astrophysics and Cosmology, the Korean Scientist Group, the Chinese Academy of Sciences (LAMOST), Los Alamos National Laboratory, the Max-Planck-Institute for Astronomy (MPIA), the Max-Planck-Institute for Astrophysics (MPA), New Mexico State University, Ohio State University, University of Pittsburgh, University of Portsmouth, Princeton University, the United States Naval Observatory, and the University of Washington.

Funding for SDSS-III has been provided by the Alfred P. Sloan Foundation, the Participating Institutions, the National Science Foundation, and the U.S. Department of Energy Office of Science. The SDSS-III web site is http://www.sdss3.org/. SDSS-III is managed by the Astrophysical Research Consortium for the Participating Institutions of the SDSS-III Collaboration including the University of Arizona, the Brazilian Participation Group, Brookhaven National Laboratory, Carnegie Mellon University, University of Florida, the French Participation Group, the German Participation Group, Harvard University, the Instituto de Astrofisica de Canarias, the Michigan State/Notre Dame/JINA Participation Group, Johns Hopkins University, Lawrence Berkeley National Laboratory, Max Planck Institute for Astrophysics, Max Planck Institute for Extraterrestrial Physics, New Mexico State University, New York University, Ohio State University, Pennsylvania State University, University of Portsmouth, Princeton University, the Spanish Participation Group, University of Tokyo, University of Utah, Vanderbilt University, University of Virginia, University of Washington, and Yale University.

The W.M. Keck Observatory is operated as a scientific partnership among the California Institute of Technology, the University of California and the National Aeronautics and Space Administration. The Observatory was made possible by the generous financial support of the W.M. Keck Foundation.

%%%%%%%%%%%%%%%%%%%%%%%%%%%%
%Tables in Appendix
%%%%%%%%%%%%%%%%%%%%%%%%%%%%
%\clearpage
\appendix

\section{Atomic data and solar abundances used in this work}
\label{appendixA}

\begin{table}
\centering
\caption{Atomic Data from \citealt{Morton2003} and \citealt{Jenkins2009}}
\begin{tabular}{lcc}
\hline
\hline
Transition & $\lambda$ (\AA)  &$f$ \\
\hline
Si\IV{} 1393  &      1393.7603  &     0.51300 \\
Si\IV{} 1402  &      1402.7729  &     0.25400 \\
Si\II{} 1526  &      1526.7070  &     0.13300 \\
C\IV{} 1548  &      1548.1949  &     0.19080 \\
C\IV{} 1550  &      1550.7700  &    0.09520 \\
C\I{} 1560    &      1560.3090  &    0.131595 \\
C\I*{} 1560   &      1560.6820  &    0.098658 \\
C\I*{} 1560a   &      1560.7090  &    0.032895 \\
C\I**{} 1561  &      1561.3400  &   0.019722 \\
C\I**{} 1561a  &      1561.3670  &   0.001312 \\
C\I**{} 1561b  &      1561.4380  &   0.110389 \\ 
Fe\II{} 1608  &      1608.4510  &    0.05770 \\
C\I*{} 1656  &       1656.2670   &  0.062068 \\
C\I{} 1656    &       1656.9280 &   0.148832 \\
C\I**{} 1657 &       1657.0080 &   0.111346 \\
C\I*{} 1657  &        1657.3790 &  0.037032 \\
C\I*{} 1657a &       1657.9070 &  0.049481 \\
C\I**{} 1658  &      1658.1210  &  0.037101 \\
Al\II{} 1670   &      1670.7886  &   1.74000\\
Ni\II{} 1709  &      1709.6042  &    0.03240 \\
Ni\II{} 1741  &      1741.5531  &    0.04270 \\
Ni\II{} 1751  &      1751.9156  &    0.02770 \\
Si\II{} 1808  &      1808.0129  &   0.00210 \\
Al\III{} 1854  &      1854.7184  &     0.55900 \\
Al\III{} 1862  &      1862.7910  &     0.27800 \\
Zn\II{} 2026  &      2026.1370  &     0.50100 \\
Cr\II{} 2026  &      2026.2690  &   0.00130 \\
Mg\I{} 2026  &      2026.4768  &     0.11300 \\
Cr\II{} 2056  &      2056.2568  &     0.10300 \\
Cr\II{} 2062  &      2062.2361  &    0.07590 \\
Zn\II{} 2062  &      2062.6604  &     0.24600 \\
Cr\II{} 2066  &      2066.1641  &    0.05120 \\
Fe\II{} 2249  &      2249.8767  &   0.00182 \\
Fe\II{} 2260  &      2260.7805  &   0.00244 \\
Fe\II{} 2344  &      2344.2139  &     0.11400 \\
Fe\II{} 2374  &      2374.4612  &    0.03130 \\
Fe\II{} 2382  &      2382.7651  &     0.32000 \\
Mn\II{} 2576  &      2576.8770  &     0.36100 \\
Fe\II{} 2586  &      2586.6499  &    0.06910 \\
Mn\II{} 2594  &      2594.4990  &     0.28000 \\
Fe\II{} 2600  &      2600.1729  &     0.23900 \\
Mn\II{} 2606  &      2606.4619  &     0.19800 \\
Mg\II{} 2796  &      2796.3552  &     0.61550 \\
Mg\II{} 2803  &      2803.5325  &     0.30580 \\
Mg\I{} 2852  &      2852.9631  &      1.83000 \\
Ti\II{} 3073  &      3073.8633  &     0.12100 \\
Ti\II{} 3242  &      3242.9185  &     0.23200 \\
Ti\II{} 3384  &      3384.7305  &     0.35800 \\
Ca\II{} 3934  &      3934.7749  &     0.62670 \\
Ca\II{} 3969  &      3969.5901  &     0.31160 \\
\hline
\end{tabular}
\label{tab:atomic}
\end{table}

\begin{table}
\centering
\caption{Solar Abundances (photosphere abundances from \citealt{Asplund2009})}
\begin{tabular}{lcc}
\hline
\hline
Element & log $X_{\sun}$&  log $(X/H)_{\sun}$ \\
\hline
H  &      12.0000  &      0.00000  \\
C  &      8.43000  &     -3.57000  \\
O  &      8.69000  &     -3.31000  \\
Mg  &      7.60000  &     -4.40000  \\
Al  &      6.45000  &     -5.55000  \\
Si  &      7.51000  &     -4.49000  \\
S  &      7.12000  &     -4.88000  \\
Cl  &      5.50000  &     -6.50000  \\
Ca  &      6.34000  &     -5.66000  \\
Ti  &      4.95000  &     -7.05000  \\
Cr  &      5.64000  &     -6.36000  \\
Mn  &      5.43000  &     -6.57000  \\
Fe  &      7.50000  &     -4.50000  \\
Ni  &      6.22000  &     -5.78000  \\
Zn  &      4.56000  &     -7.44000  \\
\hline
\end{tabular}
\label{tab:solar}
\end{table}

\section{Equivalent widths and column densities using the AODM}
\label{appendixB}

\begin{table}
\centering
\caption{Column densities: J0745+4554 at $z_{abs}$ =1.8612 }
\begin{tabular}{cccc}
\hline\hline
Line       &  $W_{\rm r}$  (\AA$ $)             & log$N_{\rm AODM}$    & log$N_{\rm adopt}^{\rm AODM}$ \\
\hline
C\IV{} 1548    & 1.2930 $\pm$ 0.0707  & 14.95 $\pm$ 0.01 & 14.98 $\pm$ 0.01\\
C\IV{} 1550    & 1.2122 $\pm$ 0.0650  & 15.02 $\pm$ 0.01 &                 \\
Si\II{} 1526   & 1.2661 $\pm$ 0.0882  & $>$ 15.73        & 15.77 $\pm$ 0.04\\
Si\II{} 1808   & 0.3130 $\pm$ 0.0348  & 15.77 $\pm$ 0.04 &                 \\
Ni\II{} 1709   & 0.1289 $\pm$ 0.0214  & 14.24 $\pm$ 0.06 & 14.24 $\pm$ 0.06\\
Ni\II{} 1741   & 0.0141 $\pm$ 0.0418  & $<$ 15.18        &                 \\
Fe\II{} 1608   & 0.9808 $\pm$ 0.0977  & 15.09 $\pm$ 0.03 & 15.09 $\pm$ 0.09\\
Fe\II{} 1611   & 0.0754 $\pm$ 0.0556  & $<$ 15.50        &                 \\
Fe\II{} 2260   & 0.1224 $\pm$ 0.0267  & 15.09 $\pm$ 0.09 &                 \\
Cr\II{} 2056   & 0.0398 $\pm$ 0.0288  & $<$ 13.90        &       $<$ 13.90          \\
Zn\II{} 2026   & 0.4384 $\pm$ 0.0480  & 13.51 $\pm$ 0.03 & 13.56 $\pm$ 0.03\\ 
Zn\II{} 2062   & 0.5046 $\pm$ 0.0539  & 13.81 $\pm$ 0.04 &                 \\       
Mg\I{} 2852    & 1.7329 $\pm$ 0.0271  & 13.38 $\pm$ 0.03 & 13.38 $\pm$ 0.03\\
Mg\II{} 2796   & 3.5434 $\pm$ 0.0351  & $>$ 14.99        & $>$ 15.22       \\
Mg\II{} 2803   & 3.3944 $\pm$ 0.0335  & $>$ 15.22        & $>$ 15.22       \\
Mn\II{} 2576   & 0.2791 $\pm$ 0.0424  & 13.17 $\pm$ 0.06 & 13.11 $\pm$ 0.05\\
Mn\II{} 2594   & 0.1767 $\pm$ 0.0316  & 13.06 $\pm$ 0.07 &                 \\
Al\II{} 1670   & 1.7889 $\pm$ 0.0488  & $>$ 14.45        & $>$ 14.45       \\
Al\III{} 1854  & 0.4540 $\pm$ 0.0458  & 13.50 $\pm$ 0.04 & 13.50 $\pm$ 0.04\\
Al\III{} 1862  & 0.2401 $\pm$ 0.0482  & 13.51 $\pm$ 0.08 &                 \\
\hline
\end{tabular}
\end{table}

\begin{table}
\centering
\caption{Column densities: J0850+5159 at $z_{abs}$ =1.3269 }
\begin{tabular}{cccc}
\hline\hline
Line       &  $W_{\rm r}$ (\AA$ $)           & log$N_{\rm AODM}$    &  log$N_{\rm adopt}^{\rm AODM}$ \\
\hline
Si\II{} 1808   & 0.8799 $\pm$ 0.0764  & 16.51 $\pm$ 0.02 & 16.51 $\pm$ 0.02\\
Ni\II{} 1741   & 0.4681 $\pm$ 0.0970  & 14.91 $\pm$ 0.05 & 14.95 $\pm$ 0.04\\  
Ni\II{} 1751   & 0.6010 $\pm$ 0.0830  & 15.01 $\pm$ 0.05 &                 \\
Fe\II{} 2249   & 0.4784 $\pm$ 0.0335  & 15.86 $\pm$ 0.02 & 15.88 $\pm$ 0.01\\
Fe\II{} 2260   & 0.6849 $\pm$ 0.0334  & 15.91 $\pm$ 0.02 &                 \\
Cr\II{} 2056   & 0.4101 $\pm$ 0.0241  & 14.14 $\pm$ 0.02 & 14.15 $\pm$ 0.02\\
Cr\II{} 2066   & 0.2587 $\pm$ 0.0195  & 14.19 $\pm$ 0.03 &                 \\
Zn\II{} 2026   & 0.8501 $\pm$ 0.0361  & 13.98 $\pm$ 0.01 & 14.04 $\pm$ 0.01\\
Zn\II{} 2062   & 0.6553 $\pm$ 0.0335  & 14.14 $\pm$ 0.01 &                 \\
Mg\I{} 2852    & 1.9164 $\pm$ 0.0266  & 13.38 $\pm$ 0.01 & 13.38 $\pm$ 0.01\\
Mg\II{} 2796   & 4.6379 $\pm$ 0.0300  & $>$ 15.07        & $>$ 15.24       \\
Mg\II{} 2803   & 4.3918 $\pm$ 0.0284  & $>$ 15.24        &                 \\
Mn\II{} 2576   & 0.8965 $\pm$ 0.0303  & 13.83 $\pm$ 0.01 & 13.84 $\pm$ 0.01\\
Mn\II{} 2594   & 0.7578 $\pm$ 0.0284  & 13.83 $\pm$ 0.01 &                 \\
Mn\II{} 2606   & 0.5990 $\pm$ 0.0279  & 13.85 $\pm$ 0.01 &                 \\
Al\III{} 1854  & 1.1734 $\pm$ 0.0621  & 14.29 $\pm$ 0.01 & 14.24 $\pm$ 0.01\\ 
Al\III{} 1862  & 0.6826 $\pm$ 0.0594  & 14.04 $\pm$ 0.03 &                 \\
Ti\II{} 3073   & 0.2397 $\pm$ 0.0259  & 13.40 $\pm$ 0.04 & 13.24 $\pm$ 0.02\\
Ti\II{} 3242   & 0.3307 $\pm$ 0.0252  & 13.23 $\pm$ 0.03 &                 \\
Ti\II{} 3384   & 0.5046 $\pm$ 0.0311  & 13.22 $\pm$ 0.02 &                 \\
Ca\II{} 3934   & 0.7506 $\pm$ 0.0407  & 13.02 $\pm$ 0.02 & 13.00 $\pm$ 0.02\\ 
Ca\II{} 3969   & 0.2970 $\pm$ 0.0425  & 12.87 $\pm$ 0.06 &                 \\
\hline
\end{tabular}
\end{table}

\begin{table}
\centering
\caption{Column densities: J0901+2044 at $z_{abs}$ =1.0191 }
\begin{tabular}{cccc}
\hline\hline
Line       &  $W_{\rm r}$ (\AA$ $)            & log$N_{\rm AODM}$    & log$N_{\rm adopt}^{\rm AODM}$ \\
\hline
Fe\II{} 2249   & 0.3082 $\pm$ 0.0117  & 15.65 $\pm$ 0.01 & 15.54 $\pm$ 0.01\\     
Fe\II{} 2260   & 0.2308 $\pm$ 0.0139  & 15.38 $\pm$ 0.02 & 	     	       \\
Cr\II{} 2056   & 0.1512 $\pm$ 0.0295  & 13.65 $\pm$ 0.07 & 13.65 $\pm$ 0.05\\
Cr\II{} 2066   & 0.0799 $\pm$ 0.0150  & 13.64 $\pm$ 0.08 & 	     	       \\
Zn\II{} 2026   & 0.3726 $\pm$ 0.0336  & 13.44 $\pm$ 0.03 & 13.16 $\pm$ 0.02\\
Zn\II{} 2062   & 0.0992 $\pm$ 0.0145  & 13.12 $\pm$ 0.02 & 	     	       \\
Mg\I{} 2852    & 0.8088 $\pm$ 0.0127  & 13.00 $\pm$ 0.01 & 13.00 $\pm$ 0.01\\
Mg\II{} 2796   & 2.1952 $\pm$ 0.0109  & $>$ 14.59        & $>$ 14.74       \\
Mg\II{} 2803   & 2.0208 $\pm$ 0.0111  & $>$ 14.74        &                 \\
Mn\II{} 2576   & 0.3214 $\pm$ 0.0144  & 13.27 $\pm$ 0.02 & 13.27 $\pm$ 0.01\\
Mn\II{} 2594   & 0.2844 $\pm$ 0.0147  & 13.30 $\pm$ 0.02 & 	     	       \\
Mn\II{} 2606   & 0.1752 $\pm$ 0.0148  & 13.23 $\pm$ 0.03 &		       \\
Ti\II{} 3242   & $<$ 0.0662           & $<$ 12.92        & 12.33 $\pm$ 0.09\\
Ti\II{} 3384   & 0.0748 $\pm$ 0.0152  & 12.33 $\pm$ 0.09 &		       \\
Ca\II{} 3934   & 0.4930 $\pm$ 0.0165  & 12.85 $\pm$ 0.01 & 12.85 $\pm$ 0.01\\
Ca\II{} 3969   & 0.2668 $\pm$ 0.0164  & 12.84 $\pm$ 0.02 & 	     	       \\
\hline
\end{tabular}
\end{table}

\begin{table}
\centering
\caption{Column densities: J0953+3225 at $z_{abs}$ = 1.2375 }
\begin{tabular}{cccc}
\hline\hline
Line       &  $W_{\rm r}$ (\AA$ $)              & log$N_{\rm AODM}$    &  log$N_{\rm adopt}^{\rm AODM}$ \\
\hline
Si\II{} 1808   & 0.1263 $\pm$ 0.0489  & 15.46 $\pm$ 0.12 & 15.46 $\pm$ 0.12\\
Fe\II{} 2249   & 0.0810 $\pm$ 0.0170  & 15.02 $\pm$ 0.09 & 14.81 $\pm$ 0.08\\ 
Fe\II{} 2260   & 0.0165 $\pm$ 0.0166  & 14.66 $\pm$ 0.14 &                 \\
Zn\II{} 2026   & $<$ 0.1958           & $<$ 13.68        & 12.98 $\pm$ 0.08\\
Zn\II{} 2062   & 0.1607 $\pm$ 0.0332  & 12.98 $\pm$ 0.08 &		       \\
Cr\II{} 2066   & $<$ 0.1511           & $<$ 13.74        & $<$ 13.74       \\ 
Mg\I{} 2852    & 0.4723 $\pm$ 0.0217  & 12.64 $\pm$ 0.02 & 12.64 $\pm$ 0.02\\
Mg\II{} 2796   & 1.5696 $\pm$ 0.0191  & $>$ 14.57        & $>$ 14.57       \\
Mg\II{} 2803   & 1.5661 $\pm$ 0.0191  & $>$ 14.51        & 	               \\
Mn\II{} 2576   & 0.1238 $\pm$ 0.0142  & 12.80 $\pm$ 0.05 & 12.62 $\pm$ 0.06\\
Mn\II{} 2594   & 0.0177 $\pm$ 0.0145  & 12.08 $\pm$ 0.31 &		       \\
Mn\II{} 2606   & 0.0451 $\pm$ 0.0249  & 12.61 $\pm$ 0.22 &		       \\
Al\III{} 1854  & 0.4066 $\pm$ 0.0378  & 13.51 $\pm$ 0.03 & 13.55 $\pm$ 0.02\\	
Al\III{} 1862  & 0.3244 $\pm$ 0.0375  & 13.66 $\pm$ 0.04 & 	     	       \\
Ti\II{} 3384   & 0.0719 $\pm$ 0.0255  & 12.32 $\pm$ 0.15 & 12.32 $\pm$ 0.15\\	
Ca\II{} 3934   & 0.1347 $\pm$ 0.0328  & 12.26 $\pm$ 0.09 & 12.28 $\pm$ 0.08\\
Ca\II{} 3969   & 0.1052 $\pm$ 0.0467  & 12.45 $\pm$ 0.17 & 	     	       \\	
\hline
\end{tabular}
\end{table}

\begin{table}
\centering
\caption{Column densities: J1006+1538 at $z_{abs}$ = 2.2062 }
\begin{tabular}{cccc}
\hline\hline
Line       &  $W_{\rm r}$ (\AA$ $)            & log$N_{\rm AODM}$    & log$N_{\rm adopt}^{\rm AODM}$ \\
\hline
Si\II{} 1526   & 1.3532 $\pm$ 0.0369  & $>$ 15.21       & 15.70 $\pm$ 0.07\\
Si\II{} 1808   & 0.2425 $\pm$ 0.0508  & 15.70 $\pm$ 0.07 &                 \\
Si\IV{} 1393   & 1.7205 $\pm$ 0.0537  & $>$ 14.91        & $>$ 15.00       \\
Si\IV{} 1402   & 1.5519 $\pm$ 0.0654  & $>$ 15.00        & 	    	       \\
Ni\II{} 1741   & $<$ 0.1365           & $<$ 13.62        & $<$ 13.62      \\
Fe\II{} 1608   & 0.2745 $\pm$ 0.0392  & 14.39 $\pm$ 0.05 & 14.34 $\pm$ 0.02\\
Fe\II{} 2344   & 0.8590 $\pm$ 0.0488  & 14.30 $\pm$ 0.02 &		       \\
Fe\II{} 2586   & 1.0064 $\pm$ 0.0780  & 14.47 $\pm$ 0.03 &		       \\
Cr\II{} 2056   & $<$ 0.1776           & $<$ 13.65        & $<$ 13.65       \\
Zn\II{} 2026   & 0.3846 $\pm$ 0.0589  & 13.38 $\pm$ 0.06 & 13.42 $\pm$ 0.05\\
Zn\II{} 2062   & 0.2698 $\pm$ 0.0543  & 13.52 $\pm$ 0.08 & 	     	       \\
Mg\I{} 2852    & 1.1197 $\pm$ 0.0602  & 13.08 $\pm$ 0.02 & 13.08 $\pm$ 0.02\\
Mg\II{} 2796   & 3.7263 $\pm$ 0.0331  & $>$ 14.45        & $>$ 14.74       \\
Mg\II{} 2803   & 3.4960 $\pm$ 0.0516  & $>$ 14.74        & 	   	       \\
Mn\II{} 2576   & $<$ 0.1204           & $<$ 12.98      & $<$ 12.98       \\
Al\II{} 1670   & 1.1464 $\pm$ 0.0497  & 13.63 $\pm$ 0.01 & 13.63 $\pm$ 0.01\\
Al\III{} 1854  & 0.5943 $\pm$ 0.0526  & 13.63 $\pm$ 0.03 & 13.60 $\pm$ 0.03\\
Al\III{} 1862  & 0.1741 $\pm$ 0.0571  & 13.37 $\pm$ 0.12  &                 \\
\hline
\end{tabular}
\end{table}

\begin{table}
\centering
\caption{Column densities: J1047+3423 at $z_{abs}$ = 1.6685 }
\begin{tabular}{cccc}
\hline\hline
Line       &  $W_{\rm r}$ (\AA$ $)             & log$N_{\rm AODM}$    & log$N_{\rm adopt}^{\rm AODM}$ \\
\hline
C\IV{} 1548    & 1.3864 $\pm$ 0.0538  & $>$ 15.55        & $>$ 15.67       \\
C\IV{} 1550    & 1.1594 $\pm$ 0.0506  & $>$ 15.67        &		       \\	
Si\II{} 1526   & 1.0529 $\pm$ 0.1400  & $>$ 15.69        & 15.82 $\pm$ 0.13\\
Si\II{} 1808   & $<$ 0.2153           & $<$ 15.94        &		       \\	
Fe\II{} 1608   & 0.7662 $\pm$ 0.1354  & 15.38 $\pm$ 0.02 & 14.69 $\pm$ 0.02\\
Fe\II{} 2374   & 0.4749 $\pm$ 0.0406  & 14.58 $\pm$ 0.03 & 	     	       \\
Cr\II{} 2056   & $<$ 0.1688           & $<$ 13.54        & $<$ 13.54       \\
Zn\II{} 2026   & 0.5067 $\pm$ 0.0497  & 13.55 $\pm$ 0.03 & 13.51 $\pm$ 0.03\\
Zn\II{} 2062   & 0.1719 $\pm$ 0.0458  & 13.32 $\pm$ 0.10 &		       \\	
Mg\II{} 2796   & 2.2178 $\pm$ 0.0265  & $>$ 14.37        & $>$ 14.37       \\
Mg\II{} 2803   & 2.0243 $\pm$ 0.0285  & $>$ 14.37        &		       \\
Mn\II{} 2594   & 0.0900 $\pm$ 0.0260  & 12.76 $\pm$ 0.12 & 12.76 $\pm$ 0.12\\
Mn\II{} 2606   & $<$ 0.0648           & $<$ 13.21        & 	               \\
Al\II{} 1670   & 1.0526 $\pm$ 0.0809  & $>$ 14.00        & $>$ 14.00       \\
Al\III{} 1854  & 0.4571 $\pm$ 0.0587  & 13.59 $\pm$ 0.04 & 13.64 $\pm$ 0.03\\
Al\III{} 1862  & 0.4176 $\pm$ 0.0570  & 13.80 $\pm$ 0.05 & 	     	       \\
\hline
\end{tabular}
\end{table}

\begin{table}
\centering
\caption{Column densities: J1127+2424 at $z_{abs}$ = 1.2110 }
\begin{tabular}{cccc}
\hline\hline
Line       &  $W_{\rm r}$ (\AA$ $)             & log$N_{\rm AODM}$    & log$N_{\rm adopt}^{\rm AODM}$ \\
\hline
Si\II{} 1808   & $<$ 1.2315           & $<$ 17.05        & $<$ 17.05       \\  
Fe\II{} 2249   & 0.1210 $\pm$ 0.0446  & 15.23 $\pm$ 0.14 & 14.84 $\pm$ 0.01\\  
Fe\II{} 2260   & 0.2162 $\pm$ 0.0419  & 15.33 $\pm$ 0.08 & 	     	       \\	
Fe\II{} 2374   & 0.8516 $\pm$ 0.0317  & 14.84 $\pm$ 0.01 &                 \\
Cr\II{} 2056   & $<$ 0.1055           & $<$ 13.87        & $<$ 13.87       \\
Zn\II{} 2026   & 0.0589 $\pm$ 0.0490  & 12.65 $\pm$ 0.12 & 12.87 $\pm$ 0.07\\
Zn\II{} 2062   & 0.2357 $\pm$ 0.0566  & 13.46 $\pm$ 0.05 & 	     	       \\
Mg\I{} 2852    & 1.1274 $\pm$ 0.0365  & 13.03 $\pm$ 0.01 & 13.03 $\pm$ 0.01\\
Mg\II{} 2796   & 3.5298 $\pm$ 0.0235  & $>$ 14.67        & $>$ 14.87       \\
Mg\II{} 2803   & 3.2420 $\pm$ 0.0244  & $>$ 14.87        &                 \\
Mn\II{} 2576   & 0.2176 $\pm$ 0.0276  & 13.04 $\pm$ 0.05 & 13.01 $\pm$ 0.04\\
Mn\II{} 2594   & 0.0972 $\pm$ 0.0277  & 12.82 $\pm$ 0.11 &		       \\	
Mn\II{} 2606   & 0.1723 $\pm$ 0.0285  & 13.18 $\pm$ 0.07 &   	       \\
Al\III{} 1854  & 0.4762 $\pm$ 0.0756  & 13.63 $\pm$ 0.05 & 13.64 $\pm$ 0.04\\ 
Al\III{} 1862  & 0.2914 $\pm$ 0.0693  & 13.66 $\pm$ 0.08 & 	     	       \\
Ti\II{} 3242   & $<$ 0.1178           & $<$ 12.70        & $<$ 12.70       \\	
Ti\II{} 3384   & $<$ 0.2363           & $<$ 12.88        &       \\
Ca\II{} 3934   & 0.6369 $\pm$ 0.0329  & 12.91 $\pm$ 0.02 & 12.92 $\pm$ 0.02\\ 
Ca\II{} 3969   & 0.3650 $\pm$ 0.0169  & 12.96 $\pm$ 0.05 & 	     	       \\
\hline
\end{tabular}
\end{table}

\begin{table}
\centering
\caption{Column densities: J1130+1850 at $z_{abs}$ = 2.0119 }
\begin{tabular}{cccc}
\hline\hline
Line       &  $W_{\rm r}$ (\AA$ $)              & log$N_{\rm AODM}$    & log$N_{\rm adopt}^{\rm AODM}$ \\
\hline
C\IV{} 1548    & 0.9257 $\pm$ 0.1271  & $>$ 15.50        & $>$ 15.60       \\
C\IV{} 1550    & 0.6976 $\pm$ 0.1272  & $>$ 15.60        &		       \\ 	
Si\II{} 1526   & 1.2465 $\pm$ 0.1276  & $>$ 15.78        & 16.60 $\pm$ 0.02\\
Si\II{} 1808   & 0.5934 $\pm$ 0.1227  & 16.60 $\pm$ 0.02 &                 \\ 
Ni\II{} 1741   & 0.4309 $\pm$ 0.0942  & 14.71 $\pm$ 0.07 & 14.71 $\pm$ 0.07\\
Fe\II{} 1608   & 0.7238 $\pm$ 0.1534  & 15.92 $\pm$ 0.01 & 15.91 $\pm$ 0.01\\ 
Fe\II{} 2260   & 0.4767 $\pm$ 0.0942  & 15.77 $\pm$ 0.06 & 	     	       \\
Cr\II{} 2056   & 0.1842 $\pm$ 0.0949  & 13.83 $\pm$ 0.16 & 13.83 $\pm$ 0.16\\
Zn\II{} 2026   & 0.5885 $\pm$ 0.0926  & 13.68 $\pm$ 0.04 & 13.70 $\pm$ 0.03\\
Zn\II{} 2062   & 0.3181 $\pm$ 0.0969  & 13.76 $\pm$ 0.06 & 	     	       \\
Mg\I{} 2852    & 1.3051 $\pm$ 0.0799  & 13.26 $\pm$ 0.02 & 13.26 $\pm$ 0.02\\
Mg\II{} 2796   & 2.8173 $\pm$ 0.1142  & $>$ 14.84        & $>$ 14.91       \\
Mg\II{} 2803   & 2.5504 $\pm$ 0.0879  & $>$ 14.91        & 	   	       \\
Mn\II{} 2576   & 0.5492 $\pm$ 0.0747  & 13.72 $\pm$ 0.03 & 13.71 $\pm$ 0.02\\   
Mn\II{} 2594   & 0.6697 $\pm$ 0.0729  & 13.71 $\pm$ 0.04 & 	     	       \\
Mn\II{} 2606   & 0.4150 $\pm$ 0.0923  & 13.67 $\pm$ 0.07 &                 \\	
Al\II{} 1670   & 1.6946 $\pm$ 0.1223  & $>$ 14.47        & $>$ 14.47       \\
Al\III{} 1854  & 0.7705 $\pm$ 0.0899  & 13.87 $\pm$ 0.03 & 13.90 $\pm$ 0.02\\ 
Al\III{} 1862  & 0.5539 $\pm$ 0.0828  & 13.98 $\pm$ 0.04 & 	     	       \\
Ti\II{} 3073   & $<$ 0.4958           & $<$ 13.38        &   $<$ 13.38               \\   
\hline
\end{tabular}
\end{table}

\begin{table}
\centering
\caption{Column densities: J1138+5245 at $z_{abs}$ = 1.1788 }
\begin{tabular}{cccc}
\hline\hline
Line       &  $W_{\rm r}$ (\AA$ $)              & log$N_{\rm AODM}$    & log$N_{\rm adopt}^{\rm AODM}$\\
\hline
Fe\II{} 2249   & $<$ 0.3880           & $<$ 16.69        & 15.88 $\pm$ 0.46\\
Fe\II{} 2260   & $<$ 0.5991           & $<$ 16.33        &		       \\
Fe\II{} 2374   & 1.5323 $\pm$ 0.0749  & $>$ 15.42        &                 \\
Mg\I{} 2852    & 1.2486 $\pm$ 0.0495  & 13.08 $\pm$ 0.01 & 13.08 $\pm$ 0.01\\
Mg\II{} 2796   & 3.9264 $\pm$ 0.0460  & $>$ 15.09        & $>$ 15.29       \\
Mg\II{} 2803   & 3.6808 $\pm$ 0.0462  & $>$ 15.29        &                 \\
Mn\II{} 2576   & 0.4007 $\pm$ 0.0660  & 13.36 $\pm$ 0.06 & 13.36 $\pm$ 0.06\\
Cr\II{} 2056   & $<$ 0.1169           & $<$ 14.13        & $<$ 14.13       \\ 
Zn\II{} 2026   & $<$ 0.4983           & $<$ 14.19        & $<$ 14.19       \\
Zn\II{} 2062   & $<$ 0.6138           & $<$ 14.35        &                 \\  
Ti\II{} 3384   & $<$ 0.2914           & $<$ 13.12        & $<$ 13.12       \\
Ca\II{} 3934   & 0.8854 $\pm$ 0.0628  & 13.07 $\pm$ 0.03 & 13.09 $\pm$ 0.03\\
Ca\II{} 3969   & 0.5906 $\pm$ 0.0848  & 13.18 $\pm$ 0.06 &		       \\ 
\hline
\end{tabular}
\end{table}

\begin{table}
\centering
\caption{Column densities: J1212+2936 at $z_{abs}$ = 1.2202 }
\begin{tabular}{cccc}
\hline\hline
Line       &  $W_{\rm r}$ (\AA$ $)              & log$N_{\rm AODM}$    &  log$N_{\rm adopt}^{\rm AODM}$\\
\hline
Si\II{} 1808   & 1.3202 $\pm$ 0.1303  & 16.93 $\pm$ 0.01 & 16.93 $\pm$ 0.01\\
Fe\II{} 2249   & 0.3241 $\pm$ 0.0388  & 15.65 $\pm$ 0.05 & 15.63 $\pm$ 0.03\\
Fe\II{} 2260   & 0.3932 $\pm$ 0.0368  & 15.62 $\pm$ 0.03 & 	     	       \\
Cr\II{} 2056   & 0.2164 $\pm$ 0.0365  & 13.82 $\pm$ 0.06 & 13.88 $\pm$ 0.05\\ 
Cr\II{} 2066   & 0.2184 $\pm$ 0.0397  & 14.10 $\pm$ 0.07 & 	     	       \\
Zn\II{} 2026   & 0.7939 $\pm$ 0.0396  & 13.78 $\pm$ 0.01 & 13.78 $\pm$ 0.01\\
Zn\II{} 2062   & 0.4427 $\pm$ 0.0368  & 13.78 $\pm$ 0.02 &		       \\	
Mg\I{} 2852    & 1.6025 $\pm$ 0.0247  & 13.29 $\pm$ 0.01 & 13.29 $\pm$ 0.01\\
Mg\II{} 2796   & 3.4739 $\pm$ 0.0181  & $>$ 14.84        & $>$ 15.17       \\ 
Mg\II{} 2803   & 3.3746 $\pm$ 0.0185  & $>$ 15.17        &                 \\
Mn\II{} 2576   & 0.6563 $\pm$ 0.0251  & 13.58 $\pm$ 0.01 & 13.62 $\pm$ 0.01\\
Mn\II{} 2594   & 0.6919 $\pm$ 0.0254  & 13.68 $\pm$ 0.01 & 	     	       \\
Mn\II{} 2606   & 0.4098 $\pm$ 0.0264  & 13.59 $\pm$ 0.02 &		       \\
Al\III{} 1854  & 1.1201 $\pm$ 0.0625  & 14.01 $\pm$ 0.02 & 14.01 $\pm$ 0.02\\
Al\III{} 1862  & 0.6743 $\pm$ 0.0612  & 14.00 $\pm$ 0.03 &	               \\
Ti\II{} 3242   & 0.2618 $\pm$ 0.0281  & 13.10 $\pm$ 0.04 & 13.01 $\pm$ 0.03\\
Ti\II{} 3384   & 0.3034 $\pm$ 0.0275  & 12.95 $\pm$ 0.04 & 	     	       \\
Ca\II{} 3934   & 0.4790 $\pm$ 0.0325  & 12.79 $\pm$ 0.03 & 12.73 $\pm$ 0.03\\
Ca\II{} 3969   & 0.0946 $\pm$ 0.0339  & 12.41 $\pm$ 0.13 & 	     	       \\
\hline
\end{tabular}
\end{table}

\begin{table}
\centering
\caption{Column densities: J1321+2135 at $z_{abs}$ = 2.1253 }
\begin{tabular}{cccc}
\hline\hline
Line       &  $W_{\rm r}$ (\AA$ $)             & log$N_{\rm AODM}$    & log$N_{\rm adopt}^{\rm AODM}$ \\
\hline
C\IV{} 1548    & 1.5006 $\pm$ 0.0988  & 15.10 $\pm$ 0.01 & 15.11 $\pm$ 0.01\\
C\IV{} 1550    & 1.3038 $\pm$ 0.1027  & 15.17 $\pm$ 0.02 & 	     	       \\
Si\II{} 1526   & 1.4952 $\pm$ 0.0768  & $>$ 15.66        & 16.26 $\pm$ 0.02\\  
Si\II{} 1808   & 0.7437 $\pm$ 0.0587  & 16.26 $\pm$ 0.02 &                 \\
Si\IV{} 1393   & 1.3669 $\pm$ 0.1420  & $>$ 15.21        & $>$ 15.27       \\ 
Si\IV{} 1402   & 0.8311 $\pm$ 0.1447  & $>$ 15.27        & 	   	       \\
Ni\II{} 1709   & $<$ 0.3985           & $<$ 14.87        & 14.62 $\pm$ 0.06\\
Ni\II{} 1741   & 0.3214 $\pm$ 0.0734  & 14.57 $\pm$ 0.08 & 	     	       \\
Ni\II{} 1751   & 0.2895 $\pm$ 0.0700  & 14.71 $\pm$ 0.08 &		       \\ 	
Fe\II{} 1608   & 1.0060 $\pm$ 0.0925  & $>$ 15.82        & 15.84 $\pm$ 0.02\\
Fe\II{} 2249   & $<$ 0.1765           & $<$ 15.86        &                 \\ 
Cr\II{} 2056   & $<$ 0.2764           & $<$ 14.35        & $<$ 14.35       \\ 
Zn\II{} 2026   & 0.4176 $\pm$ 0.0723  & 13.58 $\pm$ 0.04 & 13.61 $\pm$ 0.03\\ 
Zn\II{} 2062   & 0.3437 $\pm$ 0.0634  & 13.69 $\pm$ 0.05 & 	     	       \\
Mg\I{} 2852    & 1.9848 $\pm$ 0.1434  & 13.81 $\pm$ 0.01 & 13.81 $\pm$ 0.01\\
Mg\II{} 2796   & 3.7400 $\pm$ 0.0636  & $>$ 14.85        & $>$ 15.31       \\ 
Mg\II{} 2803   & 3.5613 $\pm$ 0.1332  & $>$ 15.31        & 	               \\
Mn\II{} 2576   & 0.6519 $\pm$ 0.0829  & 13.59 $\pm$ 0.04 & 13.56 $\pm$ 0.03\\
Mn\II{} 2594   & 0.5275 $\pm$ 0.0856  & 13.59 $\pm$ 0.06 &		       \\	
Mn\II{} 2606   & 0.2077 $\pm$ 0.0871  & 13.35 $\pm$ 0.14 &		       \\
Al\II{} 1670   & 1.8668 $\pm$ 0.0527  & $>$ 14.45        & $>$ 14.45       \\ 
Al\III{} 1854  & 0.9629 $\pm$ 0.0757  & 13.97 $\pm$ 0.02 & 13.99 $\pm$ 0.02\\
Al\III{} 1862  & 0.7160 $\pm$ 0.0791  & 14.06 $\pm$ 0.03 &                 \\
\hline
\end{tabular}
\end{table}

\begin{table}
\centering
\caption{Column densities: J1531+2403 at $z_{abs}$ = 2.0022 }
\begin{tabular}{cccc}
\hline\hline
Line       &  $W_{\rm r}$ (\AA$ $)              & log$N_{\rm AODM}$    &  log$N_{\rm adopt}^{\rm AODM}$ \\
\hline
C\IV{} 1548    & 2.4718 $\pm$ 0.0737  & $>$ 15.60        & $>$ 15.76       \\
C\IV{} 1550    & 2.3905 $\pm$ 0.0721  & $>$ 15.76        & 	   	       \\
Si\II{} 1526   & 1.9418 $\pm$ 0.0808  & 15.84 $\pm$ 0.01 & 15.83 $\pm$ 0.01\\
Si\II{} 1808   & 0.1963 $\pm$ 0.0333  & 15.55 $\pm$ 0.08 &                 \\
Si\IV{} 1393   & 2.0632 $\pm$ 0.0792  & $>$ 15.28        & $>$ 15.58       \\
Si\IV{} 1402   & 1.9184 $\pm$ 0.0912  & $>$ 15.58        &                 \\
Ni\II{} 1709   & $<$ 0.2370           & $<$ 14.30        & 14.15 $\pm$ 0.14\\
Ni\II{} 1741   & 0.1186 $\pm$ 0.0531  & 14.15 $\pm$ 0.14 & 	     	       \\
Fe\II{} 1608   & 0.8710 $\pm$ 0.0592  & 14.96 $\pm$ 0.02 & 14.90 $\pm$ 0.01\\
Fe\II{} 2249   & 0.1998 $\pm$ 0.0477  & 15.44 $\pm$ 0.09 &                 \\
Fe\II{} 2260   & $<$ 0.2250           & $<$ 15.66        &		       \\
Fe\II{} 2374   & 0.9172 $\pm$ 0.0597  & 14.86 $\pm$ 0.02 &		       \\
Cr\II{} 2056   & $<$ 0.2062           & $<$ 13.60        & $<$ 13.60       \\
Zn\II{} 2026   & 0.2624 $\pm$ 0.0568  & 13.22 $\pm$ 0.06 & 13.21 $\pm$ 0.06\\
Zn\II{} 2062   & 0.0784 $\pm$ 0.0558  & 13.12 $\pm$ 0.20 &		       \\
Mg\I{} 2852    & 1.4534 $\pm$ 0.0472  & 13.15 $\pm$ 0.01 & 13.15 $\pm$ 0.01\\
Mg\II{} 2796   & 6.0524 $\pm$ 0.0754  & $>$ 15.16        & $>$ 15.31       \\
Mg\II{} 2803   & 5.9045 $\pm$ 0.0593  & $>$ 15.31        &                 \\
Mn\II{} 2576   & $<$ 0.2680           & $<$ 13.36        & $<$ 13.36       \\
Al\II{} 1670   & 2.3483 $\pm$ 0.0625  & $>$ 14.59        & $>$ 14.59       \\
Al\III{} 1854  & 1.1982 $\pm$ 0.0500  & 13.94 $\pm$ 0.01 & 13.94 $\pm$ 0.01\\
Al\III{} 1862  & 0.6412 $\pm$ 0.0515  & 13.93 $\pm$ 0.03 & 	     	       \\
Ti\II{} 3073   & $<$0.2824               & $<$ 13.34        & $<$ 13.34       \\
\hline
\end{tabular}
\end{table}

\begin{table}
\centering
\caption{Column densities: J1737+4406 at $z_{abs}$ = 1.6135 }
\begin{tabular}{cccc}
\hline\hline
Line       &  $W_{\rm r}$ (\AA$ $)              & log$N_{\rm AODM}$    & log$N_{\rm adopt}^{\rm AODM}$ \\
\hline
C\IV{} 1548    & 1.3459 $\pm$ 0.1712  & 15.41 $\pm$ 0.01 & 15.44 $\pm$ 0.01\\
C\IV{} 1550    & 1.0590 $\pm$ 0.1665  & 15.48 $\pm$ 0.01 & 	               \\
Ni\II{} 1709   & 0.1112 $\pm$ 0.0449  & 14.22 $\pm$ 0.14 & 14.31 $\pm$ 0.06\\ 
Ni\II{} 1741   & 0.2022 $\pm$ 0.0349  & 14.33 $\pm$ 0.06 & 	               \\ 
Si\II{} 1808   & 0.3241 $\pm$ 0.0647  & 15.89 $\pm$ 0.06 & 15.89 $\pm$ 0.06\\ 
Fe\II{} 2249   & 0.1787 $\pm$ 0.0685  & 15.47 $\pm$ 0.12 & 15.48 $\pm$ 0.07\\ 
Fe\II{} 2260   & 0.2506 $\pm$ 0.0679  & 15.48 $\pm$ 0.09 & 	     	       \\
Cr\II{} 2056   & 0.2107 $\pm$ 0.0672  & 13.88 $\pm$ 0.10 & 13.88 $\pm$ 0.10\\
Zn\II{} 2026   & 0.4664 $\pm$ 0.0608  & 13.54 $\pm$ 0.04 & 13.54 $\pm$ 0.04\\
Zn\II{} 2062   & 0.2287 $\pm$ 0.0669  & 13.51 $\pm$ 0.08 &                 \\
Mg\I{} 2852    & 0.7069 $\pm$ 0.7069  & 12.92 $\pm$ 0.02 & 12.92 $\pm$ 0.02\\
Mg\II{} 2796   & 2.6606 $\pm$ 0.0607  & $>$ 14.81        & $>$ 15.06       \\ 
Mg\II{} 2803   & 2.4163 $\pm$ 0.0609  & $>$ 15.06        &                 \\
Mn\II{} 2576   & 0.4269 $\pm$ 0.0539  & 13.43 $\pm$ 0.04 & 13.42 $\pm$ 0.03\\ 
Mn\II{} 2594   & 0.3223 $\pm$ 0.0513  & 13.39 $\pm$ 0.05 & 	     	       \\
Mn\II{} 2606   & 0.2644 $\pm$ 0.0503  & 13.45 $\pm$ 0.07 &		       \\
Al\II{} 1670   & 1.1886 $\pm$ 0.0809  & $>$ 14.43        & $>$ 14.43       \\
Al\III{} 1854  & 0.3522 $\pm$ 0.0603  & 13.56 $\pm$ 0.04 & 13.64 $\pm$ 0.03\\
Al\III{} 1862  & 0.5419 $\pm$ 0.0597  & 13.91 $\pm$ 0.04 & 	     	       \\
Ti\II{} 3242   & 0.0644 $\pm$ 0.0362  & 12.53 $\pm$ 0.22 & 12.72 $\pm$ 0.07\\
Ti\II{} 3384   & 0.1873 $\pm$ 0.0352  & 12.76 $\pm$ 0.07 & 	     	       \\
\hline
\end{tabular}
\end{table}

% Don't change these lines
\bsp	% typesetting comment
\label{lastpage}
\end{document}